\documentclass[seceq,preprint]{ptptex}

\usepackage[usenames]{color}
\definecolor{MyRed}{cmyk}{0,1,1,0.15}

\def\al{\alpha}
\def\be{\beta}
\def\ga{\gamma} \def\Ga{\Gamma}
\def\ep{\epsilon}
\def\lam{\lambda}
\def\Lam{\Lambda}
\def\th{\theta}
\def\sig{\sigma}
\def\Th{\Theta}
\def\alp{{\alpha'}}
\def\sump{\sum{}'}

\def\sigp{{\sigma'}}
\def\parmedskipn        {  \par\medskip\noindent  }

\def\calA{{\cal A}}  
  
 \def\calH{{\cal H}} 
  \def\calL{{\cal L}}
\def\calM{{\cal M}}  \def\calO{{\cal O}}
\def\calP{{\cal P}}  
 \def\calT{{\cal T}} \def\calU{{\cal U}}
\def\calV{{\cal V}}  
 
\def\matrixii#1#2#3#4            {  \left(\begin{array}{cc}#1&#2\\#3&#4
                                       \end{array}\right) }
\def\vecii#1#2      {  \left(\begin{array}{c}#1\\#2\end{array}\right)  }
\def\del        {  \partial }
\def\half       {  {1\over 2}  }

\def\abs#1      {  \vert #1 \vert  }
\def\ie         {  {{\it i.e.}\  }    }

\def\comma          {\, ,}
\def\period         {\, .}
\def\lsim      {\lower .65ex \hbox{\ $\stackrel{<}{\sim}$\ } }
\def\gsim      {\lower .65ex \hbox{\ $\stackrel{>}{\sim}$\ } }

\def\ket#1{{| #1 \rangle } }
\def\com#1#2{{ \left[#1, #2\right] } }
\def\acom#1#2{{ \left\{ #1,#2\right\} } }
\def\nn               {  \nonumber  }
\newcommand{\nullify}[1]{}

\def\Ktil{{\tilde{K}}}
\def\Ntil{{\tilde{N}}}
\def\Ltil{\widetilde{L}}

\def\Rtil{\widetilde{R}}

\def\Wtil{\widetilde{W}}

\def\altil{\tilde{\alpha}}

\def\chitil{\tilde{\chi}}

\def\Pitil{\widetilde{\Pi}}

\def\adot{{\dot{a}}}
\def\bdot{{\dot{b}}}

\def\Khat{\hat{K}}

\def\Nhat{\hat{N}}
\def\Rhat{{\hat{R}}}

\def\That{{\hat{T}}}

\def\Pihat{\hat{\Pi}}

\def\zbar{\bar{z}}

\def\gabar{\bar{\gamma}}

\def\Sbar{\bar{S}}
\def\albar{\bar{\alpha}}
\def\pp{{p^+}}

\def\pplusup{p^+}
\def\pminusup{p^-}

\def\Xp{X^+}
\def\Xm{X^-}

\def\piplus#1{\pi^{+#1}}

\def\Pip{\Pi^+}

\def\Pim{\Pi^-}

\def\Pcom#1#2{\left\{#1,#2\right\}_P}
\def\Dcom#1#2{\left\{#1, #2\right\}_D}

\def\deltassp{\delta(\sig -\sigp)}
\def\deltapssp{\delta'(\sig-\sigp)}

\def\tozero{^{(0)}}

\def\sqtwo{\sqrt{2}\, }

\def\ls{\ell_s}

\def\ad{{\rm ad}}
\def\leftad#1{(\overleftarrow{\rm ad}_{#1}) }
\def\binomial#1#2{\left( #1 \atop #2 \right)}

\def\circbox{\hbox{$\scriptscriptstyle\circ$}}
\def\timesbox{\hbox{$\scriptscriptstyle\times$}}
\def\anc{{{\lower 1ex \circbox} \atop {\raise 1.5ex \circbox}}}
\def\ant{ {{\lower 1ex  \timesbox} \atop {\raise 1.5ex  \timesbox}}}
\def\ap{{\raise 0.2ex \hbox{$\scriptstyle \odot$}}}

\def\dotc{{\dot{c}}}

\def\ovsqtwo{{1\over \sqrt{2}}}
\def\bfall{\boldmath\bf }

\def\Xcirc{\overset{\circ}{X}{}}
\def\Qhat{\hat{Q}}
\def\altil{\tilde{\alpha}}
\def\Xchk{\check{X}}
\def\Pichk{\check{\Pi}}

\def\chgmunu{(\mu \leftrightarrow \nu)}
\def\zgw{{|z|>|w|}}
\def\wgz{{|w|>|z|}}

\def\cchk{\check{c}}
\def\bchk{\check{b}}
\def\mbS{\mathbb{S}}
\def\mfR{\mathfrak{R}}
\def\taubar{\bar{\tau}}
\def\Gabar{\bar{\Gamma}}
\def\chiten{\Gabar_{1,9}}

\def\chibar{\bar{\chi}}
\def\Pbar{\bar{P}}

\def\shead#1   { \parmedskipn {\bfall $\Box$\ #1}: \parmedskipn }
\def\boxit#1#2      {  \vbox{\hrule\hbox{ \hskip -4.1pt \vrule\kern3pt 
                     \vbox
                    {  \hsize #1 \strut\kern3pt #2 \kern3pt\strut  }
                       \kern3pt  \vrule} \hrule  } }
\def\centerbox#1#2  {  \mbox{  }\par\bigskip  \hfil \boxit{#1}{#2} \hfil
                       \par\bigskip\noindent }

\notypesetlogo                       
\preprintnumber[4cm]{
UT-Komaba 10-5 \\  TU-872 \\ August, 2010}

\title{
Operator Formulation of  Green-Schwarz Superstring \\  in the Semi-Light-Cone Conformal Gauge %
}

\author{
Yoichi \textsc{Kazama}$^{1,}$\footnote{kazama@hep1.c.u-tokyo.ac.jp}  and Naoto \textsc{Yokoi}$^{2,}$\footnote{nyokoi@tuhep.phys.tohoku.ac.jp}%
}

\inst{
 $^1$ 
 Institute of Physics, University of Tokyo, \\
 Komaba, Meguro-ku, Tokyo 153-8902,  Japan 
 \par\smallskip\noindent
 $^2$ 
Department of Physics, Tohoku University, \\
Sendai 980-8578, Japan 
}

\abst{
In this article we present a comprehensive account of the operator 
formulation of the Green-Schwarz superstring  in the semi-light-cone (SLC) 
gauge, where the worldsheet conformal invariance  is preserved.  
Starting from the basic action, we systematically study the symmetry structure
of the theory in the SLC gauge both in the Lagrangian and the phase space
formulations. After quantizing the theory in the latter formulation we construct 
the quantum Virasoro and the super-Poincar\'e generators and clarify 
the closure properties of these symmetry algebras. Then by making full use of 
this knowledge we will be able to construct the BRST-invariant 
vertex operators which describe the emission and the absorption of the 
massless quanta  and show that they form the appropriate representation 
of the quantum symmetry algebras. 
Furthermore,  we will construct an exact 
quantum similarity transformation  which 
connects the SLC gauge and  the  familiar  light-cone (LC) gauge. As an 
application  BRST-invariant DDF operators in the SLC gauge are obtained 
starting from the corresponding physical oscillators in  the LC gauge. 
}

\begin{document}

\maketitle

\section{Introduction }

With the advent of the idea of the D-branes\cite{Polchinski:1995mt} and the subsequent  discovery of the AdS/CFT correspondence,\cite{Maldacena:1997re, Gubser:1998bc, Witten:1998qj}  the choice of the worldsheet formalism  of the superstring theory  underwent a notable   change.  The Lorentz-covariant Ramond-Neveu-Schwarz (RNS) formalism,\cite{Ramond:1971gb, Neveu:1971rx} 
 \  which had been dominating over the alternative Green-Schwarz (GS) formalism,\cite{Green:1983wt, Green:1983sg} \  was to be used  less 
 frequently. The reason is that in the RNS formalism the spacetime spinors are described  by the composite spin fields which are not  easy to handle 
and hence  this formalism  is not so suitable   for the description of 
 the Ramond-Ramond (RR) bispinor fields characteristically  produced by the D-branes. It was thus  inevitaible that the first formulation of the superstring in the $AdS_5\times S^5$ background with the RR flux\cite{Metsaev:1998it} was made in the GS formalism, where the target space spinor fields are among the basic variables. Despite its  non-covariance at the quantum level, the GS formalism regained its raisons d'\^etre.

It was not long before another scheme containing  fundamental spinor 
variables, called pure spinor (PS) formalism,  was  invented.\cite{Berkovits:2000fe} \ The great advantage of this formalism is  that, although rigorously speaking the super-Poincar\'e covariance is broken by the underlying quantization procedure,  the rules for the computations of the amplitudes  can be made completely covariant.  Moreover, the rules for the multiloop amplitudes look quite similar to those of the bosonic or topological string.\cite{Berkovits:2005bt} \  Consequently they are formally much simpler than those of the RNS formalism and with judicious regularization procedure a number of powerful results have been obtained.\cite{Berkovits:2004px, Berkovits:2006vi} \ Also since the target space spinor variables are built in, as in the GS formalism, it is suitable for the description of a  superstring in  curved spacetimes with RR flux relevant to AdS/CFT. In fact the action in the $AdS_5\times S^5$ background was  written down already in the original work\cite{Berkovits:2000fe} that  introduced  this formalism.

Although it has many advantages as sketched above, the PS formalism is not 
entirely  without shortcomings.  One feature is that the structure of the worldsheet conformal  symmetry is not explicitly seen. The BRST operator does not contain the energy-momentum tensor and  this  presents  a difficulty in constructing  the string field theory based on this formalism.  Correspondingly,  the construction  of the`` $b$-ghost"  is quite  complex. Another problem is that due to the presence of the quadratic pure spinor constraints, the hermiticity  property of the pure spinor variables is peculiar.\cite{Aisaka:2002sd} \ This leads to the difficulty in constructing the D-brane boundary states  together with  their conjugates. In the context of the study of the AdS/CFT correspondence, PS formalism so far has not been particularly useful in actually solving the quantum dynamics  in the relevant curved background.  For instance, the spectrum of the superstring in the plane-wave background, which was obtained exactly in the light-cone gauge GS formalism,\cite{Metsaev:2001bj} \ has not been reproduced in this  formalism. PS formalism needs to be further developed for such purposes.

Let us now go back to the GS formalism and assess some of its features.  
In the past, the study 
and the use  of the GS formalism have  been made overwhelmingly in the
light-cone (LC)  gauge, where half of the light-cone components of the fermions are set to zero and the bosonic light-cone variable $X^+$ is identified with the worldsheet time. These conditions make the Lorentz invariance non-manifest and moreover break the conformal invariance.  With the lack of these symmetries the computations of the amplitudes become less organized  and cumbersome. 
 This is certainly a big disadvantage of this formalism. On the other hand, GS formalism in the LC gauge  deals directly with the physical degrees of freedome and  is powerful in studying  the physical property of the system, such as the spectrum. In this regard, we have already mentioned the celebrated exact  solution of  the spectrum for a string in the plane-wave  background using  the LC gauge,\cite{Metsaev:2001bj} \  which played a crucial  role in the understanding  of the AdS/CFT correspondence in this background.\cite{Berenstein:2002jq}

It should now be mentioned that the LC gauge is not the only gauge in which 
the GS string can be quantized. Although  there is no way to make the Lorentz symmetry manifest in the quantum GS formalism, there exists  a  more symmetric 
gauge in which  the conformal invariance can be 
retained. This is the so-called semi-light-cone (SLC) gauge,\cite{Carlip:1986cy, Carlip:1986cz, Kallosh:1987vq} \ where only the 
 fermionic gauge conditions are imposed to fix the local $\kappa$-symmetry. 
As for the worldsheet reparametrization symmetry, the usual 
conformal gauge condition is adopted so that the Virasoro 
symmetry still remains and  can be  treated by the  BRST formalism.

One of the main issues of the SLC gauge formalism when it was introduced 
was whether the theory suffers from conformal and related anomalies. 
Although early studies\cite{Carlip:1986cy, Carlip:1986cz, Kallosh:1987vq, 
Kallosh:1988wv,Gilbert:1987xp, Diaz:1988vs} claimed  that there is no anomaly, a subsequent work\cite{Kraemmer:1989af} revealed the existence of 
the 10-dimensional Lorentz anomaly. More complete study was made in Ref.~\citen{Bastianelli:1990xn, Porrati:1991ts}, which confirmed the result of Ref.~\citen{Kraemmer:1989af} as well as pointed out that the conformal and the related anomalies  can be cancelled by adding appropriate counter terms to the action and the transformation rules. 

All such studies were made in the path-integral formalism. The first operator 
 formulation in the SLC gauge was attempted in Ref.~\citen{Chu:1990jt}. In this work, to avoid dealing with the second class constraints, the Batalin-Fradkin formalism\cite{Batalin:1986aq, Batalin:1986fm} was adopted,  which makes use of 
graded brackets in the extended phase space with new fields. The BRST and the super-Poincar\'e operators were constructed with quantum modifications  but their structures were quite  complicated. Much more recently, a simpler BRST
 formalism for the GS superstring in the SLC gauge was introduced 
 in Ref.~\citen{Berkovits:2004tw} for the purpose of showing the equivalence of the 
PS and GS formalisms and it was subsequently utilized in some related works.\cite{Aisaka:2005vn, Kunitomo:2007vm, Kazama:2008as} \ 
 Although the work of Ref~\citen{Berkovits:2004tw} contained a number of  important  ideas,  it was not intended for a systematic development of the operator formalism for the GS superstring in the SLC gauge. 

A brief sketch  of the development of the  SLC gauge formulation given  above reveals that despite its long history surprisingly little has been known about 
 its fundamental structures: Among other things, quantum symmetry structure has not
been fully clarified  and no vertex operators have yet been constructed. 
The primary purpose of the present work is to fill this gap 
and lay the foundation of the GS superstring in this important and unique gauge. 

In particular, we will systematically develop the operator formalism, which is best suited for studying the quantum symmetry structure of the theory,  starting from the basic action of the GS superstring. 
We will construct the quantum Virasoro and super-Poincar\'e generators and 
 clarify  the structure of their algebras in full detail for the first time.  This knowledge 
in turn is indispensable for the construction of the BRST-invariant vertex operators. 
We will demonstrate this  by  constructing the vertex operators for the 
massless states of the open superstring in completely explicit manner.  The way 
all the quantum symmetry algebras are realized in the space of these vertex 
operators  is quite intricate and non-trivial. 

Another new result achieved in this  work is the construction of an exact 
quantum similarity transformation that connects the SLC gauge and the LC gauge. 
Among many expected applications of this mapping, in this article we will use it to construct all the so-called DDF operators,\cite{DelGiudice:1971fp} \  which generate the BRST-invariant states in the SLC gauge,  from the  simple physical oscillators in the LC gauge. 

As we wished to clarify the fundamental structures of the theory fully in a self-contained manner, 
this article has become rather  long. Therefore we will now give the outline of our work in some detail so that  the reader can  grasp the 
scope of the manuscript. 

We begin,  in section  2,  by reviewing the action and its  symmetries of the 
type II Green-Schwarz superstring  in the flat $9+1$ dimensional spacetime. 

Then in section  3 we describe the  gauge-fixing procedure for the local 
symmetries. To keep the worldsheet conformal invariance, the reparametrization symmetry is fixed by imposing the conformal gauge condition. 
As this condition is not invariant under the original local $\kappa$-symmetry transformation, one must redefine the $\kappa$-transformation by adding a
judicious compensating reparametrization transformation. Subsequently, we will
fix this modified $\kappa$-symmetry by imposing the semi-light-cone  (SLC) gauge condition. This procedure in turn breaks the global super-Poincar\'e invariance and we must add appropriate compensating 
$\kappa$-transformations to modify 
 the super-Poincar\'e transformations in order to stay in the SLC gauge. 
Finally we  check that after all this process the conformal symmetry is still preserved. 

In section  4, we develop the phase space formulation and quantize the theory. 
We will compare it to the canonical quantization approach and emphasize 
several advantages of the phase space  formulation. One practical point is 
 that  in this formulation the compensating transformations, which are often 
complicated,  are automatically taken care of  by the use of the Dirac brackets.
Another feature  is that the phase space formulation  can be useful  for
quantizing a non-linear system for which the complete classical solutions are 
not available. 

Having completed  all the necessary ground work, 
we will begin the detailed study of the quantum 
operator formulation of the GS superstring in SLC gauge. In section  5, we will 
 first clarify  the structure of the quantum symmetry algebras. Besides being important in its own right, this will be  indispensable for the construction of the 
vertex operators. The characteristic feature of the SLC gauge is that in contrast to the full light-cone gauge the conformal symmetry is retained. The corresponding Virasoro operators are constructed  first at the classical level and 
then at the quantum level. At the quantum level, one needs to add a quantum 
 correction\cite{Berkovits:2004tw} in order to cancel the conformal anomaly. The nilpotent BRST operator is obtained in the usual way with this modification.
 We then discover that  this correction, the origin of which was rather mysterious
 previously,\cite{Berkovits:2004tw} \ naturally shows itself up as one computes the  quantum supersymmetry algebra. The algebra closes only up to a  BRST-exact term, 
 where the BRST operator automatically contains the correct quantum modification. The rest of the super-Poincar\'e algebra turned out to possess similar features. With a judicious quantum correction added to a part of the Lorentz generators, the algebra precisely closes up to BRST-exact terms.

We then proceed, in section  6, to the construction of the vertex operators for the 
 massless excitations. (For simplicity we will consider the type I super-Maxwell sector.)  By definition they must be BRST-invariant and form a correct representation of the quantum super-Poincar\'e algebra established in the previous section, 
 up to BRST-exact expressions. The computations  were quite involved, 
 requiring various non-trivial $\ga$-matrix and spinorial  identities, 
but  we could obtain  the desired vertex operators which satisfy all the 
requirements consistently. 
 
In order to deepen our understanding of the GS superstring in the SLC gauge further, 
 we will study in section 7 its connection to the much-studied 
formulation  in the usual (full)  light-cone gauge. We will be able to do this in the  most direct way, namely by constructing an explicit  quantum similarity transformation which  connects the operators in the two formulations. The basic method used is the one in Ref.~\citen{Aisaka:2004ga}, which was developed to relate Green-Schwarz and an extended version of the pure spinor superstring.  
Although the presence of the extra term in the BRST operator demanded 
additional new ideas and observations, we have obtained the desired similarity 
transformation exactly.  As an application, we have been able to 
construct the 
BRST-invariant DDF operators in the SLC gauge from the physical operators 
 in the light-cone gauge by performing this similarity transformation. 

Finally, section  8 will be devoted to discussions, where we examine 
some problems which are not solved  in this work and indicate future directions. 
 Several appendices are  provided to describe our conventions and supply additional technical details. 
\renewcommand{\thefootnote}{\arabic{footnote}}
\section{Classical action and its symmetries} 
We  begin by 
 reviewing  the  classical action for  the Green-Schwarz superstring 
 and its symmetries before gauge-fixing,\cite{Green:1983wt, Green:1983sg} \  mainly to set up our notations and to make this article self-contained. 
\subsection{Action} 
The action invariant under the super-Poincar\'e transformations  is most easily constructed using the supercoset method.\cite{Henneaux:1984mh} \ It consists of the kinetic part and  the Wess-Zumino (WZ) part and can be expressed compactly 
 in terms of the worldsheet differential forms in the following way:
\begin{align}
S &= S_K + S_{WZ} \comma \\
S_K &= -{T \over 2} \int \Pi^\mu \wedge \ast \Pi_\mu \comma \\
S_{WZ} &  = -T\int \left(\Pi^\mu \wedge \Wtil_\mu +\half W^\mu \wedge \Wtil_\mu
\right) \period
\end{align}
Here, $T=1/2\pi \alp$ is the string tension\footnote{As for the fundamental 
length scale, we will use the string length  $\ls$, related to $T$ by  $\ls = 1/\sqrt{2\pi T}$.}
, and the  1-forms $\Pi^\mu, W^\mu, \Wtil^\mu$ are defined  as 
\begin{align}
\Pi^\mu &=\Pi^\mu_i d\xi^i =  dX^\mu -W^\mu  \comma \\
W^\mu & =  W^{1\mu} + W^{2\mu} \comma \qquad 
\Wtil^\mu =   W^{1\mu} - W^{2\mu} \comma \\
dX^\mu &= \del_i X^\mu d\xi^i \comma  \qquad 
 W^{A\mu} =W^{A\mu}_i d\xi^i =
 i \th^{A\al} \gabar^\mu_{\al\be} \del_i \th^{A\be} d\xi^i
\comma \quad A=1,2\period
\end{align}
$X^\mu$ are the string coordinates,  $\theta^{A\al}$ are the two sets of 16-component real chiral 
spinors\footnote{In this article we specifically deal with the type IIB case. 
Type IIA case can be easily described by adjusting certain signs. }  and  $\xi^i = (t, \sig)$ denote  the worldsheet coordinates.   The convention for the spinors and the $\gamma$-matrices\footnote{For convenience, we use different notations for the $\ga$-matrices with lower and upper  indices, namely,
 $\gabar^\mu_{\al\be}$ and $\ga^{\mu\al\be}$. See appendix A for more details.}
are elaborated  in  appendix A. 
We take the flat target space metric to be $\eta_{\mu\nu} = (-1, +1, \ldots, +1)$ and the signature of the worldsheet metric as $(-,+)$. 
The wedge product $\wedge$ and the Hodge dual ``$\ast$" with respect to the  worldsheet metric $g_{ij}$ are given  for the basic coordinate 1-form $d\xi^i$ as
\begin{align}
d\xi^i \wedge  d\xi^j &= -\ep^{ij} d^2\xi \comma \qquad \ep_{01} \equiv 1= -\ep^{01}\comma \\
\ast d\xi^i &= - \sqrt{-g}\, g^{ik} \ep_{kj} d\xi^j \period
\end{align}
This gives  $d\xi^i \wedge \ast d\xi ^j= \sqrt{-g}\, g^{ij} d^2\xi$. 
With these formulas, the Lagrangian density can be written in terms of components as%
\begin{align}
\calL_K &= -{T \over 2} \sqrt{-g} g^{ij} \Pi^\mu_i \Pi_{\mu j} \comma 
\qquad 
\calL_{WZ} =T\ep^{ij} \left(\Pi^\mu_i \Wtil_{ \mu j} 
+ \half W^\mu_i \Wtil_{\mu j} \right) \period 
\end{align}
It is sometimes useful to note that the WZ term can be written in a slightly different 
form:
\begin{align}
\calL_{WZ} &
=  T \ep^{ij}\left(  \Pi^\mu_i (W^1_{\mu j} -W^2_{\mu j} )
 -  W^{1\mu }_i W^2_{\mu j} \right) \period \label{WZtwo}
\end{align}
This is due to the identity $\half \ep^{ij}W^\mu_i \Wtil_{\mu j} 
 = - \ep^{ij} W^{1\mu}_i  W^2_{\mu j}$. 
\subsection{Symmetries of the action}
The action presented above enjoys four types of symmetries, namely, 
the worldsheet reparametrization invariance, the target space Lorentz invariance, 
the $N=2$ supersymmetry and the $\kappa$ symmetry. Since the first 
two symmetries are obvious, we will review the latter  two. 

The supercoset construction guarantees that the action is invariant under the global supersymmetry transformations 
\begin{align}
\delta_\chi \th^A &= \chi^A\comma  \qquad 
\delta_\chi   X^\mu = \sum_A i \chi^A\gabar^\mu \th^A \comma \label{susytransf}
\end{align}
where the supersymmetry(SUSY) parameters $\chi^{A\al}$ are  constant real chiral spinors\footnote{We reserve the commonly used letter $\ep$ for the $SO(8)$ anti-chiral 
components  of $\chi$, to appear later.}. 
In fact since 
$\Pi^\mu_i$   are  SUSY invariant  the kinetic term is manifestly  invariant even at the Lagrangian level.  However, the Lagrangian for the WZ part 
 is not invariant (because  $W^{A\mu}_i$ are not invariant) and  transforms into a total derivative. As this is important in deriving the Noether current, let us quickly review how this comes about. 

For this purpose, it is convenient to use the form of $\calL_{WZ}$ given in (\ref{WZtwo}). It is easy to see that although $W^{A\mu}_i$ is not an invariant, it transforms into a total derivative as
\begin{align}
\delta_\chi W^{A\mu}_i &= \del_i V^{A\mu} \comma 
\end{align}
where $V^{A\mu} = i\chi^A \gabar \theta^A $. 
Applying this to (\ref{WZtwo}) and rearranging, we readily obtain 
\begin{align}
\delta_\chi \calL_{WZ} &= T\ep^{ij}\del_j \left[\del_i X^\mu (V^1_\mu -V^2_\mu) \right]
 - T\ep^{ij} (W^{1\mu }_i \del_j V^1_\mu -W^{2\mu }_i \del_j V^2_\mu )
\period \label{susyWZ}
\end{align}
The first term  is manifestly a total derivative. To show that the second term 
 is also a total derivative, we need to use  the well-known Fierz identity. 
Define
\begin{align}
A&\equiv \ep^{ij} W^\mu_i \del_j V_\mu
= \ep^{ij} (\theta \gabar^\mu \del_i \theta) (\del_j \theta\gabar_\mu \chi) \comma \\ 
B &\equiv \ep^{ij} (\del_i \theta \gabar^\mu \del_j \theta) (\theta \gabar^\mu \chi) \period 
\end{align}
Contracting the expression $\ep^{ij} 
\th^\al \del_i \th^\be  \del_j \th^\ga \chi^\delta$ with the Fierz identity
\begin{align}
0 &= \gabar^\mu_{\al\be} \gabar_{\mu \ga\delta} + 
 \gabar^\mu_{\be\ga} \gabar_{\mu \al\delta}+
 \gabar^\mu_{\ga\al} \gabar_{\mu \be\delta} \comma 
\end{align}
we get 
$0 = 2A+B$. Using this relation,  the total derivative $\ep^{ij} \del_j 
(W^\mu_i V_\mu)$, which equals  $-B+A$,   becomes $3A$. Hence, we get a non-trivial identity
\begin{align}
\ep^{ij} W^\mu_i \del_j V_\mu &= {1\over 3} \ep^{ij} \del_j (W^\mu_i V_\mu)
\period
\end{align}
Applying this to (\ref{susyWZ}), we  readily obtain 
\begin{align}
\delta_\chi \calL_{WZ} &= \chi^{1\al}  \del_i \Lam^{1i}_\al +  \chi^{2\al}  \del_i \Lam^{2i}_\al  \comma \\
\Lam^{1i}_\al &= 
 -iT \ep^{ij} \left(\Pi^\mu_j +W^{2\mu}_j + {2\over 3} W^{1\mu}_j \right)
 (\gabar_\mu\th^1)_\al  \comma \\
\Lam^{2i}_\al &=
 iT \ep^{ij} \left(\Pi^\mu_j +W^{1\mu}_j+ {2\over 3} W^{2\mu}_j \right) (\gabar_\mu\th^2)_\al \period
\end{align}
From this result, it is easy to get  the conserved SUSY Noether currents as\footnote{
Up to an overall normalization and the interchage of $\th^1$ and $\th^2$,  this agrees with the expression given in Ref.~\citen{Green:1983sg}.}
\begin{align}
j^{1i}_\al &=  4iT (\gabar^\mu \th^1)_\al \left( \sqrt{-g} P^{ij}_- 
\Pi_{\mu j} -{2\over 3}   \ep^{ij} W^1_{\mu j} \right)  \comma 
\label{susycur1}\\
j^{2i}_\al &=   4iT (\gabar^\mu \th^2)_\al \left( \sqrt{-g} P^{ij}_+
\Pi_{\mu j} +{2\over 3}   \ep^{ij} W^2_{\mu j} \right) \period
\label{susycur2}
\end{align}

Another important symmetry is the $\kappa$-symmetry. The 
action is invariant under the 
off-shell symmetry transformations given by 
\begin{align}
\delta_\kappa X^\mu &= i\th^A \gabar^\mu \delta_\kappa \th^A \comma \qquad \delta_\kappa \th^{A\al} = (\ga_i)^{\al\be} \kappa^{A i}_\be \comma 
\label{kappaxth}\\
\delta_\kappa (\sqrt{-g}\, g^{ij}) &= 8i \sqrt{-g}\, 
\left( P^{ki}_+ \del_k \th^1\kappa^{1j} +P^{ki}_- \del_k \th^2 \kappa^{2j}
\right) \period\label{kappag}
\end{align}
Here  $\ga_i \equiv \Pi_i^\mu \ga_\mu$,  $\kappa^i_A$ are    local fermionic parameters,  which are  anti-chiral spinors  in 
spacetime and vectors  on the worldsheet,  and the  projection operators $P^{ij}_\pm$ are given by 
\begin{align}
P^{ij}_\pm &= \half \left( g^{ij} \pm {\ep^{ij} \over \sqrt{-g}} \right) \period
\label{Pij} 
\end{align}
For invariance, the $\kappa$ parameters must satisfy the conditions 
\begin{align}
P^{ij}_+ \kappa^1_j &=0 \comma \qquad P^{ij}_- \kappa^2_j =0 
\comma 
\label{condkappa}
\end{align}
where $\kappa^A_i \equiv  g_{ik} \kappa^{A k}$. 
\section{Gauge-fixing and compensating transformation}
In this section, we perform the gauge-fixing of the local symmetries. We will do this 
 in two steps:  First, we pick  the 
 conformal gauge to fix the reparametrization symmetry.  Since the original $\kappa$-transformation acts on the worldsheet metric as in (\ref{kappag}), we must modify the $\kappa$-transformation rule by adding  an appropriate compensating reparametrization  in order to stay in the conformal gauge.  We will then fix the $\kappa$-symmetry by adopting 
 the so-called semi-light-cone (SLC)  gauge.  As this gauge is not invariant under 
the supersymmetry transformation nor  the Lorentz transformation,  we must 
include suitable compensating $\kappa$-transformations  in these transformations 
 in order to keep  the SLC gauge.  Below we  describe these procedures 
 in some detail. 
\subsection{Conformal gauge fixing }
As we wish to keep the worldsheet conformal invariance,  we will 
choose the conformal gauge,  where  the Weyl-invariant combination 
$\sqrt{-g}\, g^{ij}$ takes the flat form $\eta^{ij}$.  Since the original 
$\kappa$-transformation (\ref{kappag}) changes this value, to remain 
 in the conformal gauge we must 
 make a judicious compensating reparametrization transformation 
$\delta_f \xi^i = f^i(\xi)$ so  that 
\begin{align}
(\delta_\kappa + \delta_f) \sqrt{-g}\, g^{ij} =0  \label{repcompeq}
\end{align}
holds at the conformal gauge point. 
Under the reparametrization transformation, $X^\mu, \th^{A\al}$ 
are  scalars while  $g_{ij}$ and $g^{ij}$ are 
 covariant and  contravariant tensors. So they transform like 
\begin{align}
\delta_f X^\mu &= f^i \del_i X^\mu \comma \qquad 
\delta \th^{A\al} = f^i \del_i \th^{A\al}\comma  \\
\delta_f g_{ij} 
&= \nabla_i f_j + \nabla_j f_i \comma  \qquad 
\delta_f g^{ij} = -(\nabla^i f^j + \nabla^j f^i)  \comma 
\end{align}
where $ f_k \equiv g_{kl} f^l$.  As for $\sqrt{-g}$, it transforms as
$\delta \sqrt{-g} =\half \sqrt{-g}\, g^{kl} \delta g_{kl} 
 = \sqrt{-g}\, (\nabla^k f_k)$. Combining these results, one easily finds that 
at the conformal gauge point, $\sqrt{-g} g^{ij}$ transforms like 
\begin{align}
\left. \delta_f  (\sqrt{-g} g^{ij})\right|_{conf} &= 
- (\del^i f^j + \del^j f^i -\eta^{ij} \del^k f_k )  \period
\end{align}
Now let us denote  the $\kappa$-transform of $\sqrt{-g} g^{ij}$ at the 
conformal gauge point by 
$h^{ij} \equiv \delta_\kappa  \left.(\sqrt{-g} g^{ij})\right|_{conf}$. 
This quantity  is symmetric and traceless. 
Then  (\ref{repcompeq}) reduces to 
\begin{align}
\del^i f^j + \del^j f^i -\eta^{ij} \del_k f^k &= h^{ij}  \period 
\end{align}
Applying $\del_i$ on both sides, we immediately obtain 
$\Box f^j = \del_i h^{ij}$ (where $\Box \equiv \del_i \del^i$) and hence we can solve for $f^j$ as 
\begin{align}
f^j &= \Box^{-1} \del_i h^{ij} \period  \label{fsol}
\end{align}
In the present case, $h^{ij}$ is given by (see (\ref{kappag}))
\begin{align}
h^{ij}  &=  8i 
\left( P^{ki}_+ \del_k \th^1\kappa^{1j} +P^{ki}_- \del_k \th^2 \kappa^{2j}
\right) \comma \label{hij}
\end{align}
where    $P^{ki}_\pm \equiv \half (\eta^{ki} \pm \ep^{ki})$.
The conditions (\ref{condkappa}) on the $\kappa$ parameters $\kappa^{A i}$ 
become
\begin{align}
-\kappa^{1, 0} + \kappa^{1,1} =0 \comma \qquad \kappa^{2,0}+\kappa^{2,1} =0  \period \label{confcondkappa}
\end{align}
Putting (\ref{hij})  into (\ref{fsol}), we get the appropriate 
compensating reparametrization transformation\footnote{In the 
 historic   paper,\cite{Green:1983sg} \ where this was first discussed,  only the case of a very special
$\kappa$ transformation was considered. Namely the authors imposed
the extra conditions $\del_-\kappa^{1j} = \del_+\kappa^{2j}=0$.
In this case, the expression for $f^j$ simplifies to 
$f^j=4i \sum_A \theta^A \kappa^{Aj}$. In general, however, one should 
 not (and need not) impose such {\it dynamical} restrictions on the $\kappa$-parameters.} . 
With the function  $f^i$ obtained in (\ref{fsol}),
 the modified $\kappa$-transformations
for $\th^{A\al}$ and $X^\mu$ are given by 
\begin{align}
\delta_\kappa \th^A &= \delta_\kappa^0 \th^A  + f^i  \del_i \th^A 
\comma \label{cgktheta}
 \\
\delta_\kappa X^\mu &=\sum i\th^A \gabar^\mu \delta^0_\kappa \th^A  + f^i \del_i X^\mu \comma \label{cgkX}
\end{align}
where $\delta_\kappa^0$  denotes the original transformation
(\ref{kappaxth}).  These  transformations leave
 the conformal-gauge-fixed action invariant. 
\subsection{Semi-light-cone gauge fixing}
\subsubsection{SLC gauge condition and the Lagrangian}
Next, let us fix the $\kappa$-symmetry by imposing the 
SLC  gauge conditions given by 
\begin{align}
\gabar^+_{\al\be}  \th^{A\be} &=0  \quad\Leftrightarrow\quad  \theta^{A\adot}=0\comma 
\end{align}
where $\theta^{A\adot}$ denotes the $SO(8)$ anti-chiral components. 
Here and hereafter, we will often make use of the $SO(8)$ decompositions  of  the spinors and the $\ga$-matrices. Our conventions and the properties of the $\ga$-matrices are summarized in  appendix A. 
In this gauge various terms in the action simplify considerably. First,  the only non-vanishing components  of $W^{A\mu}_i$ are  $W^{A-}_i= i\sqrt{2} \theta^{Aa}
\delta_{ab} \del_i \theta^{Ab}$ 
 because $\gabar^\mu_{ab}$ is non-vanishing only for  $\mu=-$.  This immediately  leads to the formulas 
\begin{align}
\Pi^+_i &= \del_i X^+ \comma \quad \Pi_i^I = \del_i X^I 
\comma \quad \Pi^-_i  = \del_i X^- -W^-_i \comma \quad 
W_i^{A\mu} W^B_{\mu j} = 0 \period
\end{align}
The kinetic and the WZ parts of the Lagrangian  become 
\begin{align}
\calL_K &= -{T \over 2} \left[ 2 \del_i X^+\left(\del^i X^- -\sum_A i\th^A \gabar^- \del^i \th^A\right) 
 + \del_i X^I \del^i X^I \right] \comma \label{LK}\\
\calL_{WZ} &= i  T \ep^{ij} \del_iX^+ \sum_A  \eta_A  \th^A \gabar^- \del_j \th^A  \comma  \label{LWZ} \\
\eta_1 &= -\eta_2=1 \period \label{etaA}
\end{align}
Although there are still cubic terms in $\calL$, the equations of motion for all the fields  reduce to ``free field forms", namely  $\del_i \del^i X^\mu =0$, $\del_+\theta^1=0$ and $\del_-\theta^2=0$, because the interaction terms vanish on-shell. 
\subsubsection{ Supersymmetry  in the SLC gauge}
Let us now begin the discussion of the modification of the symmetry transformations 
needed in the SLC gauge.  
First consider the SUSY transformations. As we need to distinguish between the 
 $SO(8)$ chiral and anti-chiral parts, we will split the SUSY parameters as  $\chi^{A\al} = (\eta^{Aa}, \ep^{A\adot})$ and write the SUSY transformations in the form
\begin{align}
\delta_\eta \th^{Aa} &= \eta^{Aa} \comma \qquad 
\delta_\ep  \th^{A\adot} = \ep^{A\adot} \period  \label{etaepSUSY}
\end{align}
Since only the anti-chiral $\ep^{A\adot}$ transformations violate the 
gauge conditions, the compensating $\kappa$-transformations will involve 
 only half of the $\kappa$-parameters. Indeed we will find that it is enough 
 to keep  $\kappa^{A\adot}$ and set $\kappa^{Aa}$ to zero. 

It is useful to note that in such a case we can ignore 
 the reparametrization part of the modified $\kappa$ transformations 
obtained in (\ref{cgktheta}, \ref{cgkX}) and simply use the original 
 $\kappa$ transformations. The reason is as follows: The expression $h^{ij}$ 
given in (\ref{hij}) is made up of the expressions of the form 
\begin{align}
\del_k \theta^{A\al}\kappa^{Ai}_\al &= \del_k \theta^{A a} \kappa^{Ai}_a 
 + \del_k \theta^{A\adot} \kappa^{Ai}_\adot \period  
\label{vanishf}
\end{align}
This vanishes in the SLC gauge  for $\kappa^{Ai}_a=0$ and hence the reparametrization parameter $f^i$ vanishes. 

Thus, the parameter for the compensating $\kappa$-transformation 
 should be  determined by the requirement
\begin{align}
\delta_\ep \th^{A\adot} &\equiv ( \delta^0_\ep + \delta_\kappa)\theta^{A\adot} =  \ep^{A\adot} + (\Pi^\mu_i  \ga_\mu 
\kappa^{A i})^\adot =0 \comma 
\end{align}
where $\delta^0_\ep$ denotes the original transformation (\ref{etaepSUSY}). 
Writing this out explicitly with $\kappa^{Ai}_a$ set to zero, we obtain the equation
\begin{align}
\ep^{A\adot} -\sqrt{2} \Pi^+_i \delta^{\adot\bdot} \kappa_\bdot^{Ai} 
 =0 \period
\end{align}
Recalling that only half of $\kappa^{Ai}_\bdot$ are independent,
as  in (\ref{confcondkappa}),  we can write this as 
\begin{align}
\ep^{1\adot} &= 2\sqrt{2}\, \del_+ X^+ \delta^{\adot\bdot} 
\kappa^{1,0}_\bdot \comma \\
\ep^{2\adot} &= 2\sqrt{2}\, \del_- X^+ \delta^{\adot\bdot} 
\kappa^{2,0}_\bdot \comma 
\end{align}
where we introduced the worldsheet light-cone coordinates and the derivatives as 
$\sig^\pm =\xi^0 \pm \xi^1$ and  $\del_\pm = \half (\del_0 \pm \del_1)$. 
Therefore, as long as $\del_\pm X^+$ do not vanish, we can solve for 
the $\kappa$ parameters:
\begin{align}
\kappa^{1, 0}_\adot  &=\kappa^{1,1}_\adot =  {\delta_{\adot\bdot} \ep^{1 \bdot} 
 \over 2\sqrt{2} \del_+ X^+} \comma \label{kapone} \\
\kappa^{2,0}_\adot &= -\kappa^{2,1}_\adot
 =  {\delta_{\adot\bdot} \ep^{2 \bdot} 
 \over 2\sqrt{2} \del_- X^+} \period \label{kaptwo}
\end{align}
Throughout we assume that the zero-mode part of $T\del_0 X^+$, {\it i.e.} the momentum $p^+$, is non-vanishing and define the operators 
 such as $1/\del_+ X^+$ by expanding around $p^+$.

We can now write down the $\ep$-SUSY transformations  for the fields  $\theta^{Aa}$ and $X^\mu$, modified by the $\kappa$-transformations. The $\ep$-transformation for  $\theta^{Aa}$ consists solely of the $\kappa$-transformation and is given by  $\delta_\ep \theta^{Aa}= (\Pi^\mu_i \ga_\mu \kappa^{Ai})^a = (\ga^I)^{a\bdot} \del_i X^I \kappa^{Ai}_\bdot$. Substituting 
 (\ref{kapone}) and (\ref{kaptwo}), this becomes 
\begin{align}
\delta_\ep \th^{1a} &=(\ga^I)^{a\bdot} \delta_{\bdot\dotc} {\del_+ X^I \over \sqrt{2} \del_+ X^+}
\ep^{1 \dotc}  \comma \label{epthetaone} \\
\delta_\ep \th^{2a} &=(\ga^I)^{a\bdot} \delta_{\bdot\dotc}
 {\del_- X^I \over \sqrt{2} \del_- X^+}
\ep^{2 \dotc} \period \label{epthetatwo}
\end{align}
As for the $\ep$-transformation of $X^\mu$, it is given by
\begin{align}
\delta_\ep X^\mu &= i \ep^{A\adot} (\gabar^\mu)_{\adot b} \theta^{Ab} 
 + i \theta^{Aa} (\gabar^\mu)_{ab} \delta_\ep  \theta^{Ab} \period
\end{align}
From the property of $\gabar^\mu$, the first term is non-vanihsing 
 only for $\mu=I$, while  the second term exists only for $\mu=-$.  Using the 
 results (\ref{epthetaone}) and (\ref{epthetatwo}),  we get
\begin{align}
\delta_\ep X^I &=  i \ep^{A} \gabar^I  \theta^{A} 
\comma \nn\\
\delta_\ep X^- &= i(\theta^1 \gabar^I \ep^1) {\del_+ X^I \over \del_+ X^+} 
+  i(\theta^2 \gabar^I \ep^2) {\del_- X^I \over \del_- X^+} 
\period
\end{align}

We will end the discussion of the SUSY  in the SLC gauge by giving the 
form of the supercharge densities  in this gauge. 
Before gauge-fixing, they are given as the time-components of the currents 
(\ref{susycur1}) and (\ref{susycur2}) as 
\begin{align}
j^{10}_\al &= 4i T (\gabar^\mu \th^1)_\al \left(\sqrt{-g}\, P^{0j}_- \Pi_{\mu j}
 + {2\over 3} W^1_{\mu 1} \right) \comma \\
j^{20}_\al &= 4i T (\gabar^\mu \th^2)_\al \left(\sqrt{-g}\, P^{0j}_+ \Pi_{\mu j}
 - {2\over 3} W^2_{\mu 1} \right) \period
\end{align}
The expressions in the conformal gauge are obtained by substituting 
$\sqrt{-g} = 1$, $ P^{0j}_- \Pi_{\mu j} = -\half (\Pi_{\mu 0} -\Pi_{\mu 1})$ and $ P^{0j}_+ \Pi_{\mu j} = -\half (\Pi_{\mu 0} +\Pi_{\mu 1})$. 
As we further impose the SLC gauge condition $\gabar^+\th^A =0$,  the following 
 simplifications occur. First, for any vector $A_\mu$ we have 
$(\gabar^\mu \th) A_\mu = (\gabar^-\th)A^+ + (\gabar^I \th) A^I $. 
Next,  $\Pi^\mu_i$ are simplified as 
$\Pi^+_i = \del_i X^+$ and $\Pi^I_i = \del_i X^I$. Finally, one can check the 
useful identity $(\gabar^\mu \th)W^A_{\mu i}  = 0$. Becaue of this, $W^A_{\mu i}$ all disappear and we obtain 
\begin{align}
j^{01}_\al &= -4i T \left( (\gabar^- \th^1)_\al \del_- X^+ + (\gabar^I \th^1)_\al \del_-
X^I \right) \comma \\
j^{02}_\al &= -4i T \left( (\gabar^- \th^2)_\al \del_+ X^+ + (\gabar^I \th^2)_\al \del_+
X^I \right) \period 
\end{align}
For later convenience, let us make  the $SO(8)$  decomposition. 
 Using $(\gabar^-)_{ab} = \sqrt{2}\, \delta_{ab}$, $(\gabar^-)_{\adot\bdot} =0$, we get
\begin{align}
j^{01}_a &= -4\sqrt{2}\, i \th^1_a T \del_- X^+ \comma  \label{j01a}\\
j^{02}_a &=  -4\sqrt{2}\, i \th^2_a T \del_+ X^+  \comma \\
j^{01}_\adot &= -4i (\gabar^I)_{\adot b} \th^{1b} T \del_- X^I  \comma \\
j^{02}_\adot &= -4i (\gabar^I)_{\adot b} \th^{2b} T \del_+ X^I \comma 
\label{j02adot}
\end{align}
where we defined\footnote{Once  decomposed into $SO(8)$ components, the
 spinor  indices can be trivially raised and lowered.}  $\th^A_a \equiv \delta_{ab} \th^{Ab}$. 
We will later quantize them in the phase space formulation and study the quantum 
 algebra of the supercharges. 
\subsubsection{Lorentz symmetry  in the SLC gauge}
The next subject  is  the form of the Lorentz transformations
  in the SLC gauge. 
Before gauge-fixing, the infinitesimal Lorentz transformations for $X^\mu$
and $\theta^{A\al}$ are given by 
\begin{align}
\delta  X^\mu &= \half \xi _{\rho\sigma} 
\left( \eta^{\mu \rho}X^\sigma 
-\eta^{\mu \sigma} X^\rho \right)  \comma \label{LorX}\\
\delta  \th^{A\al}  &= {1\over 4} \xi_{\rho\sigma} (\ga^{\rho\sigma})^\al{}_\be
\th^{A\be} \comma \label{Lortheta}
\end{align}
where $\xi_{\rho\sig}$ are  infinitesimal antisymmetric 
parameters and $\ga^{\rho\sig}\equiv \half (\ga^\rho\gabar^\sig -\ga^\sig\gabar^\rho)$.  The SLC gauge conditions are broken in general  by the Lorentz  transformations, which act on $\theta^{A\adot}$ as  $\delta \theta^{A\adot} = (1/4)\xi_{\rho\sig} (\ga^{\rho\sig})^\adot{}_b \theta^{Ab} \ne 0$.  
 From the $SO(8)$ structure of 
the $\ga$-matrices discussed in appendix A, it is not difficult to see that $ (\ga^{\rho\sig})^\adot{}_b $ is non-vanishing only for $(\rho\sig) = (I-)$ (and of course $(-I)$). 
Thus for such transformations, we must add  compensating $\kappa$ transformations
 in order to stay in  the SLC gauge.  It turns out that, just as in the case of the supersymmetry, 
one can find such transformations using  $\kappa^{Ai}_\bdot$ only,  with 
  $\kappa^{Ai}_b$ set to zero.  Therefore, as explained around (\ref{vanishf}),  the modification  of the $\kappa$-transformation due to reparametrization can be
ignored. Hence 
the condition to fix the compensating $\kappa$ parameters takes the form 
\begin{align}
0&=\delta_{\xi_{I-}} \th^{A\adot} =(\delta_{\xi_{I-}}^0  + \delta_{\xi_{I-}}^\kappa ) \th^{A\adot}
=  \xi_{I-} {1\over 2} (\ga^{I-})^\adot{}_b \th^{Ab} 
+ (\Pi^\mu_i \ga_\mu \kappa^{Ai})^\adot   \nn\\
&= \xi_{I-} {1\over 2} (\ga^{I-})^\adot{}_b \th^{Ab} 
- \Pi^+_i \delta^{\adot\bdot} \kappa^{Ai}_\bdot \period
\end{align}
where $\delta_{\xi_{I-}}^0$ denotes  the original Lorentz transformation with the parameter $\xi_{I-}$. 
 Recalling $\Pi^+_i \kappa^{1i}_\bdot
 = 2 \del_+ X^+ \kappa^{1,0}_\bdot$ and $\Pi^+_i \kappa^{2i}_\bdot
 = 2 \del_- X^+ \kappa^{2,0}_\bdot$, we easily find the solution:
\begin{align}
\kappa^{1,0}_\adot (\xi_{I-}) &= { \delta_{\adot\bdot} \xi_{I-} (\ga^{I-})^\bdot{}_c 
\theta^{1c} \over 4\del_+X^+} \comma  \\
\kappa^{2,0}_\adot (\xi_{I-}) &= { \delta_{\adot\bdot} \xi_{I-} (\ga^{I-})^\bdot{}_c 
\theta^{2c} \over 4\del_-X^+} \period 
\end{align}
Accordingly  we must change the transformation for $\th^{Aa}$ and $X^\mu$ 
 by adding the $\kappa$-transformation with these field-dependent parameters. 

Consider first $\delta_{\xi_{I-}}  \th^{Aa}$. The direct  Lorentz transformation 
$\xi_{I-} {1\over 2} (\ga^{I-})^a{}_\bdot \th^{A\bdot}$ vanishes  since $\th^{A\bdot}=0$. The transformation for $\th^{1a}$ induced by the $\kappa$ transformation is 
\begin{align}
\delta_{\xi_{I-}}^\kappa  \th^{1a} &= (\Pi^\mu_i \ga_\mu \kappa^{1i}(\xi_{I-}))^a 
= 2\del_+ X^J  (\ga^J)^{a\bdot} \kappa_\bdot^{1,0}(\xi_{I-}) \nn\\
&={1\over 2}{\del_+X^J \over \del_+X^+}  (\ga^J)^{a\bdot}   \delta_{\bdot\dotc}  (\ga^{I-})^\dotc{}_d \theta^{1d} \xi_{I-} \period
\end{align}
This can be slightly simplified by noting $(\gabar^+)_{\bdot \dotc} = -\sqtwo \delta_{\bdot\dotc}$ and $\ga^+\gabar^- \th = 2 \th$ ( since $\gabar^+\th =0$).  After 
 the simplification, together with the similar result  for $\th^{2a}$, we obtain  

\begin{align}
\delta_{\xi_{I-}}^\kappa  \th^{1a} &={1\over \sqtwo} {\del_+X^J \over \del_+X^+}
(\ga^J \gabar^I \th^1)^a  \xi_{I-} \comma \\
\delta_{\xi_{I-}}^\kappa  \th^{2a} &={1\over \sqtwo} {\del_-X^J \over \del_-X^+} 
(\ga^J \gabar^I \th^2)^a  \xi_{I-} \period
\end{align}

Next consider the transformation of $X^\mu$. Since the $\kappa$-transformation for 
$X^\mu$ in the SLC gauge is $\delta_\kappa X^\mu
= i \sum_A \th^{Aa} (\gabar^\mu)_{ab} \delta^0_\kappa \th^{Ab}$, only
 the $X^-$ component is affected by this  transformation since $(\gabar^-)_{ab}
=\sqtwo \delta_{ab}$ is the only non-vanishing component.  For the case with the 
parameter $\xi_{I-}$, this induced piece $\delta_{\xi_{I-}}^\kappa X^-$ is  in fact the only 
 contribution to $\delta_{\xi_{I-}} X^-$, because the original Lorentz transformation 
 $\delta_{\xi_{I-}}^0X^-$ vanishes. The explicit expression for $\delta_{\xi_{I-}}^\kappa X^-$ is simplified by using the following relation 
\begin{align}
 \th ^a (\gabar^-)_{ab} (\ga^J)^{b\bdot} \delta_{\bdot \dotc} (\ga^{I-})^\dotc{}_d 
\th^d 
&=-\ovsqtwo  \th \gabar^- \ga^J \gabar^+ \ga^{I-}\th  
 = \ovsqtwo  \th \gabar^J \ga^I \gabar^- \ga^+  \gabar^-  \th \nn\\
& = \sqtwo \th \gabar^J \ga^I \gabar^- \th  = \sqtwo \th \gabar^{JI} 
\gabar^- \th \comma 
\end{align}
where we used the Clifford algebra and the identity $\th \gabar^- \th =0$.  In this way we obtain 
\begin{align}
\delta^\kappa_{\xi_{I-}} X^- &={ i \over \sqtwo  }
\left( {\del_+X^J \over \del_+X^+}\th^1 \gabar^{JI} \gabar^- \th^1
+  {\del_-X^J \over \del_-X^+}\th^2 \gabar^{JI} \gabar^- \th^2 
\right)\xi_{I-} \period
\end{align}
\subsubsection{Conformal symmetry in the SLC gauge}
Finally, let us briefly touch upon  the classical conformal symmetry,  
which is still preserved in the SLC gauge.  This can easily be seen 
if  we express the Lagrangian given in (\ref{LK}) and (\ref{LWZ}) in terms 
 of the worldsheet light-cone coordinates as 
\begin{align}
\calL &= 2T \biggl[ \del_+ X^+ 
\left(\del_- X^- -\sum_A i\th^A \gabar^- \del_-\th^A\right) 
 \nn\\
&+   \del_- X^+ \left(\del_+ X^- - \sum_A i\th^A \gabar^- \del_+\th^A\right) 
 +\del_+ X^I  \del_- X^I \biggr] \nn\\
& +2iT \left[ \del_+ X^+ \sum_A \eta_A \th^A \gabar^-\del_-\th^A
+ \del_- X^+ \sum_A \eta_A \th^A \gabar^-\del_+\th^A \right]
\comma 
\end{align}
where the sign $\eta_A$ is as defined in (\ref{etaA}).  Because of the derivative structure, 
 under the conformal transformation 
 $\sig_\pm \rightarrow f_\pm(\sig_\pm)$, the action is invariant 
 provided that $X^\mu$ and $\th^A$ are considered as conformal scalars. 
The Virasoro algebra  formed  by the conformal transformations 
 will be discussed in detail at the classical and the quantum level 
 in section 5.1. 
\section{Phase space formulation and quantization}
Having discussed the classical symmetries and how some of 
the transformation laws are modified 
 in the SLC gauge,  we now wish to quantize the system and study
  the quantum realization  of these symmetries. In this section 
we will develop  the phase space formulation,  which will be 
 most suited  for that purpose. 
This is mainly because the complicated compensating transformations 
will be automatically taken into account by the use of the Dirac bracket. 
 This will be spelled out  in section 4.2. 
\subsection{Poisson-Dirac brackets   and quantization}
We begin by setting  up the Poisson-Dirac bracket between the basic fields and their conjugates. 
We will denote the momenta conjugate to $(X^+, X^-, X^I)$ as 
$(P^-, P^+, P^I)$. From the Lagrangian given in (\ref{LK}) and (\ref{LWZ}), they are readily obtained as 
\begin{align}
P^+ &= T \del_0 X^+ \comma \\
P^- &= T \bigl[\del_0 X^- 
 -2\sqrt{2} i (\th^1_a\del_+ \th^1_a +\th^2_a\del_-\th^2_a)\bigr] \comma 
\\
P^I &= T \del_0 X^I\period 
\end{align}
As for the fermionic fields, the momenta $p^A_a$ conjugate to $\th^{Aa}$ take the form
\begin{align}
p^A_a &=i\sqrt{2}T (\del_0X^+ -\eta_A \del_1 X^+) \th^A_a 
= i\piplus{A}\th^A_a \comma \label{pA} 
\end{align}
where 
\begin{align}
\piplus{A}&\equiv \sqrt{2} (P^+-\eta_A T\del_1 X^+)
 \period \label{defpipA}
\end{align}
As is well-known, these equations actually give the constraints
\begin{align}
d^A_a &\equiv p^A_a - i\piplus{A}\th^A_a =0 \period
\end{align}
We define the Poisson brackets  as 
\begin{align}
\Pcom{X^I(\sig,t)}{P^J(\sigp,t)} &= \delta^{IJ} \deltassp \comma 
\label{pcomXIPJ} \\
\Pcom{X^\pm(\sig, t)}{P^\mp(\sigp, t)} &= \deltassp \comma
\label{pcomXpmPmp} \\
\Pcom{\th^A_a(\sig,t)}{p^B_b(\sigp,t)} &= -\delta^{AB} \delta_{ab}
\deltassp \comma \\
\mbox{rest} &=0 \period 
\end{align}
Under this bracket, the fermionic constraints $d^A_a$ form the second class algebra
\begin{align}
\Pcom{d^A_a(\sig,t)}{d^B_b(\sigp,t)} &= 2 i \delta^{AB} \delta_{ab} 
\piplus{A}(\sig,t) \deltassp\period
\end{align}
Defining   the Dirac bracket in the standard way, $\th^A$'s  become self-conjugate 
 and satisfy
\begin{align}
\Dcom{\th^A_a(\sig,t)}{\th^B_b(\sigp,t)} &= {i \delta^{AB} \delta_{ab}
\over 2 \piplus{A}(\sig,t)}\deltassp \period
\end{align}
By going to the Dirac bracket,  the relations (\ref{pcomXIPJ}) and 
(\ref{pcomXpmPmp}) conitinue to hold, but the brackets between $(X^-, P^-)$ and $\th^A$, which vanished under Poisson bracket,  become non-trivial\footnote{The bracket $\Dcom{X^-}{P^-}$
 still vanishes due to $\th^A_a \th^A_a =0$.}. One finds
\begin{align}
\Dcom{X^-(\sig,t)}{\th^A_a(\sigp,t)} &= -{1\over \sqrt{2} \piplus{A}(\sig,t)}
 \th^A_a 
\deltassp \comma \\
\Dcom{P^-(\sig,t)}{\th^A_a(\sigp,t)} &= -{1\over \sqrt{2} \piplus{A}(\sigp,t)} \th^A_a (\sigp,t)
\deltapssp \period
\end{align}
However, if we define the combination 
\begin{align}
\Theta^A_a \equiv \sqrt{2\piplus{A}} \th^A_a \comma \label{defTh}
\end{align}
it is not difficult to check that 
they satisfy 
\begin{align}
\Dcom{\Theta^A_a(\sig,t)}{\Theta^B_b(\sigp,t)} &= i\delta^{AB} \delta_{ab}
\deltassp \comma 
\end{align}
and commute with all the other fields. So the fields to be used 
 are $(X^\pm, X^I, P^\pm, P^I, 
\Theta^A_a)$, which satisfy the canonical form of the Dirac  bracket relations. 

For later convenience, we introduce the following {\it dimensionless} fields  $\{A, B, \Pi, \Pitil, S \}$, 
 with  appropriate sub- and super-scripts:
\begin{align}
A &= \sqrt{2\pi T}\, X  \comma \qquad B= \sqrt{2\pi \over T} P
\comma \nn\\
\Pi&= {1\over \sqrt{2}} (B +\del_1 A) \comma \qquad \Pitil 
 = {1\over \sqrt{2}} (B -\del_1 A) \comma  \label{defABetc} \\
S &= i \sqrt{2\pi}\,  \Th   \period  \label{defS}
\end{align}
The non-vanishing Diract brackets  among them are
\begin{align}
\Dcom{A^+(\sig,t)}{B^-(\sigp,t)} &= \Dcom{A^-(\sig,t)}{B^+(\sigp,t)}=2\pi \deltassp \comma \\
\Dcom{A^I(\sig,t)}{B^J(\sigp,t)} &= 2\pi \delta^{IJ} \deltassp \comma \\
\Dcom{S^A_a(\sig,t)}{S^B_b(\sigp,t)} &= {2\pi \over i} \delta^{AB} \delta_{ab} \deltassp \period 
\end{align}

The quantization can be performed in the standard way, namely by replacing the Dirac brackets by 
the (anti-)commutators and $\deltassp \rightarrow i \deltassp$. 
 We will do this  at time $t=0$ and hence drop 
$t$ for all the fields from now on. 
 In particular, the non-vanishing (anti-)commutators among the basic dimensionless fields are then given by 
\begin{align}
\com{A^+(\sig)}{B^-(\sigp)} &= 
\com{A^-(\sig)}{B^+(\sigp)} =2\pi i \deltassp \comma  \\
\com{A^I(\sig)}{B^J(\sigp)} &= 2\pi i \delta^{IJ} \deltassp \comma \\
\acom{S^A_a(\sig)}{S^B_b(\sigp)} &= 2\pi \delta^{AB}\delta_{ab} \deltassp \period
\end{align}
We take the Fourier mode expansions  of $A^\star, B^\star$ ($\star = (\pm, I)$) 
and $S^A_a$ to be 
\begin{align}
A^\star(\sig) &= \sum_n A^\star_n e^{-in\sig} \comma \qquad B^\star(\sig) = 
\sum_n B^\star_n e^{-in\sig} \comma \qquad S^A_a(\sig) = \sum_n S^A_{a,n} e^{-in\sig} \period
\end{align}
Then, these modes satisfy the simple (anti-)commutation relations:
\begin{align}
\com{A^\pm_m}{B^\mp_n} &= i \delta_{m+n,0} \comma \quad 
\com{A^I_m}{B^J_n} = i \delta^{IJ} \delta_{m+n,0} \comma  \\
\acom{S^A_{a,m}}{S^B_{b,n}} &= \delta^{AB}\delta_{ab} \delta_{m+n,0}
\comma \qquad \mbox{rest=0} \period \label{acomSS}
\end{align}
It should be noted that the commutator of the modes of the original fields $X^\mu(\sig)=\sum_n X^\mu_ne^{-in\sig}$ and $P^\nu(\sig)=\sum_n P^\nu_ne^{-in\sig}$ 
has an extra factor of $1/2\pi$ as 
\begin{align}
\com{X^\mu_m}{P^\nu_n} &= {i \over 2\pi} \eta^{\mu\nu} \delta_{m+n,0} 
\period
\end{align}
For this reason, we will often use the following notations for the zero modes, which 
satisfy the canonical commutation relations:
\begin{align}
x^\mu &\equiv X^\mu_0 \comma \qquad  p^\nu \equiv 2\pi P^\nu_0 \comma 
\qquad \com{x^\mu}{p^\nu} = i \eta^{\mu\nu} \period
\end{align}
As for the $\Pi$ and $\Pitil$ fields, we have 
\begin{align}
\com{\Pi^+(\sig)}{\Pi^-(\sigp)} &=  
-\com{\Pitil^+(\sig)}{\Pitil^-(\sigp)} =2\pi i \deltapssp   \comma \nn\\
\com{\Pi^I(\sig)}{\Pi^J(\sigp)} &= -\com{\Pitil^I(\sig)}{\Pitil^J(\sigp)} =
2\pi i \delta^{IJ} \deltapssp \comma \nn\\
\com{\Pi^I(\sig)}{A^J(\sigp)} &= \com{\Pitil^I(\sig)}{A^J(\sigp)}
= -{i \over \sqrt{2}} \delta^{IJ} \deltassp \period
\end{align}
We will write the mode expansion of $\Pi^\mu$ and $\Pitil^\mu$ in the 
following way:
\begin{align}
\Pi^\mu(\sig) &\equiv \sum_n \al^\mu_n e^{-in\sig} \comma \\
\Pitil^\mu(\sig) & \equiv \sum_n \altil^\mu_n e^{-in\sig} \period 
\end{align}
In terms of the modes of $A^\mu$ and $B^\mu$,   the oscillators  $\al^\mu_n$ and 
$\altil^\mu_n$ are given by 
\begin{align}
\al^\pm_n &\equiv {1\over \sqrt{2}}(B^\pm_n -inA^\pm_n) \comma \quad 
\al^I_n \equiv {1\over \sqrt{2}}(B^I_n -inA^I_n) \comma \\
\altil^\pm_n &\equiv {1\over \sqrt{2}}(B^\pm_n +inA^\pm_n) \comma \quad 
\altil^I_n \equiv {1\over \sqrt{2}}(B^I_n +inA^I_n) \comma 
\end{align}
and they satisfy the following commutation relations:
\begin{align}
\com{\al^\pm_m}{\al^\mp_n} &= m \delta_{m+n,0}
\comma \quad \com{\al^I_m}{\al^J_n} = m \delta^{IJ} \delta_{m+n,0} \comma \\
\com{\altil^\pm_m}{\altil^\mp_n} &= -m \delta_{m+n,0}
\comma \quad \com{\altil^I_m}{\altil^J_n} = -m \delta^{IJ} \delta_{m+n,0}
\comma \label{ComPi}
\period
\end{align}
Note that $\al$ and $\altil$ commutators are opposite in sign. 

Now in constructing the quantum symmetry generators in the subsequent section, 
we will have to  specify the 
normal-ordering of operators, in particular of  non-zero modes.  In the present 
 situation,  it is easy to see that the Hamiltonian is quadratic in the modes and the 
appropriate  normal-ordering  is uniquely dictated  by the criterion that the modes with positive (negative) energy should be regarded as 
creation (annihiliation) operators.  Specifically, up to a constant,   our Hamiltonian has the structure
\begin{align}
H &=\ls^2 p^2  + \sum_{n \ge 1} \bigl( \al^+_{-n} \al^-_n + \al^-_{-n} \al^+_n  
+ \al^I_{-n} \al^I_n 
+  \altil^-_n \altil^+_{-n} + \altil^+_n   \altil^-_{-n}
+  \altil^I_n \altil^I_{-n}\nn\\
& + n S^2_{a, -n} S^2_{a, n} 
+n S^1_{a, n} S^1_{a, -n} \bigr)\period \label{Hamilzero}
\end{align}
From the (anti)-commutation relations of the modes, we immediately find that, for $n>0$,  
$\al^\pm_{-n}, \al^I_{-n}, \altil^\pm_n, \altil^I_n, 
S^2_{a,-n}, S^1_{a,n}$ are eigen-operators of $H$ with  positive energy and hence should be regarded  as creation operators. {\it Note that for $\altil^\mu$ and $S^1_a$ 
the moding is opposite to the usual convention}.  For this reason,  hereafter 
 we will introduce the following notations (we will leave $\al^\mu_{-n}$ intact):
\begin{align}
\albar^\mu_{-n}  &\equiv \altil^\mu_n \comma \qquad 
S_{a,-n} \equiv S^2_{a,-n}\comma \qquad \Sbar_{a,-n} \equiv S^1_{a,n}
\period
\end{align}
In this notation,  the modes with  negative mode number are uniformly regarded 
 as creation operators. Then, the properly normal-ordered Hamiltonian 
 is given by 
\begin{align}
H &=\ls^2 p^2  + \sum_{n \ge 1} \bigl( \al^+_{-n} \al^-_n + \al^-_{-n} \al^+_n  
+ \al^I_{-n} \al^I_n 
+  \albar^-_{-n} \albar^+_{n} + \albar^+_{-n}   \albar^-_{n}
+  \albar^I_{-n} \albar^I_{n}\nn\\
& + n S_{a, -n} S_{a, n} 
+n \Sbar_{a,- n} \Sbar_{a, n} \bigr)\period \label{Hamiltonian}
\end{align}
\subsection{Compensating transformation in  the phase space formulation}
In section 3, we gave  a detailed discussion of the gauge-fixing  and the  associated compensating transformations needed for the super-Poincar\'e transformations in the Lagrangian formulation.  
One of the great advantages of the phase space formulation is  that in this formulation  it is not necessary to work out the compensating transformations,  which are 
 often complicated. Conceptually, the reason is very simple: The Dirac bracket is so designed to remove any flow out of the chosen gauge slice. 
Since the role of the compensating transformation is precisely to keep the system on 
the gauge slice, the Dirac bracket automatically fulfills  this role. 

It is instructive to see this  explicitly.  Let $\phi(x)$ be a field and $\pi(x)$ be its conjugate,  with the Poisson bracket relation $\Pcom{\phi(x)}{\pi(y)} = \delta(x-y)$. Suppose that the system is simultaneously invariant under a   gauge transformation   generated by a first class constraint $\Phi_1(\phi,\pi)(x)$ and  under a gauge-invariant global symmetry transformation generated  by $U(\phi,\pi)$. They satisfy $\Pcom{\Phi_1}{U} =0$. Now let us  fix  a gauge by imposing a gauge condition $\Phi_2(\phi, \pi)(x) =0$. This means that we have $\Pcom{\Phi_1(x)}{\Phi_2(y)}
= \ep_{ij} C(x) \delta(x-y) \ne 0$, where $i,j=1,2$ and $\ep_{ij}$ is the antisymmetric $\ep$-symbol with $\ep_{12}=1$.  We are interested in the case where $U$ breaks the gauge condition, that is,  $\delta_\ep \Phi_2(x) \equiv  \Pcom{\Phi_2(x)}{\ep U} \ne 0$, 
 where $\ep$ is an infinitesimal parameter. 
In such a case, to stay in the original gauge, we must   modify the generator $U$ by 
 adding a compensating gauge generator  $\Delta U$, with a judicious gauge parameter function $\al(x)$,  of the form 
\begin{align}
\Delta U &= \int dy \al(y) \Phi_1(y) \period 
\end{align}
The function $\al(x)$ should be  determined by requiring that the total transformation leaves $\Phi_2$ invariant. In other words, 
\begin{align}
(\delta_\ep +\delta^{gauge}_\ep) \Phi_2(x) &=
\ep \Pcom{\Phi_2(x)}{U+\Delta U} 
 =\ep \left(\Pcom{\Phi_2(x)}{U}   - \al(x) C(x)\right)=0 \period
\end{align}
This gives 
\begin{align}
\al(x) &= C^{-1}(x) \Pcom{\Phi_2(x)}{U} \period 
\end{align}
Thus the combined total transformation for an arbitrary field $F(\phi,\pi)(x)$ becomes
\begin{align}
\delta_\ep^{total} F(x) &= \ep \Pcom{F(x)}{U} 
 + \ep \int dy\Pcom{F(x)}{\Phi_1(y)} C^{-1}(y) \Pcom{\Phi_2(y)}{U} 
\period 
\end{align}
Since $\Pcom{\Phi_1}{U}=0$, we can write this  as 
\begin{align}
\delta_\ep^{total} F(x) &= \ep \left[ \Pcom{F(x)}{U} 
 - \int dy\Pcom{F(x)}{\Phi_i(y)} C^{-1}(y)_{ij} \Pcom{\Phi_j(y)}{U} 
\right]\comma 
\end{align}
where  $(C^{-1})_{ij} = -\ep_{ij} C^{-1}$. This however is nothing but the 
 Dirac bracket $\ep \Dcom{F(x)}{U}$.  So we confirm that  the compensating transformation is automatically included  by the use of  the Dirac bracket. 
\subsection{Relation to the canonical quantization scheme} 
Starting from the next section, we will study the structure of the quantum symmetry algebras and construct the vertex operators.  For these purposes, 
it will be convenient to make use of the powerful technique of operator product expansion (OPE). Since this technique is usually based on the canonically quantized fields 
 defined on the complex plane,  it is appropriate here 
to clarify the relation between the 
fields in the phase space formulation and those  in the canonical formulation. 

In  the canonical quantization scheme, one begins by obtaining the general complete 
solutions of  the equations of motion at the classical level. From this dynamical 
information,  one tries to express the {\it time-independent modes} in terms of the fields
 using the completeness relation. Once this is achieved, the quantum commutation 
relations among these modes can be obtained  from the basic 
 quantization rule   for the fields  at equal time. 
This then allows one to compute the commutation relations between quantum fields 
 at unequal  times.  So,  the characteristic feature of this scheme is that 
 the dynamics must be solved completely {\it at the classical level 
 before the quantization}. 

In contradistinction, the phase space quantization scheme does not require the solution 
 of the equations of motion. The quantization of fields and thier conjugates 
is done at a particular fixed time, say $t=0$,  by simply replacing the Poisson-Dirac brackets by the quantum  brackets. To construct the quantum fields at arbitrary time $t$, 
 one needs to carry out the time evolution by the quantum Hamiltonian $H$
as $\phi(t) = e^{iHt} \phi(0) e^{-iHt}$  for every field $\phi(t)$ (with spatial coordinates suppressed).  The field $\phi(t)$ constructed this way 
then  satisfies the equation of motion $\del_t \phi(t) = -i \com{\phi(t)}{H}$. 
So, the characteristic feature of this scheme is that the dynamics must be worked out 
{\it at the  quantum level  after the quantization}. 

The summary  of the foregoing discussion is expressed, for our two dimensional 
case,  by  the  equation 
\begin{align}
\phi_{can} (\sig,t ) &= e^{iHt} \phi_{phase}(\sig, 0) e^{-iHt} \period 
\label{relcanphase}
\end{align}
This formula is completely general and  is particularly  useful for quantizing a non-linear  system for which the complete solutions of the classical equations of motions are difficult  to obtain.  In such a case, the canonical quantization cannot be performed but 
one can still quantize  in  the phase space formulation and then try to evaluate 
the right hand side of (\ref{relcanphase}) by some appropriate means. As a matter of fact, for the computation of  $t$-independent quantities, such as various symmetry charges, 
even that  computation is not necessary and one can obtain quantum information
 by using the  phase space fields quantized at a common fixed time\footnote{The advantage of the phase space quantization scheme was emphasized in Ref.~\citen{Kazama:2008as} and utilized for the  quantization of superstring in the plane-wave background in the SLC gauge}. 

Now for the present system, although the Lagrangian 
(\ref{LK}, \ref{LWZ}) contains cubic interactions, the Hamiltonian expressed in terms 
the phase space variables is quadratic, as  given in (\ref{Hamiltonian}), 
 and one can compute the right hand side  of (\ref{relcanphase}) explicitly. 
The mode expansions of the basic phase space fields are given by 
\begin{align}
X^\mu (\sig,0) &=\sum_n  X^\mu _n e^{-in\sig} =  x^\mu + i\ls \sum_{n \ne 0} \left( {1\over n} \al^\mu _n e^{-in\sig}
+ {1\over n} \albar^\mu _n e^{in\sig} \right) \comma \\
P^\mu(\sig,0) &= \sum_n  P^\mu _n e^{-in\sig} 
= {p^\mu \over 2\pi}  + {1\over4\pi \ls}  \sum_{n \ne 0} \left( \al^\mu _n e^{-in\sig}
+  \albar^\mu _n e^{in\sig} \right)  \comma \\
S_a(\sig,0)&= \sum_n S_{a,n} e^{-in\sig} \comma \qquad 
\Sbar_a(\sig,0) =  \sum_n \Sbar_{a,n} e^{in\sig} \period
\end{align}
By time-developing via  the Hamiltonian (\ref{Hamiltonian}),  we obtain 
 the canonical fields at time $t$ as 
\begin{align}
X^\mu (\sig, t) &= e^{iHt} X^\mu(\sig,0) e^{-itH} 
= x^\mu + 2 \ls^2 p^\mu  t 
+ i \ls \sum_{n \ne 0} \left({1\over n}  \al^\mu _n e^{-in(t+\sig)}
+ {1\over n} \albar^\mu _n e^{-in(t-\sig)} \right) \comma  \\
P^\mu (\sig, t) &= e^{iHt} P^\mu(\sig,0) e^{-itH} 
= {p^\mu \over 2\pi} + {1\over4\pi \ls} \sum_{n \ne 0} \left( \al^\mu _n e^{-in(t+\sig)}
+ \albar^\mu _n e^{-in(t-\sig)} \right) \comma \\
S_a(\sig, t) &= e^{iHt} S_a(\sig,0) e^{-itH} = \sum_n S_{a,n} e^{-in(t+\sig)} \comma \\
 \Sbar_a(\sig,t) &= e^{iHt} \Sbar_a(\sig,0) e^{-itH} = \sum_n \Sbar_{a,n} e^{-in(t-\sig)} \period
\end{align}
By introducing the Euclidean time $\tau \equiv it$ and the variables 
 $z\equiv \exp(\tau+i\sig), \zbar\equiv \exp(\tau-i\sig)$, $X^\mu(\sig, t)$ can 
 be expressed as
\begin{align}
X^\mu(z,\zbar) &= x^\mu -i\ls^2 
p^\mu (\ln z + \ln \zbar) + i\ls \sum_{n \ne 0}
 \left({1\over n} \al_n z^{-n} + {1\over n} \albar_n  \zbar^{-n} \right) \period \label{defXzzbar}
\end{align}
Similarly, other canonical fields can be expressed  as functions of  $z$ and $\zbar$. 
Later, when we regard  them as conformal fields on the complex plane, 
we will have to take  into account the effect of  the  conformal transformation from the cylinder to the complex plane, which shifts the power of $z$ (and $\zbar$) according to the conformal weights. 
\section{Structure of the quantum symmetry algebras}
This section will be devoted to the study of  the structure of the quantum symmetry algebras of the  system.  Besides being important in its own right, 
this  is an indispensable  prerequisite for the construction of the 
vertex operators, to be performed in the subsequent section. 
\subsection{Virasoro algebra and BRST symmetry}
We will first focus on the Virasoro algebra 
and the associated BRST symmetry. The reason is that, as we shall see, 
{\it the super-Poincar\'e algebra closes  only up to 
 BRST transformations.} 
\subsubsection{Classical Virasoro algebra}
Let us begin with the classical Virasoro algebra. 
Because the Lagrangian in the SLC gauge is classically conformally invariant, 
the $++$ and the $--$ components of the energy-momentum tensor, to be denoted 
 by  $T_\pm$,  become the Virasoro constraints. Through the  standard procedure  one obtains 
\begin{align}
T_\pm &= 
T\left(2\del_\pm  \Xp \del_\pm \Xm +  (\del_\pm  X_I)^2  
\right)
 -{i 2\sqrt{2}\, T} \del_\pm \Xp (\th^1_a \del_\pm \th^1_a + \th^2_a \del_\pm \th^2_a) \period 
\end{align}
Note that at this stage  $T_\pm$ contain non-linear coupling between $X^+$ and the fermions $\theta^A_a$. 
However, upon expressing them in terms of the dimensionless variables defined in (\ref{defABetc}) satisfying 
the canonical Dirac brackets, 
 $T_\pm$ become quadratic in fields as 
\begin{align}
T_+ &= \half (\calH + \calP) 
= {1\over 2\pi} \biggl(\Pi^+\Pi^- + \half \Pi_I^2 +{i \over 2} S^2 \del_1 S^2 
\biggr) \comma \label{calTplus}\\
T_- &= \half (\calH - \calP) 
={1\over 2\pi} \biggl( \Pitil^+\Pitil^- + \half \Pitil_I^2 -{i \over 2} S^1 \del_1 S^1 
 \biggr)\period \label{calTminus}
\end{align}
$\calP$ and $\calH$ are, respectively, (the dimensionless version of) the momentum density and the Hamiltonian density.  
To compute the Dirac bracket relations among 
 $\calP$ and $\calH$ (and
 hence among $T_\pm$),  we make use of the following formulas
for the derivatives of the $\delta$-function  for a general field $\calO$:
\begin{align}
\calO(\sigp)\deltapssp &= \calO(\sig)\deltapssp + \del_1 \calO(\sig) \deltassp  \comma \\
\calO(\sigp)\delta''(\sig-\sigp) &= \calO(\sig)\delta''(\sig-\sigp) + 2\del_1 \calO(\sig)
\deltapssp + \del_1^2 \calO(\sig)\deltassp \comma 
\end{align}
which must be understood in the sense of distributions. After  a straightforward calculation, we obtain the expected results:
\begin{align}
\Dcom{\calH(\sig,t)}{\calH(\sigp,t)} &= \Dcom{\calP(\sig,t)}{\calP(\sigp,t)} \nn\\
& = 2\calP(\sig,t) \deltapssp + \del_1 \calP(\sig,t) \deltassp \comma \\
\Dcom{\calP(\sig,t)}{\calH(\sigp,t)} &= \Dcom{\calH(\sig,t)}{\calP(\sigp,t)}   \nn\\
& =2\calH(\sig,t) \deltapssp + \del_1 \calH(\sig,t) \deltassp  \comma \\
\Dcom{T_\pm(\sig,t)}{T_\pm(\sigp,t)} 
&= \pm 2 T_\pm(\sig,t)\deltapssp \pm 
\del_1 T_\pm(\sig,t) \deltassp\comma  \label{DcomT}\\
\Dcom{T_\pm(\sig,t)}{T_\mp(\sigp,t)} &=0 \period
\end{align}
The last two lines show that $T_\pm$ form mutually commuting  Virasoro algebras\footnote{$\pm$ signs on the RHS are just right for the Virasoro mode operators $L^\pm_n$ to satisfy the same algebra. }. 
Integrating the first two equations with respect to  $\sigp$ and identifying $\int d\sigp \calH(\sigp,t)$ to be  the Hamiltonian $H$, which generates the time-development 
of a field $A(\sig,t)$  as $\del_0 A = \Dcom{A}{H}$, one readily finds
\begin{align}
\del_0\calH &= \Pcom{\calH}{H} = \del_1 \calP \comma \qquad 
\del_0\calP = \Pcom{\calP}{H} = \del_1 \calH \period
\end{align}
Combining them, we get $\del_\mp T_\pm =0$,  showing that  $T_\pm= T_{\pm}(\sigma_\pm)$. Therefore, we can define the Virasoro mode operators 
$L^\pm_n$ by 
\begin{align}
T_\pm &= {1\over 2\pi} \sum_n L^\pm_n e^{-in(t \pm \sig)} \period
\end{align}
Putting this into (\ref{DcomT}), one verifies that $L^\pm_n$  satisfy the usual 
 form of the classical Virasoro algebra, namely
\begin{align}
\Dcom{L^\pm_m}{L^\pm_n} &= {1\over i} (m-n) L^\pm_{m+n} 
\comma \qquad \Dcom{L^\pm_m}{L^\mp_n} =0 \period
\end{align}
What is important here is that since $L^\pm_n$ are independent of $t$ and $\sigma$ 
 they can be obtained from $T_\pm$ at one time slice, which we take to be $t=0$:
\begin{align}
L^\pm_n &= \int_0^{2\pi}  d\sig e^{\pm in\sig} T_{\pm}(\sig, t=0) \period
\label{defTpmn}
\end{align}
But at $t=0$, we know the exact Dirac brackets for the fields composing $T_\pm$ and hence we can quantize them in the standard way. Consequently, any 
quantum properties of the
 system which are dictated by $L^\pm_n$ can be calculable\footnote{In the present system $L_n^\pm$ are quadratic in canonical fields and hence this statement may sound trivial. However, this feature is valid even when $L_n^\pm$ contain non-linear terms.}. 
\subsubsection{Quantum Virasoro  algebra}
We are now ready to study the quantum  Virasoro algebra. Since $T_+$ and $T_-$, 
being made up of mutually independent fields, commute, we will concentrate on 
$T_+$. 
To compute the algebra, the most convenient way is to form the generating function
\begin{align}
T(z) &\equiv \sum_n L^+_n z^{-n-2}
\end{align}
on the complex $z$-plane, express it in terms of suitable chiral  fields and 
make use of the familiar operator product expansion (OPE) method. The 
 chiral fields which make up $T(z)$  are given by 
\begin{align}
\Pi^\mu(z) &\equiv i\ls^{-1}\del X^\mu(z) =  \sum_n \al^\mu_n z^{-n-1} \comma \qquad 
S_a(z) \equiv \sum_n S_{a,n} z^{-n -1/2}  \comma  \label{defPiS}
\end{align}
where the basic chiral field $X^\mu(z)$ is  defined, as usual, by 
\begin{align}
X^\mu(z) &\equiv x^\mu -i\ls^2 p^\mu \ln z + i\ls \sum_{n \ne 0} 
 {1\over n} \al_n^\mu z^{-n} \period \label{defXz}
\end{align}
This is obtained from (\ref{defXzzbar}) by dropping the $\zbar$-dependent parts. 
With the normal-ordering prescription discussed previously, these fields  satisfy the 
OPE's
\nullify{\footnote{The normal-ordering for $S^a(z)$ is such that 
$S^a_n$ for $n\ge 1$ are regarded as annihilation oprators. As for the zero modes $S^a_0$, we use the rule $S^a_0S^a_0=\half$ ($a$ not summed) that  follows from the anticommutation relations (\ref{acomSS}). 
This gives the OPE below where, in particular,  $S^a(z) S^a(w) = 1/(z-w) +
\calO(z-w)$. 
}
}
\begin{align}
X^\mu(z) X^\nu(w) &\sim -\eta^{\mu\nu} \ls^2 \ln (z-w) \comma  \label{XXOPE}\\
X^\mu(z) \Pi^\nu(w) &\sim {i \ls \eta^{\mu\nu}  \over z-w} \comma 
\qquad \Pi^\mu(z) \Pi^\nu(w) \sim  {\eta^{\mu\nu} \over (z-w)^2} \comma 
\label{XPiOPE} \\
S_a(z) S_b(w) &\sim  {\delta_{ab} \over z-w} \period \label{SSOPE} 
\end{align}
In terms of these fields  $T(z)$ is  expressed as 
\begin{align}
T(z) &= \Pi^+(z) \Pi^-(z) + \half (\Pi^I(z))^2 -\half S_a \del S_a (z) 
\comma 
\end{align}
where the products are understood to be normal-ordered.  It is straightforward to 
 compute the $T(z) T(w)$ with the result
\begin{align}
T(z) T(w) &= {7 \over (z-w)^4} + {2T(w) \over (z-w)^2}
 + {\del T(w) \over z-w} \period
\end{align}
This shows   that $L^+_n$'s  satisfy  the  Virasoro algebra with the central charge 
$14$, of which 10 comes from the bosons and 4 from the self-conjugate fermions.  In order to 
construct a consistent string theory with conformal symmetry, we need to supply 12 more units of central charge. This was  achieved  in Ref.~\citen{Berkovits:2004tw} by adding a quantum correction 
of the form $\half \del^2 \ln \Pi^+$.   The modified $T(z)$ is therefore given by\footnote{The Virasoro operator $\bar{T}$ for the right-going sector 
is obtained by putting a bar on all the fields. }
\begin{align}
T(z) &= \Pi^+(z) \Pi^-(z) + \half (\Pi^I(z))^2 -\half S_a \del S_a (z) 
+ \half \del^2 \ln \Pi^+ \comma  \label{Tz}
\end{align}
where explicitly
\begin{align}
\del^2 \ln \Pi^+ &= {\del^2 \Pi^+ \over \Pi^+} -\left( {\del \Pi^+ \over \Pi^+}\right)^2 \period
\end{align}
One can readily verify that $T(z)$ continues to satisfy the usual form of the Virasoro 
algebra, this time with $c=26$.  
At this stage, the addition of the peculiar term 
$\half \del^2 \ln \Pi^+$ appears to be  rather ad hoc and its quantum origin 
 is unclear. One might even wonder whether it is unique. A  natural and clarifying 
  answer 
 to these questions will emerge  in the next subsection, as
 we examine the quantum closure of the supersymmetry algebra. 

It is easy to see that $\Pi^+$ and $\Pi^I$ are primary fields of dimension 1 and 
 $S_a$ is a primary of dimension $1/2$.  As for $\Pi^-$, it  is no longer 
 a primary field because of the presence of  the $\half \del^2 \ln \Pi^+$ term. 
 However,  one can modify $\Pi^-$ to  construct a primary of  dimension 1, 
 to be called $\Pihat^-$,  in the following way:
\begin{align}
\Pihat^- &\equiv \Pi^- + \del^2 \left( {1\over 2 \Pi^+}\right)
 = \Pi^- -{1\over 2 \Pi^+} \left(( \del \ln\Pi^+)^2 +\del^2 \ln \Pi^+\right) \nn\\
&=\Pi^{-}(z) - \frac{1}{2}\frac{\partial^{2} \Pi^{+}(z)}{\left(\Pi^{+}\right)^{2}} + 
\frac{\left(\partial \Pi^{+}\right)^{2}(z)}{\left(\Pi^{+}\right)^{3}} \period
\label{modPiminus}
\end{align}
This fact will be relevant  when we construct the quantum Lorentz generator involving $\Pi^-$. 
\subsubsection{BRST operator}
With the Virasoro operator (\ref{Tz}) with $c=26$ at hand, 
 the nilpotent BRST operator is constructed  in the usual way.  For the left sector on which  we focus, it is given by 
\begin{align}
Q &\equiv \int[dz] \left( cT + bc \del c\right)(z) \comma  \label{Q}
\end{align}
where $c(z)$ and $b(z)$ are the standard reparametrization ghost and anti-ghost
 operators satisfying  the OPE
\begin{align}
c(z) b(w) &= b(z) c(w)  \sim {1\over z-w}  \period \label{cbcor}
\end{align}
If we define the total Virasoro operator including the ghost sector by $T^{tot}
= \acom{Q}{b}$, $b$ and $c$ carry conformal weight  $2$ and $-1$ with respect 
 to $T^{tot}$. 
In the next subsection, we will encounter another BRST operator $\Qhat$, which is 
 related to $Q$ by a similarity transformation. Further,  in section 6, 
we will show how  $Q$ can be reduced, again by a similarity transformation, to the 
form relevant in the full  light-cone gauge formulation.  
\subsection{Super-Poincar\'e algebra }
Next we study how the super-Poincar\'e algebra is realized,  in particular quantum mechanically,  in the SLC gauge. Although  the SLC gauge condition  breaks the part of the super-Poincar\'e symmetry, the symmetry is still  preserved  in the physical  Hilbert space. Accordingly, we will see that the algebra  closes up to  BRST exact expressions. 
\subsubsection{Algebra of supercharges}
We begin with the algebra of supercharges. The supercharge densities in the Lagrangian formulation were given  in (\ref{j01a}) $\sim$ (\ref{j02adot}). 
Using various relations  already given, it is straightforward to express them in 
terms of the phase space variables. The result is
\begin{align}
j^{0A}_a(\sig) &= -\sqrt{{\pi^{+A}(\sig) \over \pi}} S_a^A(\sig) \comma \\
j^{0A}_\adot(\sig) &=- {1\over \sqrt{\pi \pi^{+A}(\sig)}} \gabar^I_{\adot b} S^{A b}(\sig) (P^I -\eta_A \del_1 X^I )(\sig) \comma \qquad \eta_1=-\eta_2=1 \period
\end{align}
where $\pi^{+A}(\sig)$ was defined in (\ref{defpipA}). From these expressions 
 we see that the charge densities in the left ($A=2$) and the right ($A=1$) sectors are independent, and hence we will deal only with the left sector  and 
 suppress the index $A=2$. 

To facilitate the computation, we again use the OPE method. In terms of the 
chiral fields,  the supercharges  $Q_\al \equiv  \int d\sig j^0_\al(\sig)$   can be written as
\begin{align}
Q_\al &= \int[dz] j_\al(z) \comma \qquad [dz] \equiv {dz \over 2\pi i} \comma  \\
j_a(z) &=  - \rho  \sqrt{\Pi^+(z)}\, S_a(z) \comma  \qquad 
j_\adot = -\rho {\Pi^I(z) \over \sqrt{2\Pi^+(z)}} \gabar^I_{\adot b} S^b(z) \comma \label{SUSYcharge}\\
\rho &= (32\pi T)^{1/4} = 2^{3/4} \ls^{-1/2} 
\period
\end{align}
The expression such as $\sqrt{\Pi^+(z)}$ is defined,  as usual,  by the 
 expansion around the zero mode in the manner
\begin{align}
\sqrt{\Pi^+(z)} &= \left({\al^+_0 \over z} 
+ \sum_{n \ne 0} \al^+_n z^{-n-1}\right)^{1/2} 
= \sqrt{\al^+_0} z^{-1/2} + {z^{1/2} \over 2\sqrt{\al^+_0} }
 \sum_{n \ne 0} \al^+_n z^{-n-1} + \cdots  \period
\end{align}
Therefore the currents $j_\al(z)$ above do not have cuts and can be integrated 
over $z$ to give $Q_\al$ properly.  Also it is easy to check that $j_\al(z)$ 
 are primary fields of dimension 1 so that $Q_\al$ are conformally invariant. 

The computations of   $\acom{Q_a}{Q_b}$ and $\acom{Q_a}{Q_\bdot}$  
are trivially done by using the OPE $S^a(z) S^b(w) \sim  \delta^{ab}/(z-w)$ and we
 get the standard answer 
\begin{align}
\acom{Q_a}{Q_b} &= 2\sqrt{2}\delta_{ab} p^+ \comma \label{Qab}\\
\acom{Q_a}{Q_\bdot}&= 2\gabar^I_{a\bdot} p^I \comma \label{Qadotb}
\end{align}
where we used $\int[dw] \Pi^\mu(w) = \ls p^\mu$. 

The calculation of  the anti-commutator $\acom{Q_\adot}{Q_\bdot}$ 
is more involved and  it receives an important  quantum correction. 
It is convenient to split the contributions into  three parts as
\begin{align}
\acom{Q_\adot}{Q_\bdot}&=A_{SS} +A_{\Pi\Pi}+A_{SS\Pi\Pi}  \period
\end{align}
Here $A_{SS} , A_{\Pi\Pi}$ and $A_{SS\Pi\Pi} $ are  due, respectively,  to the  contraciton of $S^aS^b$, the contraction of $\Pi^I\Pi^J$ and the double contractions. 
Using the OPE technique, they are given by
\begin{align}
A_{SS} &= \half \rho^2\delta_{\adot\bdot}\int[dw]{ (\Pi^I)^2\over \Pi^+}(w) \comma \\
A_{\Pi\Pi}&= - \half \rho^2 \delta_{\adot\bdot} \int [dw] { S^a\del S^a \over \Pi^+}(w)
\comma \\
A_{SS\Pi\Pi} &= - \half \rho^2 \delta_{\adot\bdot} \int [dw]  {(\del \Pi^+)^2 \over (\Pi^+)^3}(w)  \period
\end{align}
We can rewrite $A_{SS\Pi\Pi}$ by using the identity
\begin{align}
 {(\del \Pi^+)^2 \over (\Pi^+)^3} &= \del^2 \ln \Pi^+ -
 \del \left( {\del \Pi^+ \over (\Pi^+)^2} \right) \comma 
\end{align}
where the second term on the right hand side vanishes upon integration. Therefore we can write the total contribution as 
\begin{align}
\acom{Q_\adot}{Q_\bdot} &= \rho^2 \delta_{\adot\bdot} \int [dw] {1\over \Pi^+}
\left( \half (\Pi^I)^2 - \half S^a \del S^a - \half \del^2 \ln \Pi^+ \right)  \nn\\
&= -\rho^2 \delta_{\adot\bdot} \int[dw] \Pi^- 
 + \rho^2\delta_{\adot\bdot} \int[dw] \calT \comma \label{QadQbd}
\end{align}
where we have defind the quantity 
\begin{align}
\calT &\equiv \Pi^-+  {1\over \Pi^+} 
\left(  \half (\Pi^I)^2 - \half S^a \del S^a - \half 
\del^2 \ln \Pi^+ \right) \period \label{defcalT}
\end{align}
Therefore, we have found that $ \acom{Q_\adot}{Q_\bdot}$ closes into 
the usual momentum term plus an extra contribution proportional to 
$\int[dw] \calT$. 

Let us now clarify the nature of this extra term. 
First note that $\calT$ is almost  identical  to $T/\Pi^+$, where 
$T$ is the Virasoro operator  given in (\ref{Tz}), {\it except for the sign of the last 
logarithmic term}. This difference is very important. With the minus sign as  in (\ref{defcalT}) above, one can verify, with  a slightly tedious calculation,  that  $\calT(z) \calT(w) \sim 0$, {\it i.e.} its OPE with itself  is non-singular. Because of this property, we can define  a nilpotent BRST-like  operator $\Qhat$ by 
\begin{align}
\Qhat &\equiv \int[dz] \calT(z) c(z) \comma 
\end{align}
and write $\int[dw] \calT$  in  a ``BRST-exact" form 
\begin{align}
 \int[dw] \calT  &= \int[dw] \acom{\Qhat}{b(w)} \comma 
\end{align}
where $c(z)$ and $b(z)$ are the ghost-antighost pair satisfying 
 the OPE (\ref{cbcor}).  It is interesting to introduce the ``anti-ghost for $\Qhat$" 
as $B \equiv b\Pi^+$ and define the corresponding energy-momentum tensor, 
 including the ghost sector, by 
\begin{align}
\That^{tot}(z) &\equiv \acom{\Qhat}{B(z)} =\Pi^+\Pi^-  + \half (\Pi^I)^2 - \half S^a \del S^a - \half \del^2 \ln \Pi^+ -b \del c \period
\end{align}
With respect to $\That^{tot}$, the $(b,c)$ ghosts carry dimensions $(1,0)$. 

We now show that $\Qhat$ is related to the usual BRST opeartor $Q$ 
given in (\ref{Q})  by a quantum similarity  transformation\footnote{This was 
first mentioned in Ref.~\citen{Berkovits:2004tw}. Here we  display  the slightly  non-trivial details.}.  Consider an operator 
\begin{align}
R &\equiv \int[dz] cb \ln \Pi^+ \period 
\end{align}
Then it is easy to see that 
\begin{align}
e^R ce^{-R} &= \Pi^+ c \comma \qquad e^R b e^{-R} = {b \over \Pi^+} \period
\end{align}
Similarly, it is straightforward to show that 
\begin{align}
e^R \calT e^{-R} &= \calT + \com{R}{\calT} + \half \com{R}{\com{R}{\calT}} \nn\\
&= \calT + \del \left( {cb \over \Pi^+}\right) + {1\over 2\Pi^+}
\left( \del^2 \ln \Pi^+ -(\del \ln \Pi^+)^2 \right) \period
\label{simtrcalT}
\end{align}
To compute $e^R \Qhat e^{-R}$ using  these results, one must be careful since the 
products  of $\Pi^+c$ and the operators $\Pi^-$ and $\del (cb/\Pi^+)$ contained in (\ref{simtrcalT}) are singular. To multiply them properly, one must point-split the 
arguments of the products and then take the coincident limit.  Since originally there is  no singularity in the product of $\calT$ and $c$, one expects that the singularities 
 would  cancel. Indeed, we find 
\begin{align}
\Pi^-(z) (\Pi^+c)(w) &= {c(w) \over (z-w)^2} + :\Pi^-\Pi^+: c \comma \\
\del \left( {cb \over \Pi^+}(z)\right) (\Pi^+c)(w) 
&= -{c(w) \over (z-w)^2} + \half \del^2 \left( {c \over \Pi^+}(w) \right) \Pi^+(w) + :bc\del c:(w)  \comma 
\end{align}
and the singularities do cancel each other.  The rest of the calculations are straightforward and we obtain
\begin{align}
e^R \Qhat e^{-R} &= Q = \int[dz] (cT + bc\del c) \comma 
\end{align}
where $T$ is precisely the expression of the Virasoro operator given in (\ref{Tz}). 
Note that through the similarity transformation  the conventional non-linear term  $bc\del c$ is generated and moreover {\it the sign in front of $\del^2 \ln\Pi^+$ is reversed.} 

Getting back to the closure $ \acom{Q_\adot}{Q_\bdot}$, we can now express 
 the  extra term as 
\begin{align}
 \int[dw] \acom{\Qhat}{b(w)}  &= e^{-R} \int[dw] \acom{Q}{{b \over \Pi^+}(w)}e^{R} \period
\end{align}
Since $Q_\adot$ and the zero mode part  $ \int[dw] \Pi^-$ 
commute with the operator $R$,  we can make a similarity transformation 
of the form  $e^R (\star) e^{-R}$ without affecting them. In this way, we finally 
 obtain 
\begin{align}
\acom{Q_\adot}{Q_\bdot} &= -2\sqrt{2} \delta_{\adot\bdot} 
 p^-+\acom{Q}{ {2\sqrt{2}\over \ls} \delta_{\adot\bdot}  \int[dz] {b \over \Pi^+}(z)} 
\comma  \label{Qdotadotb}
\end{align}
showing that the algebra closes up to a BRST-exact contribution. 

We would like to emphasize that from the  direct computation of the 
algebra $\acom{Q_\adot}{Q_\bdot}$, the structure of the Virasoro 
operator  $T(z)$ including  the correct extra $+\half \del^2 \ln \Pi^+$ contribution naturally emerged. The recognition that the proper implementation 
 of the global algebra can dictate the correct structure of the local constraint algebra 
can be a very useful and powerful observation in other similar circumstances. 

\subsubsection{Lorentz algebra}
We now construct the Lorentz part of the super-Poincar\'e generators and examine the structure of the closure.  

First,  consider  the classical charge densities. By using the Noether method\footnote{For fermionic variables, we use left derivative throughout.} with the transformation rules given in (\ref{LorX}) and (\ref{Lortheta}), we readily obtain  the charge dentisities in terms of the phase space variables in the following form:
\begin{align}
j^{0,\mu\nu} &= j^{0,  \mu\nu}_B + j^{0, \mu\nu}_F \comma \\
  j^{0,  \mu\nu}_B &= X^\mu P^\nu - X^\nu P^\mu \comma \\
 j^{0, \mu\nu}_F &= -\half (\ga^{\mu\nu})^\al{}_\be \th^{A\be} p_\al^A
\period \label{lorentz current}
\end{align}
In the SLC gauge, where $\th^{A\adot}=0$, the fermionic part $ j^{0, \mu\nu}_F$ simplifies to 
$ j^{0, \mu\nu}_F = -\half ((\ga^{\mu\nu})^a{}_b \th^{A b} p_a^A
 +(\ga^{\mu\nu})^\adot{}_b \th^{Ab} p_\adot^A)$. The only non-vanishing 
components of $(\ga^{\mu\nu})^a{}_b$ and $(\ga^{\mu\nu})^\adot{}_b $ are $(\ga^{IJ})^a{}_b, (\ga^{+-})^a{}_b$ and $(\ga^{I-})^\adot{}_b$. 
Now we must substitute the form of the momenta $p_a^A$ and $p_\adot^A$ 
in terms of $\theta^{Aa}$.  As for $p_a^A$, we already have the appropriate expression (\ref{pA}). On the other hand,  for  $p^A_\adot$  we 
need to  use the general definition of $p^A_\al$ before $\kappa$ symmetry 
gauge-fixing.  It is given by 
\begin{align}
p^A_\al &= i(P^\mu -\eta_A T(\Pi^\mu_1 -W^{A\mu}_1))(\gabar_\mu \theta^A)_\al  \period \label{defpAal}
\end{align}
For $\al=\adot$ case,  in the SLC gauge $(\gabar_\mu \theta^A)_\adot$ is non-vanishing only for $\mu=I$  and for these transverse components we have 
$\Pi^I_1=\del_1X^I$ and $W^{AI}_1=0$. Therefore $p^A_\adot$ simplifies to
\begin{align}
p^A_\adot &= i(P^I -\eta_A T \del_1 X^I) (\gabar^I)_{\adot b} \theta^{Ab} 
\period
\end{align}
When these formulas for  $p^A_\al$ are substituted,  $j^{0,\mu\nu}_F$ become
\begin{align}
j_F^{0,IJ} &= {i \over\sqrt{2}}  \th^A_a (\ga^{IJ})^a{}_b \th^{Ab} 
(P^+ -\eta_A T \del_1 X^+) \comma \\
j_F^{0,+-} &= 0 \comma \\
j_F^{0,I-}&= -{i \over \sqrt{2}} (\gabar^I\th^A)_\adot 
(\gabar^J\th^A)^\adot (P^J -\eta_A T \del_1 X^J) \period
\end{align}
 Finally expressing  $\theta^{Aa}$ in terms of the canonical variable $\Th^{Aa}$ 
using the definition  (\ref{defTh}) and further converting them into $S^A_a$ 
via (\ref{defS}), we obtain  
\begin{align}
J_F^{IJ} &
= -{i \over 8\pi} \left(  S^1_a (\ga^{IJ})^a{}_b S^{1b} + S^2_a (\ga^{IJ})^a{}_b S^{2b} \right) \comma \\
J_F^{+-} &= 0 \comma \\
J^{I-}_F 
&= {i \over 8\pi} \left( (\gabar^IS^1)_\adot \delta^{\adot\bdot}
(\gabar^J S^1)_\bdot {\Pitil^J \over \Pitil^+} 
+ (\gabar^IS^2)_\adot \delta^{\adot\bdot}
(\gabar^J S^2)_\bdot {\Pi^J \over \Pi^+} \right) \period \label{compensation part}
\end{align}
The Lorentz generators obtained by integrating  these densities will be denoted by 
$M^{\mu\nu} \equiv \int d\sig j^{0,\mu\nu}(\sig)$.

We now wish to examine the quantum closure of the algebra of the operators $M^{\mu\nu}$. 
To compute the commutators of $M^{\mu\nu}$, we again wish to make use of the powerful OPE method. However, contrary to the case of the supersymmetry algebra, one encounters  several technical problems in doing so. 
 Since these problems 
are of general nature, we will illustrate  them in the context of free bosonic string in order to make the points clear. 

A generic   problem 
is that  $M_B^{\mu\nu}$ does not split cleanly into 
 the left and the right going parts,
 due to the presence of the common zero mode part. 
  Explicitly, we have the structure
\begin{align}
M^{\mu\nu}_B &= M_{0,B}^{\mu\nu} + \check{M}^{\mu\nu}_L 
 + \check{M}^{\mu\nu}_R \comma 
\end{align}
where 
\begin{align}
M_{0,B}^{\mu\nu} &= x^\mu p^\nu -x^\nu p^\mu \comma \\
 \check{M}^{\mu\nu}_L  &= {1\over i} \sum_{n \ge 1}{1\over n}
(\al^\mu_{-n} \al^\nu_n - \al^\nu_{-n} \al^\mu_n ) \comma \\
\check{M}^{\mu\nu}_R &= {1\over i} \sum_{n \ge 1}{1\over n}
(\albar^\mu_{-n} \albar^\nu_n - \albar^\nu_{-n} \albar^\mu_n )
\period
\end{align}
Since each type of the generators $M_{0,B}^{\mu\nu}$, $ \check{M}^{\mu\nu}_L$ and  $\check{M}^{\mu\nu}_R$ separately satisfies the same form of the Lorentz algebra, if we wish to  focus just on the ``left-going part", we should not  consider 
$\half M_{0,B}^{\mu\nu} +\check{M}^{\mu\nu}_L$ but rather 
the combination $M_{0,B}^{\mu\nu} +\check{M}^{\mu\nu}_L$. This unfortunately  is not easy to describe in terms of the chiral  fields on the complex $z$-plane. 
The candidate expression would be\footnote{This expression appears in some of the literatures, but it does not make sense, as we explain below.} 
\begin{align}
\half \ls^{-1}\int[dz] (X^\mu \Pi^\nu -X^\nu \Pi^\mu )(z)\comma \label{naiveXPi}
\end{align}
where $\Pi^\mu(z)$ and $X^\mu(z)$ are defined in (\ref{defPiS}) and (\ref{defXz}).  However,  this expression 
 is not suitable for the followng reasons:  First, $X^\mu(z)$ 
  is not a genuine  conformal primary field. It contains $\ln z$ and the integral over $z$ is not well-defined. Second, even if we cure it with the regularization $\ln z \rightarrow \ln (z+\ep)$, with a small non-vanishing constant $\ep$, what we obtain 
 is  $\half M_{0,B}^{\mu\nu} +\check{M}^{\mu\nu}_L$, which  does not satisfy the proper Lorentz algebra. 

Despite these difficultiles, we can still make use of the OPE method by modifying the zero mode part of $X^\mu$ and carefully follow the proper definitions of the  commutators. The new coordinate field $\Xcirc^\mu$ is defined by 
\begin{align}
\Xcirc^\mu(z) &\equiv 2x^\mu + \Xchk^\mu(z) 
\comma \qquad \Xchk^\mu(z) =  i   \sum_{n \ne 0}
{1\over n}\al^\mu_n z^{-n} \comma 
\end{align}
where  we put the over-check mark to denote the non-zero-mode part. 
Since the dependence on the scale $\ls$ is rather trivial, we have  set $\ls=1$ for simplicity. We will  continue to do so for the rest of this subsection. 
The field  $\Pi^\mu(z)$ is unchanged and  its relation to the coordinate field now reads
\begin{align}
 i\del \Xcirc^\mu &= i \del \Xchk^\mu =\Pi^\mu -{p^\mu \over z}
\equiv  \Pichk^\mu\period \label{defPichk}
\end{align}
With these definitions,  one can check that  the correct generator $L^{\mu\nu}=M_{0,B}^{\mu\nu} +\check{M}^{\mu\nu}_L$  is reproduced, including the zero mode part,  in the form
\begin{align}
L^{\mu\nu} &= \half \int[dz] (\Xcirc^\mu \Pi^\nu(z) -\chgmunu ) \period
\label{Lmunu}
\end{align}

Let us now sketch  how one can compute the Lorentz algebra $\com{L^{\mu\nu}}{L^{\rho\sig}}$ by  the OPE method. 
The basic OPE's among  the fundamental variables are 
\begin{align}
\Xcirc^\mu(z) \Xcirc^\nu(w) &\sim -\eta^{\mu\nu} \ln \left( 1-{w\over z} \right) \comma  \label{prodXX}\\
\Xcirc^\mu(z) \Pi^\nu(w) &\sim  {i \eta^{\mu\nu} 
\over z-w} \comma \label{prodXcircPi} \\
\Pi^\mu(z) \Xcirc^\nu(w) &= \Pichk^\mu(z) \Xchk^\nu(w) + {2\over z} \com{p^\mu}{x^\nu}
\sim -i \eta^{\mu\nu} \left( {1\over z-w} +{1\over z} \right) \comma 
\label{prodPiXcirc}\\
\Pi^\mu(z) \Pi^\nu(w) &\sim {\eta^{\mu\nu} \over (z-w)^2} \period
\label{prodPiPi}
\end{align}
As these rules are slightly different from the usual ones for  $X^\mu(z)$, we will have to be careful at certain  points. To recognize this, it is instructive to recall how the OPE method works for computing the commutators  in the usual situation. Suppose we 
 wish to compute the commutator\footnote{For illustration, we consider the case of bosonic operators.}  $\com{A}{B}$, where the operators $A, B$ are 
 given by contour integrals around $z=0$ over holomorphic operators
$A(z), B(z)$ defined on $z$-plane as 
$A=\int[dz] A(z)$ and $B=\int[dz]B(z)$.  One normally deals with the situation 
 where the following two conditions are satisfied: (i)\, The product $A(z)B(w)$ defined  for $\zgw$ with  renormal-ordering has only the pole singularities at the coincident points  of the form  $A(z) B(w) = \sum_k C(w)/(z-w)^k$. (ii)\, In the domain
 $\wgz$, 
 the product $B(w)A(z)$ is given also by $ \sum_k C(w)/(z-w)^k$, {\it i.e}  formally by the same expression as $A(z)B(w)$. When these two conditions are satisfied, the 
 commutator $\com{A}{B}$ is defined and computed as
\begin{align}
\com{A}{B} &= \int_\zgw A(z)B(w) 
 - \int_\wgz B(w) A(z)  \label{comAB} \\
&= \left(\int_\zgw -\int_\wgz\right)[dz][dw] \sum_k  {C_k(w) \over (z-w)^k}   \nn\\
&= \int [dw] \int [dz]_w \sum_k  {C_k(w) \over (z-w)^k} 
= \int[dw] C_1(w)  \comma
\end{align}
where $\int [dz]_w$ means the integral circling  $w$. This discussion shows that 
we have to be careful when the conditions (i) and/or  (ii) are not satisfied. 
Looking at the basic OPE's given in (\ref{prodXX}) $\sim$ (\ref{prodPiPi}), we see 
 that in the present case there are indeed cases where these conditions are not quite 
 satisfied. In such cases, what we have to do is to simply go back to the 
 basic definition, {\it i.e.} the first line  (\ref{comAB}) above,  and treat
those parts which violate (i) and/or (ii) separately.  Once this is done then the 
 rest  that  satisfies (i) and (ii) can be computed easily by the 
usual OPE method as above.  In appendix B.1  we shall illustrate this method 
 explicitly for the computation of $\com{L^{\mu\nu}}{L^{\rho\sig}}$, 
 where $L^{\mu\nu}$ is given by (\ref{Lmunu}). 

It should now be clear that by using the above trick  we can study the closure of the Lorentz algebra by effectively separating the analysis into that of the  holomorphic sector and the  anti-holomorphic sector. For our system, the holomorphic  Lorentz currents are given by 
\begin{align}
J^{I J}(z) &= \frac{1}{2} (\Xcirc^{I} \Pi^{J}(z) - \Xcirc^{J} \Pi^{I}(z)) 
- \frac{i}{4} S^{a} \left(\gamma^{I J}\right)_{a b} S^{b}(z) , \notag \\  
J^{\mu +}(z) &= \frac{1}{2} (\Xcirc^{\mu} \Pi^{+}(z) - \Xcirc^{+} \Pi^{\mu}(z))  \comma  \\
J^{I -}(z) &= \frac{1}{2} (\Xcirc^{I} \Pi^{-}(z) - \Xcirc^{-} \Pi^{I}(z))
+ \frac{i}{4}\frac{(\bar{\gamma}^I S)_{\dot{a}}(\bar{\gamma}^{K} S)_{\dot{a}} 
\Pi^{K}(z)}{\Pi^{+}(z)}\comma   \label{holomorphic current}
\end{align}
and the anti-holomorphic currents are quite similar. 
Note that the fermionic contributions in $J^{IJ}$ and the 
last term of $J^{I-}$, which was generated by the compensating 
$\kappa$-transformation,  are naturally holomorphic as they do not involve  $X^\mu$.  

We now describe  the actual analysis of the  closure of the quantum Lorentz 
algebra. Since only the boost generator $M^{I-}=\int [dz]  J^{I-}(z)$ 
is affected non-trivially by the SLC gauge condition, we will focus on this 
 generator. Before we begin the computation of the commutators, however, 
we must examine if  the generators  are physical, \ie if they commute with the 
 BRST charge $Q$. By using our OPE technique, it is easily found  that,  while 
 $M^{IJ}$ and $M^{\mu +}$ are physical,   $M^{I-}$ is not. 
(This phenomenon was  noted previously in  Ref.~\citen{Berkovits:2004tw, Kunitomo:2007vm}.)
This problem is cured by adding a simple quantum correction. 
The modified generator ${\cal M}^{I -}$ is given by 
\begin{align}
{\cal M}^{I-} &= M^{I-} +\Delta M^{I-} \comma \label{quantum boost}\\
M^{I-} &= \int[dz]~ \left\{\frac{1}{2} (\Xcirc^{I} \Pi^{-}(z) - \Xcirc^{-} \Pi^{I}(z)) 
+~ \frac{i}{4}\frac{(\bar{\gamma}^I S)_{\dot{a}}(\bar{\gamma}^{K} S)_{\dot{a}}
\Pi^{K}(z)}{\Pi^{+}(z)} \right\} \comma \\
\Delta M^{I-} &= -\int [dz] ~ \frac{i}{2} \frac{\partial \Pi^{I}(z)}{\Pi^{+}(z)} \period
\label{quantum correction} 
\end{align} 
which can be checked to satisfy the desired physical condition $[ Q,\, {\cal M}^{I -} ] = 0$. The necessity of the added term $\Delta M^{I-}$ can be understood in the following way. Up to a zero mode part,  $\int[dz] \frac{1}{2} (\Xcirc^{I} \Pi^{-} - \Xcirc^{-} \Pi^{I}) $ equals  $\int [dz] \Xcirc^{I} \Pi^{-}$ and to promote $\Pi^-$ to a primary field we should make a replacement 
$\Pi^- \rightarrow \Pihat^-$, where $\Pihat^-$ is given in (\ref{modPiminus}).This adds to $M^{I-}$ the extra term $\int [dz] \Xcirc^I \del^2 (1/2\Pi^+)$,  which upon integration  by parts twice becomes precisely $\Delta M^{I-}$ above. 

Let us now compute the commutator $\left[ {\cal M}^{I -},\, {\cal M}^{J -} \right]$. From the experience of the computation in the full light-cone gauge,\cite{Green:1987sp} \ 
 it is expected to yield a non-trivial  result.  Using the 
 OPE technique, we obtain
\begin{align} 
&\left[ {\cal M}^{I -},\, {\cal M}^{J -}\right] = \int[dw]~\Biggl[ \frac{1}{2}\frac{\Pi^{+} \Pi^{-}(w)
(\bar{\gamma}^{I} S)_{\dot{a}}(\bar{\gamma}^{J} S)_{\dot{a}}(w)}{\left(\Pi^{+}(w)\right)^{2}} \notag \\
& \quad -~\frac{1}{4}\frac{\Pi^{I} \Pi^{K}(w)
(\bar{\gamma}^{J} S)_{\dot{a}}(\bar{\gamma}^{K} S)_{\dot{a}}(w)}{\left(\Pi^{+}(w)\right)^{2}}
+~\frac{1}{4}\frac{\Pi^{J} \Pi^{K}(w)
(\bar{\gamma}^{I} S)_{\dot{a}}(\bar{\gamma}^{K} S)_{\dot{a}}(w)}{\left(\Pi^{+}(w)\right)^{2}} \notag \\
& \quad +~\frac{1}{2}\frac{\Pi^{I} \partial \Pi^{J}(w) - 
\Pi^{J} \partial \Pi^{I}(w)}{\left(\Pi^{+}(w)\right)^{2}}
+~\frac{1}{4}\left\{\partial^{2} \left(\frac{1}{\Pi^{+}(w)}\right)\right\}
\left(\frac{(\bar{\gamma}^{I}S)_{\dot{a}}(\bar{\gamma}^{J}S)_{\dot{a}}(w)}{\Pi^{+}(w)}\right) \notag \\
& \quad -~\frac{1}{16} \left\{\partial 
\left(\frac{(\bar{\gamma}^I S)_{\dot{a}}
(\bar{\gamma}^{K} S)_{\dot{a}}(w)}{\Pi^{+}(w)}\right)\right\}
\left(\frac{(\bar{\gamma}^J S)_{\dot{b}}(\bar{\gamma}^{K} S)_{\dot{b}}(w)}{\Pi^{+}(w)}\right) \notag\\
& \quad -~\frac{1}{4}\left(\frac{\Pi^{K} \Pi^{L}(w)}{\left(\Pi^{+}(w)\right)^2}\right)
\left\{S_{a}(w)\left(\gamma^{IK}\,\gamma^{JL}\right)^{a}{}_{b}S^{b}(w)\right\} \notag \\
& \quad +~\frac{1}{8}\left\{\partial \left(\frac{S_{a}(w)}{\Pi^{+}(w)}\right)\right\}
\left(\gamma^{IK}\,\gamma^{JK}\right)^{a}{}_{b}\left\{\partial\left(\frac{S^{b}(w)}
{\Pi^{+}(w)}\right)\right\} \notag \\
& \quad -~\frac{1}{8}\left\{\partial \left(\frac{\Pi^{K}(w)}{\Pi^{+}(w)}\right)\right\}
\left(\frac{\Pi^{L}(w)}{\Pi^{+}(w)}\right) {\rm Tr}\left(\gamma^{IK}\,\gamma^{JL}\right) 
\Biggr]. \label{boost com1}
\end{align}
In the course of this calculation,  we have performed appropriate  integration  by parts and discarded the resultant total-derivative terms. This type of manipulation will often  be made  in the subsequent calculations as well. 
By using various $\ga$-matrix formulas given in appendix C.1, this can be simplified to 
\begin{align}
&\left[ {\cal M}^{I -},\, {\cal M}^{J -}\right] = \int[dw]~ \Biggl[ \frac{1}{2}\frac{\Pi^{+} \Pi^{-}(w)
(\bar{\gamma}^{I} S)_{\dot{a}}(\bar{\gamma}^{J} S)_{\dot{a}}(w)}{\left(\Pi^{+}(w)\right)^{2}} \notag \\
& \quad +~\frac{1}{4}\frac{\Pi^{K} \Pi^{K}(w)(\bar{\gamma}^{I} S)_{\dot{a}}
(\bar{\gamma}^{J} S)_{\dot{a}}(w)}{\left(\Pi^{+}(w)\right)^{2}} 
-~\frac{1}{4}\frac{S_{a}\partial S_{a}(\bar{\gamma}^{I}S)_{\dot{a}}(\bar{\gamma}^{J}S)_{\dot{a}}(w)}
{\left(\Pi^{+}(w)\right)^{2}} \notag \\
& \quad +~\left\{\partial^{2}\left(\frac{1}{\Pi^{+}(w)}\right)\right\} 
\left(\frac{\left(\bar{\gamma}^{I}S\right)_{\dot{a}}\left(\bar{\gamma}^{J}S\right)_{\dot{a}}(w)}
{\Pi^{+}(w)}\right) \notag \\
& \quad +~\frac{3}{4} \frac{\left(\bar{\gamma}^{I} \partial^{2}S\right)_{\dot{a}}
\left(\bar{\gamma}^{J}S\right)_{\dot{a}}(w)}{\left(\Pi^{+}(w)\right)^{2}} 
-~\frac{3}{2} \frac{\partial \Pi^{+}(w) \left(\bar{\gamma}^{I}\partial S\right)_{\dot{a}}
\left(\bar{\gamma}^{J}S\right)_{\dot{a}}(w)}{\left(\Pi^{+}(w)\right)^{3}} \Biggr]. \label{boost com2}
\end{align}
Some detail of this computation  is given in  appendix B.2.

So the commutator does not vanish. However, we can now show that 
 it is actually a BRST-exact term. To this end, consider the fermionic operator 
\begin{align}
\Psi \equiv \frac{1}{2} \int[dw]~\left(\frac{b(w) 
(\bar{\gamma}^{I} S)_{\dot{a}}(\bar{\gamma}^{J} S)_{\dot{a}}(w)}{\left(\Pi^{+}(w)\right)^2}\right).
\end{align}
The anti-commutator of $\Psi$ and the BRST charge $Q$ can be calculated straightforwardly and yields 
\begin{align}
&\left\{ Q,\,\Psi\right\} = 
\int[dw]~\Biggl[ \frac{1}{2}\frac{\Pi^{+} \Pi^{-}(w)
(\bar{\gamma}^{I} S)_{\dot{a}}(\bar{\gamma}^{J} S)_{\dot{a}}(w)}{\left(\Pi^{+}(w)\right)^{2}} \notag \\
& \quad +~\frac{1}{4}\frac{\Pi^{K} \Pi^{K}(w)(\bar{\gamma}^{I} S)_{\dot{a}}
(\bar{\gamma}^{J} S)_{\dot{a}}(w)}{\left(\Pi^{+}(w)\right)^{2}} 
-~\frac{1}{4}\frac{S_{a}\partial S_{a}(\bar{\gamma}^{I}S)_{\dot{a}}(\bar{\gamma}^{J}S)_{\dot{a}}(w)}
{\left(\Pi^{+}(w)\right)^{2}} \notag \\
& \quad -~\frac{1}{4}\frac{\left(\partial \Pi^{+}(w)\right)^{2}}{\left(\Pi^{+}(w)\right)^{4}} 
(\bar{\gamma}^{I} S)_{\dot{a}}(\bar{\gamma}^{J} S)_{\dot{a}}(w) -~\frac{1}{4}
\frac{\partial^{2} \Pi^{+}(w)}{\left(\Pi^{+}(w)\right)^{3}} (\bar{\gamma}^{I} S)_{\dot{a}}
(\bar{\gamma}^{J} S)_{\dot{a}}(w) \notag \\
& \quad +~\frac{3}{8}\frac{(\bar{\gamma}^{I}\partial^{2}S)_{\dot{a}}
(\bar{\gamma}^{J} S)_{\dot{a}}(w)}{\left(\Pi^{+}(w)\right)^2} 
+~\frac{3}{8}\frac{(\bar{\gamma}^{I} S)_{\dot{a}}
(\bar{\gamma}^{J}\partial^{2}S)_{\dot{a}}(w)}{\left(\Pi^{+}(w)\right)^2}\Biggr]. 
\label{qpsi}
\end{align}
(Although  ghost-dependent terms also appear in the OPE, such terms become 
total-derivatives
in $\partial_{w}$ and drop out\footnote{This is consistent with the fact that the Lorentz current 
and their commutator do not include ghosts.}. )  Now by performing partial integration 
twice for the last term of  (\ref{qpsi}), one finds that this expression 
 becomes  identical to  the commutator (\ref{boost com2}) we found previously. Thus, we have proved the desired relation 
\begin{align}
\left[ {\cal M}^{I -},\, {\cal M}^{J -} \right] = 0+
\left\{ Q,\, \Psi\right\} \period \label{finalboostcom}
\end{align}
This constitutes a highly non-trivial check of the closure of the Lorentz algebra 
in the BRST-cohomology of the superstring in the SLC gauge. 
We have also calculated  the other commutators involving  ${\cal M}^{I -}$ in a similar way and found that they  are consistent with the Lorentz algebra {\it without}  BRST-exact terms. 
 
\subsubsection{Spinorial property of supercharges}
As the final step  of the study of the realization of the super-Poicar\'e algebra, let us examine if the supercharges transform properly  under the 
Lorentz transformations. Since only the boost generators  ${\cal M}^{I -}$ are 
non-trivially affected by the SLC gauge condition, we again  concentrate on 
the commutators  $\com{{\cal M}^{I -}}{Q_a}$ and $\com{{\cal M}^{I -}}{Q_\adot}$. 
From the expressions of the  supercurrents (\ref{SUSYcharge}), the supercharges are given by (we continue to set $\ls=1$ for simplicity) 
\begin{align}
Q_{a} = - 2^{3/4} \int[dz]~ \sqrt{\Pi^{+}(z)}\, S_{a}(z), \quad 
Q_{\dot{a}} = - 2^{1/4} \int[dz]~ \frac{(\bar{\gamma}^{I} S)_{\dot{a}}\, \Pi_{I}(z)}
{\sqrt{\Pi^{+}(z)}} , \label{SUSYls1}
\end{align}
We begin with the commutator with $Q_{a}$ \ie  the $\eta$-SUSY generator which is not affected by gauge-fixing. 
The basic OPE formulas (\ref{prodPiPi}) and (\ref{SSOPE}) lead to 
the following result:
\begin{align}
\left[ {\cal M}^{I -},\, Q_{a} \right] &= i 2^{-1/4} \int[dw]~ \left\{ \frac{S_{a}\, \Pi^{I}(w)}
{\sqrt{\Pi^{+}(w)}} - 
\left(\frac{\Pi^{K}(w)}{\sqrt{\Pi^{+}(w)}}\right) \left(\gamma^{KI}\right)_{a b} S_{b}(w) \right\} \notag \\
&= i 2^{-1/4} \int[dw]~ \left\{ \gamma^{I}_{a \dot{a}} \frac{(\bar{\gamma}^K S)_{\dot{a}}\, \Pi^{K}(w)}
{\sqrt{\Pi^{+}(w)}}\right\} = -\frac{i}{\sqrt{2}} \gamma^{I}_{a \dot{a}}~ Q_{\dot{a}} \period 
\end{align}
Since $\bar{\gamma}^{I-}_{a\adot} = - \sqrt{2} \gamma^{I}_{a \dot{a}}$ in our convention, 
this result gives the correct commutator
\begin{align}
\left[ {\cal M}^{I-},\, Q_{a} \right] = \frac{i}{2} \left(\bar{\gamma}^{I -}\right)_{a \dot{a}} Q_{\dot{a}}.
\end{align}

Next, we proceed to compute the commutator of ${\cal M}^{I -}$ and $Q_{\dot{a}}$. 
Since both generators contain non-trivial compensation terms due to the SLC gauge-fixing, 
this  commutator is expected to yield the correct algebra  up to  BRS exact terms, just as in the case of 
$\left[ {\cal M}^{I -},\, {\cal M}^{J -}\right]$.
With the expressions of $\calM^{I-}$ and $Q_\adot$ given in 
(\ref{quantum boost}) and (\ref{SUSYls1}), we can straightforwardly compute the commutator and obtain 
\begin{align}
&\left[ {\cal M}^{I -},\, Q_{\dot{a}} \right] = 
(-i\, 2^{1/4}) \int[dw]~\biggl\{ \frac{\Pi^{-} \left(\bar{\gamma}^{I}S\right)_{\dot{a}}(w)}
{\sqrt{\Pi^{+}}} + \frac{1}{2} \frac{\Pi^{K} \Pi^{K} 
\left(\bar{\gamma}^{I} S\right)_{\dot{a}}(w)}{\left(\Pi^{+}\right)^{3/2}} \nn\\
&+ \frac{21}{16} \frac{\left(\partial \Pi^{+}\right)^{2} \left(\bar{\gamma}^{I} S\right)_{\dot{a}}(w)}
{\left(\Pi^{+}\right)^{7/2}} 
  -~ \frac{7}{8} \frac{\left(\partial^{2} \Pi^{+}\right) 
\left(\bar{\gamma}^{I} S\right)_{\dot{a}}(w)}{\left(\Pi^{+}\right)^{5/2}} 
+ \frac{1}{2} \partial^{2}\left(\frac{1}{\Pi^{+}(w)}\right) 
\frac{(\bar{\gamma}^{I} S)_{\dot{a}}(w)}{\sqrt{\Pi^{+}}}  \nn\\
&+ \frac{1}{4} \partial\left(\frac{(\bar{\gamma}^I S)_{\dot{b}}(\bar{\gamma}^{K} S)_{\dot{b}}(w)}
{\Pi^{+}}\right)\,\left(\frac{(\bar{\gamma}^{K} S)_{\dot{a}}(w)}
{\sqrt{\Pi^{+}}}\right) \biggr\} \period \label{boostsusycom}
\end{align}
In getting the first line,  the identity (\ref{cubicgamma com}) has been utilized. Now we can rewrite  the last term,   cubic in $S^a$,  by 
judiciously performing integration by parts and get 
\begin{align}
&\partial\left(\frac{(\bar{\gamma}^{I} S)_{\dot{b}}(\bar{\gamma}^{K} S)_{\dot{b}}}
{\Pi^{+}}\right)\,\left(\frac{(\bar{\gamma}^{K} S)_{\dot{a}}}{\sqrt{\Pi^{+}}}\right) = 
\frac{1}{3} \frac{\left\{ \partial \left[(\bar{\gamma}^{I} S)_{\dot{b}}(\bar{\gamma}^{K} S)_{\dot{b}}\right] 
(\bar{\gamma}^{K} S)_{\dot{a}} - 2 (\bar{\gamma}^{I} S)_{\dot{b}}(\bar{\gamma}^{K} S)_{\dot{b}} 
\partial \left[(\bar{\gamma}^{K} S)_{\dot{a}}\right] \right\}}{\left(\Pi^{+}\right)^{3/2}} . \label{cubicS Id1}
\end{align}
Further, by using the $\ga$-matrix identities (\ref{GammaId1}) and (\ref{GammaId2}), the numerator of this expression can be simplified to 
\begin{align}
\partial \left[(\bar{\gamma}^I S)_{\dot{b}}(\bar{\gamma}^{K} S)_{\dot{b}}\right] 
(\bar{\gamma}^{K} S)_{\dot{a}} - 2 (\bar{\gamma}^I S)_{\dot{b}}(\bar{\gamma}^{K} S)_{\dot{b}} 
\partial \left[(\bar{\gamma}^{K} S)_{\dot{a}}\right] = - 6 S_{b} \partial S_{b} 
\left(\bar{\gamma}^{I} S\right)_{\dot{a}}\period  \label{cubicS Id2}
\end{align}
Using these formulas, we finally obtain
\begin{align}
& \left[ {\cal M}^{I -},\, Q_{\dot{a}} \right] = (-i\, 2^{1/4}) \int[dw]~ \left\{ \frac{\Pi^{-} 
\left(\bar{\gamma}^I S\right))_{\dot{a}}(w)}{\sqrt{\Pi^{+}}} + \frac{1}{2} \frac{\Pi^{K} \Pi^{K} 
\left(\bar{\gamma}^{I} S\right)_{\dot{a}}(w)}{\left(\Pi^{+}\right)^{3/2}} \right. \notag\\
& \left. \quad + \left(\frac{37}{16}\right) \frac{\left(\partial \Pi^{+}\right)^{2} 
\left(\bar{\gamma}^{I} S\right)_{\dot{a}}(w)}{\left(\Pi^{+}\right)^{7/2}} 
- \left(\frac{11}{8}\right) \frac{\left(\partial^{2} \Pi^{+}\right) 
\left(\bar{\gamma}^{I} S\right)_{\dot{a}}(w)}{\left(\Pi^{+}\right)^{5/2}} 
- \frac{1}{2} \frac{S_{b} \partial S_{b} \left(\bar{\gamma}^{I} S\right)_{\dot{a}}(w)}
{\left(\Pi^{+}\right)^{3/2}} \right\} .
\label{boostsusycom2}
\end{align}

Now, analogously to the case of the boost commutator, 
we define the following bosonic operator $\Phi$:
\begin{align}
\Phi = (- i\, 2^{1/4}) \int[dw]~ \left\{\frac{b(w) 
(\bar{\gamma}^{I} S)_{\dot{a}}(w)}{\left(\Pi^{+}(w)\right)^{3/2}} \right\}. \label{Phi}
\end{align}
Then, the commutator with the BRST operator yields
\begin{align}
\left[Q,\, \Phi \right] =& 
(- i\, 2^{1/4}) \int[dw]~ \left\{ \frac{\Pi^{-} \left(\bar{\gamma}^I S\right)_{\dot{a}}(w)}
{\sqrt{\Pi^{+}}} + \frac{1}{2} 
\frac{\Pi^{K} \Pi^{K} \left(\bar{\gamma}^{I} S\right)_{\dot{a}}(w)}
{\left(\Pi^{+}\right)^{3/2}} 
- \frac{1}{2} \frac{S_{b} \partial S_{b} \left(\bar{\gamma}^{I} S\right)_{\dot{a}}(w)}
{\left(\Pi^{+}\right)^{3/2}} \right. \notag \\
& \left. - \frac{1}{2}\frac{\left(\partial \Pi^{+}\right)^{2} 
\left(\bar{\gamma}^{I} S\right)_{\dot{a}}(w)}
{\left(\Pi^{+}\right)^{7/2}} - \frac{1}{4} \frac{\partial^{2} \Pi^{+} 
\left(\bar{\gamma}^{I} S\right)_{\dot{a}}(w)}{\left(\Pi^{+}\right)^{5/2}} 
+ \frac{3}{4} \frac{\left(\bar{\gamma}^{I} \partial^{2} S\right)_{\dot{a}}(w)}
{\left(\Pi^{+}\right)^{3/2}} \right\} . \label{qphi}
\end{align}
By rewriting the last term of (\ref{qphi}) by performing partial integration twice, we find that this expression coincides with 
the commutator (\ref{boostsusycom2}). Thus, we have obtained 
the desired result 
\begin{align}
\left[ {\cal M}^{I-},\, Q_{\dot{a}} \right] =0+ \left[ Q,\, \Phi \right],
\end{align}
which shows that $Q_\adot$ transforms correctly as a spinor in the physical Hilbert space.  In checking these non-trivial relations, one notes that the quantum correction terms in $T(z)$ and in $\calM^{I-}$ play crucial roles. 

This completes the demonstration that the super-Poincar\'e algebra is 
properly represented  in the quantum 
Green-Schwarz superstring in the SLC gauge in the sense of BRST cohomology. 
\section{Vertex operators for massless states} 
Having clarified the structure of the quantum Virasoro and super-Poincar\'e algebras in the SLC gauge, we are now ready  to construct the vertex operators of the theory.  Although we have been dealing with  the  type II closed superstring, to avoid unnecessary technical complications we will study the type I open string vertices at the massless level for illustration. Extension to the closed string case is  straightforward. 

As is well-known, the vertex operators at  the massless level of the open superstring should describe the ``photon" and the ``photino" excitations of the super- Maxwell theory in 10 dimensions\footnote{
This of course can be regarded as the linear part of the super-Yang-Mills 
 theory,  with the Chan-Paton factor suppressed.} 
The basic principle for the construction is that these vertex operators must 
be BRST invariant and  form the appropriate representation of the super-Poincar\'e algebra,  up to BRST-exact terms. As we shall see, these requirements 
will indeed fix the form of the vertex operators albeit in a fairly  intricate manner.
\subsection{General form of the BRST-invariant vertex operators}
In the following, we will be dealing with the integrated vertex operator of the 
  form $V =\int [dz] \calV(z)$. (The unintegrated vertex operator $\calU(z)$ is related to $\calV(z)$  by $Q\calV(z) = \del \calU(z)$, where $Q$ is the BRST operator\footnote{This is the general  
definition of the unintegrated vertex operator applicable to any theories. 
In the present case it reduces to  the familiar relation $\calU = c\calV$. 
}.) 
 For $V$ to be BRST invariant, $\calV(z)$ must be a primary 
 operator of dimension $\Delta =1$ on the mass-shell, where the momentum 
$k^\mu$ satisfies $k^\mu k_\mu=0$ and the wave functions satisfy appropriate equations of motion. 
In addition, for a technical reason to be explained shortly, 
 we need to  consider  the vertex operators in the  Lorentz frame in which  $k^+=0$.  
$\calV(z)$ must also be consistent with the symmetries which remain manifest in the SLC gauge. One is of course the $SO(8)$ symmetry.  Another useful symmetry is  the boost symmetry  generated by  $M^{+ -}$. From the transformation property under $M^{+ -}$, 
one can read off  the boost charges for various quantities as 
\begin{align}
X^{+},  \zeta^{+}~  &:~ + 1 \comma \quad X^{-},  k^{-},  \zeta^{-}~:~ - 1 \comma
\quad X^{I},  k^{I},  \zeta^{I}~:~ 0 \comma \\
u^{a}~&:~ - 1/2 , \quad u^{\dot{a}}~:~ + 1/2 , \quad S_{a}~:~ 0 \period
\end{align}

Consider first the candidates for the boson emission vertex operator, which must be
 linear in the polarization vector $\zeta^\mu$.  If we temporarily disregard the 
common  structure  $\exp(i k_\mu X^\mu)$,  there are two types of operators which satisfy all the  requirements above. 
One type consists of\footnote{$\zeta^J k^J k^I \Pi^I$ is omitted 
since  $\zeta^{I}k^{I} 
= -\zeta^{+} k^{-}$.}
\begin{align}
\zeta^{-}\Pi^{+},~ \zeta^{+} \hat{\Pi}^{-},~ 
\zeta^{I} \Pi^{I},~ \zeta^{I} R^{I},~ \zeta^{+}k^{-} k^{I} \Pi^{I} \comma 
\end{align}
where $R^I$ is a fermion bilinear defined by 
\begin{align}
R^{I}&= k_{J}\,S\gamma^{IJ}S \period \label{defRI}
\end{align}
Another type of operators are of the form $(\zeta^+/\Pi^+)\times$ ($\Delta=2$
transverse primaries), which are\footnote{Note that $\Pi^I \Pi^I$ and $k^I \del \Pi^I$ are not included in the list since  they produce 4th and 3rd order poles  in the OPE with $T(z)$ and are not primary operators.}
\begin{align}
\frac{\zeta^{+}}{\Pi^{+}} \times \left(\Pi^{I} R^{I},~R^{I}R^{I},~(k^{I}\Pi^{I})(k^{J}\Pi^{J}) 
\right) \period
\end{align}

Now we have to combine these operators with the exponential factor  $\exp(i k_\mu X^\mu)$. 
If $k^+ \ne 0$, a problem arises. As we already mentioned 
 at the end of section 5.1.2, the field $\Pi^-(z)= i\del X^-(z)$ and hence $X^-(z)$ as well do not have good conformal properties and consequently the factor 
 $\exp(i k^+X^-)$ is not a primary operator. One might try to use the modified field $\hat{X}^- \equiv X^- -\del(1 / 2 \del X^+)$, which is the integral of $(1/i)\Pihat^- $ and has a good conformal property, but unfortunately  $\hat{X}^-(z)$ has a severe singularity with itself and it is difficult to define the  operator $\exp(ik^+\hat{X}^-)$.  Therefore in this work we will only deal 
with the configuration  (or the frame) where $k^+=0$  and 
$\exp(ik_\mu X^\mu)$ will mean $\exp(i k^-X^+ + ik^I X^I)$. How one might be able to improve on this situation will be discussed in the discussion 
section. 
 Combined with this special type of exponential factor, 
it is easy to check that,  while   the terms  $ \zeta^{+} \hat{\Pi}^{-}$ and $\zeta^{I} \Pi^{I}$ must 
appear together as $( \zeta^{+} \hat{\Pi}^{-} +\zeta^{I} \Pi^{I})e^{ik_\mu X^\mu}$ in order to produce a genuine primary operator of 
dimension one,  each of the remaining structures  continues  to be  an independent  primary operator of dimension one. 
Therefore,  the general form of the BRST-invariant  vertex operator
 for the photon  is 
\begin{align}
&V_{B}(\zeta) = \int [dz]~ e^{i k_{\mu}X^{\mu}(z)}\, \Biggl\{A\, \zeta^{-}\, \Pi^{+}(z) + B\,\bigl(\zeta^{I}\,\Pi^{I}(z) +\zeta^+\hat{\Pi}^{-}(z)\bigr)
+ C\,\zeta^{I}\, R^{I}(z) \Biggr. \notag\\
&\left. +~\zeta^{+}\left( D\,\frac{\Pi^{I}\,R^{I}(z)}{\Pi^{+}} + 
E\,\frac{R^{I}\,R^{I}(z)}{\Pi^{+}} + Y\, k^{-}\, k_{I}\Pi^{I}(z) + Z\, \frac{\left(k_{I} \Pi^{I}\right)
\left(k_{J}\Pi^{J}\right)(z)}{\Pi^{+}}\right) \right\}, \label{generalVb}
\end{align} 
where $\zeta^\mu$ satisfies  the transverse condition $\zeta^\mu k_\mu =0$.  The relative magnitude of the coefficients $A, B, C, D, E, Y, Z$, which are aribtrary at this stage, should be determinable solely from the requirement 
 of the primarity,  had we been able to deal with the general $k^+\ne 0$
 case.  
For lack of such means,  we will show that these 
 coefficients can alternatively be uniquely determined from the requirement of super-Poincar\'e covariance. 

Next, consider the candidates for the fermion vertex,  describing the photino, linear in the fermionic wave function $u^{\alpha}$. Again there are two types. 
The one containing $u^a$ is unique and is given by $u^a \sqrt{\Pi^{+}} S_{a}$. The other type proportional to  $u^\adot$ consists of two operators 
\begin{align}
\frac{u^\adot \left(\bar{\gamma}^{I} S\right)_{\dot{a}} \Pi^{I}}{\sqrt{\Pi^{+}}}, \quad
\frac{u^\adot \left(\bar{\gamma}^{I} S\right)_{\dot{a}} R^{I}}{\sqrt{\Pi^{+}}} \period  
\end{align}
Thus the general form of the fermion vertex (which is a bosonic operator as a whole)  is
\begin{align}
V_{F}(u) &= \int[dz]~ e^{i k_{\mu}X^{\mu}(z)}\, \left\{u^{a}\,
\left(G\,\sqrt{\Pi^{+}}\,S_{a}(z)\right) \right. \notag \\ 
&\left. +~ u^{\dot{a}}\,\left(K\,\frac{\left(\bar{\gamma}^{I} S\right)_{\dot{a}}\,\Pi^{I}(z)}{\sqrt{\Pi^{+}}} 
+ L\,\frac{\left(\bar{\gamma}^{I} S\right)_{\dot{a}}\,R^{I}(z)}{\sqrt{\Pi^{+}}}\right) \right\} \period 
\label{generalVf}
\end{align} 
$G,K, L$ are coefficients to be determined and $u^\al$ must satisfy the 
 on-shell condition $k_{\mu}\gamma^{\mu}\,u = 0$. 

These vertex operators must transform properly under the supersymmetry 
 transformations.  This requires 
\begin{align}
\left[ \eta^{a}Q_{a},\,V_{B}(\zeta) \right] = V_{F}(\tilde{u}), &\quad
\left[ \eta^{a}Q_{a},\,V_{F}(u) \right] = -V_{B}(\tilde{\zeta}), \label{susyreleta} \\ 
\left[ \epsilon^{\dot{a}}Q_{\dot{a}},\,V_{B}(\zeta) \right] = V_{F}(\tilde{\tilde{u}}), &\quad
\left[ \epsilon^{\dot{a}}Q_{\dot{a}},\,V_{F}(u) \right] = -V_{B}(\tilde{\tilde{\zeta}})\comma 
\label{susyreleps} 
\end{align}
where $\tilde{\zeta}, \tilde{\tilde{\zeta}}$ and 
 $\tilde{u}, \tilde{\tilde{u}}$ are, respectively, the polarization vectors
 and the spinor wave functions obtained by the  supersymmetry 
 transformations of the corresponding super-Maxwell fields. Their explicit 
form will be given in the next subsection. 
Note the minus signs in front of the $V_B$'s on the right hand side. 
We will see below that it is dictated by the consistent realization of the 
super-Poincar\'e symmetries on the vertex operators. 
%
\subsection{SUSY transformation of the wave functions}
To obtain the transformation rules for the wave functions $\zeta^\mu$ and 
 $u^\al$ under the $\ep$- and $\eta$-supersymmetries, we must examine the super-Maxwell theory in 10 dimensions. The action is given by 
\begin{align}
S = \int d^{10}x \left(-\frac{1}{4} F_{\mu\nu}F^{\mu\nu} + 
\frac{i}{2} \bar{\psi}\Gamma^{\mu} \partial_{\mu} \psi \right), 
\end{align}
where  $\psi^{\alpha}(x)$ is a sixteen-component Majorana-Weyl spinor.
It  is invariant under the SUSY transformation with a spinor parameter $\epsilon^{\alpha}$:
\begin{align}
\delta A^{\mu} &= i \bar{\epsilon} \Gamma^{\mu} \psi 
= i \epsilon^{\alpha} \left(\bar{\gamma}^{\mu}\right)_{\alpha\beta} \psi^{\beta}, \label{susytrfA}\\
\delta \psi^{\alpha} &= \frac{1}{2} F_{\mu\nu} \Gamma^{\mu\nu} \epsilon 
= \frac{1}{2} F_{\mu\nu} \left(\gamma^{\mu\nu}\right)^{\alpha}{}_{\beta} \epsilon^{\beta} . \label{susytrfpsi}
\end{align}  
The equation of motion for $A^{\mu}$,  in the Lorentz gauge $\partial_{\mu} A^{\mu} = 0$,  is 
\begin{align}
\Box A^{\mu} = 0 \comma \label{eomA}
\end{align}
while that of the fermion $\psi^{\alpha}$ is 
\begin{align}
\Gamma^{\mu} D_{\mu} \psi = 0 \period  \label{eompsi}
\end{align}
The  SUSY transformation above 
 is compatible with these equations of motion and 
the Lorentz gauge condition.

To separate the $\ep$- and $\eta$-SUSY transformations, we make 
 the $SO(8)$ decomposition of spinors and $\ga$-matrices described in appendix A.2. In particular, 
the spinor field and the SUSY parameter  are decomposed as 
$\psi^{\alpha} = (\psi^{a}, \psi^{\dot{a}})$ and  $\epsilon^{\alpha} = (\eta^{a}, \epsilon^{\dot{a}})$. Then, the SUSY transformation (\ref{susytrfA}) for $A^{\mu}$ is decomposed as 
\begin{align}
\delta A^{+} &= i \epsilon \bar{\gamma}^{+} \psi = 
i \epsilon^{\dot{a}}\bar{\gamma}^{+}_{\dot{a}\dot{b}}\psi^{\dot{b}} = 
- i \sqrt{2} \epsilon^{\dot{a}}\psi_{\dot{a}}, \\ 
\delta A^{-} &= i \epsilon \bar{\gamma}^{-} \psi = 
i \eta^{a}\bar{\gamma}^{-}_{a b}\psi^{b} = i \sqrt{2} \eta^{a}\psi_{a}, \\
\delta A^{I} &= i \epsilon \bar{\gamma}^{I} \psi = i \epsilon^{\dot{a}} \gabar^{I}_{\dot{a} b} \psi^{b} 
+ i \eta^{a} \gabar^{I}_{a \dot{b}} \psi^{\dot{b}} .
\end{align} 
Similarly,  the SUSY transformation (\ref{susytrfpsi}) for 
  $\psi^{\alpha}$ becomes 
\begin{align}
\delta \psi^{a} &= \frac{1}{2}F_{\mu\nu}\left(\left(\gamma^{\mu\nu}\right)^{a}{}_{b} \eta^{b} +
\left(\gamma^{\mu\nu}\right)^{a}{}_{\dot{b}} \epsilon^{\dot{b}} \right) \notag\\
&= \frac{1}{2} F_{I J}\left(\gamma^{I J}\right)^{a}{}_{b} \eta^{b} + F_{+ -} \eta^{a} 
+ \sqrt{2} F_{+ I} \left(\gamma^{I}\right)^{a}{}_{\dot{b}} \epsilon^{\dot{b}} \comma \\
\delta \psi^{\dot{a}} &= \frac{1}{2}F_{\mu\nu}\left(\left(\gamma^{\mu\nu}\right)^{\dot{a}}{}_{b} \eta^{b} +
\left(\gamma^{\mu\nu}\right)^{\dot{a}}{}_{\dot{b}} \epsilon^{\dot{b}} \right) \notag\\
&= \frac{1}{2} F_{I J}\left(\gamma^{I J}\right)^{\dot{a}}{}_{\dot{b}} \epsilon^{\dot{b}} + F_{- +} \epsilon^{\dot{a}} 
+ \sqrt{2} F_{I -} \left(\gamma^{I}\right)^{\dot{a}}{}_{b} \eta^{b} \period
\end{align}

Now the polarization vector $\zeta^{\mu}(k)$ and the fermionic wave function $u^{\alpha}(k)$ are defined  through  the Fourier transform  of $A^{\mu}$ and $\psi^{\alpha}$  as
\begin{align}
A^{\mu}(x) &= \int [dk]~ \zeta^{\mu}(k) e^{i k x} , \\
\psi^{\alpha}(x) &= \int [dk]~ u^{\alpha}(k) e^{i k x} .
\end{align}
We emphasize that since  $\psi^\al$ is Grassmann odd 
 $u^\al$ should also be regarded as Grassmann odd quantity for consistency. 
From the transformation rules for $A^\mu$ and $\psi^\al$, 
it is now straightforward to read off the SUSY transformation properties of
 the components of $\zeta^\mu$ and $u^\al$. 
For the SUSY transformation with the parameter $\eta^{a}$ ($\eta$--SUSY), 
we get
\begin{align}
\delta_{\eta} \zeta^{+} &= 0, \quad \delta_{\eta} \zeta^{-} = i \sqrt{2} \eta^{a} u_{a}, \quad  
\delta_{\eta} \zeta^{I} = i \eta^{a} \gabar^{I}_{a \dot{b}} u^{\dot{b}} , \label{etasusyzeta} \\
\delta_{\eta} u^{a} &= i k_{I} \zeta_{J} \left(\gamma^{I J}\right)^{a}{}_{b} \eta^{b} 
+ i \left(k^{-} \zeta^{+} - k^{+} \zeta^{-}\right) \eta^{a}, \notag \\
\delta_{\eta} u^{\dot{a}} &= i \sqrt{2} \left(k^{I} \zeta^{+} - k^{+} \zeta^{I}\right) 
\left(\gamma^{I}\right)^{\dot{a}}{}_{b} \eta^{b} . \label{etasusyu}
\end{align}   
Similarly, the SUSY transformation with the parameter $\epsilon^{\dot{a}}$ ($\epsilon$--SUSY) is given by 
\begin{align}
\delta_{\epsilon} \zeta^{+} &= -i \sqrt{2} \epsilon^{\dot{a}} 
u_{\dot{a}}\comma 
\quad \delta_{\epsilon} \zeta^{-} = 0\comma  \quad  
\delta_{\epsilon} \zeta^{I} = i \epsilon^{\dot{a}} \gabar^{I}_{\dot{a} b} u^{b} \comma  \label{epssusyzeta} \\
\delta_{\epsilon} u^{a} &= i \sqrt{2} \left(k^{-} \zeta^{I} - k^{I} \zeta^{-}\right) 
\left(\gamma^{I}\right)^{a}{}_{\dot{b}} \epsilon^{\dot{b}} \comma \\
\delta_{\epsilon} u^{\dot{a}} &= i k_{I} \zeta_{J} \left(\gamma^{I J}\right)^{\dot{a}}{}_{\dot{b}} \epsilon^{\dot{b}} 
- i \left(k^{-} \zeta^{+} - k^{+} \zeta^{-}\right) \epsilon^{\dot{a}} \period \label{epssusyu}
\end{align} 
Then $\tilde{\zeta}, \tilde{u}$ etc.  introduced in the previous subsection 
 are identified as $\tilde{\zeta} \equiv \delta_{\eta} \zeta$, $\tilde{u} \equiv \delta_{\eta} u$ and  
$\tilde{\tilde{\zeta}} \equiv \delta_{\epsilon} \zeta$, $\tilde{\tilde{u}} \equiv \delta_{\epsilon} u$. 

The on-shell conditions for the polarization and fermionic wave function can be also obtained by the Fourier transformation of the equations of motion, (\ref{eomA}) and (\ref{eompsi}), 
and the Lorentz gauge condition. 
For the polarization, we have
\begin{align}
k_{\mu}k^{\mu} = 2 k^{+} k^{-} + k^{I} k^{I} &= 0 \comma \\
k_{\mu} \zeta^{\mu} = k^{+} \zeta^{-} + k^{-} \zeta^{+} + k^{I} \zeta^{I} &= 0 \period
\end{align}
As for the fermionic wave function, using the $SO(8)$ decomposition, we get
\begin{align}  
\sqrt{2} k^{+} u_{a} + k^{I} \gabar^{I}_{a \dot{b}} u^{\dot{b}} &= 0 \comma  \\
- \sqrt{2} k^{-} u_{\dot{a}} + k^{I} \gabar^{I}_{\dot{a} b} u^{b} &= 0 \period
\end{align}
In the frame where $k^{+} = 0$, these equations  become 
\begin{align}
k^{I} k^{I} = 0, \quad k^{-} \zeta^{+} + k^{I} \zeta^{I} = 0
\comma  \label{onshellzeta}  \\
k^{I} \gabar^{I}_{a \dot{b}} u^{\dot{b}} = 0, \quad 
\quad \sqrt{2} k^{-} u_{\dot{a}} = k^{I} \bar{\gamma}^{I}_{\dot{a} b} u^{b} \period \label{onshellu}
\end{align}
\subsection{Transformation of the vertex operators under $\eta$-SUSY}
In order to  determine the coefficients in the vertex operators, we first 
 compute  the commutator $\com{\eta^a Q_a}{V_B(\zeta)}$ by the OPE method, where the generator $Q_a$  of the $\eta$-SUSY is 
 is given in  (\ref{SUSYls1}).
Using the OPE formulas (\ref{XPiOPE}) and (\ref{SSOPE}) with 
$\ell_{s} = 1$ and partial integration, we get the following result:
\begin{align}
&\left[ \eta^{a}Q_{a},\, V_{B}(\zeta) \right] = 
\int[dw]~ e^{i k\cdot X(w)} \left\{ \left(2^{\frac{7}{4}}\,C\right)\,k_{I}\,\zeta^{J} \sqrt{\Pi^{+}}\,
\left(\eta\,\gamma^{I J} S\right)(w) \right. \notag\\
&\quad~~ + \left( 2^{-\frac{1}{4}}\, B\right)\,\zeta^{+}\,k^{-} 
\sqrt{\Pi^{+}} \left(\eta^{a} S_{a}(w)\right) + \left( 2^{-\frac{1}{4}}\,B \right)\,\zeta^{+}\,
\frac{k^{I} \Pi^{I}(w)}{\sqrt{\Pi^{+}}} \left(\eta^{a} S_{a}(w)\right) \notag\\
&\left. \quad~~ + \left(2^{\frac{7}{4}}\,D\right)\,\zeta^{+} 
\frac{k_{I} \Pi^{J} \left(\eta\,\gamma^{I J} S\right)(w)}{\sqrt{\Pi^{+}}} + 
\left(2^{\frac{11}{4}}\,E\right)\,\zeta^{+} 
\frac{k_{I}\,R^{J} \left(\eta\,\gamma^{I J} S\right)(w)}{\sqrt{\Pi^{+}}}\right\} \period  
\end{align} 
Here and hereafter, $\exp\left(i k\cdot X \right)$ means 
$\exp\left(i k^{-} X^{+} + i k^{I} X^{I} \right)$ \ie with $k^+=0$. 
 
This should be compared with the fermionic vertex $V_F(\tilde{u})$. 
Inserting $\tilde{u}^{\alpha}$ given in 
(\ref{etasusyu}) (with $k^{+}=0$)  into the general form of $V_F(u)$  
given in (\ref{generalVf}),  we get
\begin{align}
V_{F}\left(\tilde{u}^{a},\,\tilde{u}^{\dot{a}}\right) &= \int[dz]~ e^{i k\cdot X(z)} \left\{
(- i G) k_{I} \zeta^{J} \sqrt{\Pi^{+}} \left(\eta\,\gamma^{I J} S\right)(z) + (i G) k^{-} \zeta^{+} \sqrt{\Pi^{+}} 
\left(\eta^{a} S_{a}(z)\right) \right. \notag\\
&+ \left. \left(i \sqrt{2} K\right) \zeta^{+} k_{J} \left(\eta\,\gamma^{J}\right)^{\dot{a}}
\frac{\left(\bar{\gamma}^{I} S\right)_{\dot{a}} \Pi^{I}(z)}{\sqrt{\Pi^{+}}} 
+ \left(i \sqrt{2} L\right) \zeta^{+} k_{J} \left(\eta\,\gamma^{J}\right)^{\dot{a}}
\frac{\left(\bar{\gamma}^{I} S\right)_{\dot{a}} R^{I}(z)}{\sqrt{\Pi^{+}}}\right\} \notag\\
&= \int[dz]~ e^{i k\cdot X(z)} \left\{
(- i G) k_{I} \zeta^{J} \sqrt{\Pi^{+}} \left(\eta\,\gamma^{I J} S\right)(z) 
+ (i G) k^{-} \zeta^{+} {\sqrt{\Pi^{+}}} \left(\eta^{a} S_{a}(z)\right) \right. \notag\\
&+  \left(i \sqrt{2} K\right) \frac{k_{I} \Pi^{I}(z)}{\sqrt{\Pi^{+}}}\left(\eta^{a} S_{a}(z)\right) 
+ \left(i \sqrt{2} K\right) \frac{k_{I} \Pi^{J}(z)}{\sqrt{\Pi^{+}}} \left(\eta\,\gamma^{IJ} S\right)(z) \notag\\ 
&+ \left. \left(i \sqrt{2} L\right) \frac{k_{I} R^{J}(z)}{\sqrt{\Pi^{+}}} 
\left(\eta\,\gamma^{IJ} S\right)(z)\right\} \period
\end{align}
For the second equality  we have used the $\gamma$-matrix identities (\ref{GammaId2}) and the relation 
$k_{I} R^{I}(z) = k_{I} k_{J} \left(S \gamma^{IJ} S\right)(z) = 0$.     

With these results, the SUSY relation $\left[ \eta^{a}Q_{a},\,V_{B}(\zeta) \right] = V_{F}\left(\tilde{u}\right)$  yields the following relations among the 
 coefficients: 
\begin{align}
& 2^{\frac{7}{4}} C = - i G, \quad 2^{-\frac{1}{4}} B = i G, \quad 2^{-\frac{1}{4}} B = i \sqrt{2} K, 
\notag \\
& 2^{\frac{7}{4}} D = i \sqrt{2} K, \quad 2^{\frac{11}{4}} E = i \sqrt{2} L.
\label{SUSYboson}
\end{align}

Next, we consider the second part of the SUSY relation (\ref{susyreleta}), 
namely $\com{\eta^a Q_a}{V_F(u)} = - V_B(\tilde{\zeta})$. 
The commutator can be calculated  straightforwardly as
\begin{align}
\left[ \eta^{a}Q_{a},\,V_{F}(u) \right] &= \quad \int[dw]~ e^{i k\cdot X(w)} 
\left\{ \left(2^{\frac{3}{4}} G\right) \Pi^{+}(w) \left(\eta^{a} u_{a}\right) 
+ \left(2^{\frac{3}{4}} K\right) \Pi^{I}(w) \left(\eta\,\gamma^{I} u\right) \right. \notag \\
&\left. + \left(2^{\frac{3}{4}} L\right) \left( R^{I}(w) \left(\eta\,\gamma^{I} u\right) - 2\, k_{J} 
\left(u\,\bar{\gamma}^{I} S\right) \left(\eta\,\gamma^{I J} S\right)\right) \right\} \period
\end{align}
Reshuffling the spinor indices by the identity (\ref{GammaId1}) and 
using the on-shell condition (\ref{onshellu}) for the fermion,
the last term can be rearranged  into 
\begin{align}
k_{J} \left(u\,\bar{\gamma}^{I} S\right) \left(\eta\,\gamma^{I J} S\right) = 
- \left(\eta\,\gamma^{I} u\right) k_{J} \left(S\,\gamma^{I J} S\right) = - R^{I} \left(\eta\,\gamma^{I} u\right).
\end{align}
Then, the  commutator simplifies to 
\begin{align}
\left[ \eta^{a}Q_{a},\,V_{F}(u) \right] &= \int[dw]~ e^{i k\cdot X(w)} 
\left\{ \left(2^{\frac{3}{4}} G\right) \left(\eta^{a} u_{a}\right) \Pi^{+}(w) 
+ \left(2^{\frac{3}{4}} K\right) \left(\eta\,\gamma^{I} u\right) \Pi^{I}(w)\right. \notag\\ 
&\quad \left. + \left(2^{\frac{3}{4}}~3~L\right) \left(\eta\,\gamma^{I} u\right) R^{I}(w) \right\}.
\end{align}

On the other hand,  with the form of the polarization $\tilde{\zeta}^{\mu}$ given in (\ref{etasusyzeta}) inserted into $V_B(\zeta)$ in 
(\ref{generalVb}),  $V_B(\tilde{\zeta})$ is obtained as 
\begin{align}
V_{B}(\tilde{\zeta}) = \int[dw]~ e^{i k\cdot X(w)} \left\{ i \sqrt{2} \left(\eta^{a} u_{a}\right)
A~\Pi^{+}(w) + i \left(\eta\,\gamma^{I} u\right) \left(B~\Pi^{I}(w) + C~R^{I}(w)\right)\right\} \period
\end{align}
Comparing these results, we obtain further relations among the coefficients as
\begin{align}
-i \sqrt{2} A = 2^{\frac{3}{4}} G, \quad -i B = 2^{\frac{3}{4}} K, \quad -i C = 2^{\frac{3}{4}}~3~L \period
\label{SUSYfermion}
\end{align}   

Now from the set of relations  (\ref{SUSYboson}) and   (\ref{SUSYfermion})
we can determine many of the coefficients. Let us fix the overall normalization 
 by setting  $B = 1$, which is the conventional  choice  in the case of  the full light-cone gauge vertex. Then the unique solution satisfying the above relations is 
\begin{align}
A = 1, \quad & B = 1, \quad C = - \frac{1}{4}, 
\quad D = \frac{1}{4}, \quad E = - \frac{1}{96} \comma \notag \\
& G = - i\, 2^{- \frac{1}{4}}, \quad K = -i\, 2^{- \frac{3}{4}}, \quad
L = i\, \frac{2^{-\frac{3}{4}}}{12} \period \label{CorrectAns}
\end{align}
It  should be remarked 
that the coefficients $A = B = 1, C = - \frac{1}{4}$ for  $V_{B}(\zeta)$ and $G, K, L$ in (\ref{CorrectAns}) for $V_{F}(u)$ are consistent with 
the result in the full light-cone gauge.\cite{Green:1981xx, Green:1987sp} 

On the other hand, the coefficients  $Y$ and $Z$ in  $V_B(\zeta)$
 have not been determined by the forgoing analysis. We will see that they will  be fixed by the analysis of the property  under $\ep$-SUSY. 
\subsection{Transformation of the vertex operators under $\epsilon$-SUSY}
From the $\eta$--SUSY relations, we have been able to fix 
 the form of the massless vertex operators except for a few terms. 
We will now check if  these vertex operators also satisfy the $\epsilon$--SUSY relations and fix the remaining coefficients. 

This analysis is rather non-trivial since the $\epsilon$--SUSY transformation contains the conpensating $\kappa$ transformation induced by the SLC gauge fixing condition.  As a result, there will appear certain BRST-exact terms 
 in the consistency  relations. Explicitly, we will find that the following relations 
are realized:
\begin{align}
& \left[ \epsilon^{\dot{a}}Q_{\dot{a}},\,V_{F}(u) \right] = -\,V_{B}(\tilde{\tilde{\zeta}})~ 
+~ \left\{ Q ,~\Psi_{F}(\ep, u) \right\} \comma \label{epscomVf} \\  
& \left[ \epsilon^{\dot{a}}Q_{\dot{a}},\,V_{B}(\zeta) \right] = V_{F}(\tilde{\tilde{u}})~ 
+~ \left\{ Q,~\Psi_{B}(\ep, \zeta) \right\} \period \label{epscomVb}
\end{align}
Here  $Q$ is the BRST charge defined in (\ref{Q}) and $\Psi_F$ and $\Psi_B$
 are appropriate fermionic operators.  Note that since $V_F(u)$ has already been  completely fixed by the analysis of the previous subsection, the minus sign 
 in front of $V_{B}(\tilde{\tilde{\zeta}})$ is dictated by the calculation 
 of the commutator $\left[ \epsilon^{\dot{a}}Q_{\dot{a}},\,V_{F}(u) \right]$. 

First, let us  derive  the  relation (\ref{epscomVf}). 
The supercharge $Q_{\dot{a}}$ for $\epsilon$--SUSY is given by (\ref{SUSYls1}) and the fermion vertex operator $V_{F}(u)$, which has been 
 determined  completely from the previous analysis,  is of  the form 
\begin{align}
V_{F}(u) &= \int[dz]~ e^{i k\cdot X(z)}\, \left\{u^{a}\,\left( -i\,2^{-1/4} 
\sqrt{\Pi^{+}}\,S_{a}(z)\right) \right. \notag\\
& \left. +~ u^{\dot{a}}\,\left( -i\,2^{-3/4} \frac{\left(\bar{\gamma}^{I} S\right)_{\dot{a}}\,\Pi_{I}(z)}
{\sqrt{\Pi^{+}}} 
+ i \left(\frac{2^{-3/4}}{12}\right)\frac{\left(\bar{\gamma}^{I} S\right)_{\dot{a}}\,
R_{I}(z)}{\sqrt{\Pi^{+}}}\right) \right\} \period \label{FVertex} 
\end{align}   
Using the OPE technique, the relevant commutator 
$\left[ \epsilon^{\dot{a}} Q_{\dot{a}} ,\, V_{F}(u) \right]$ 
is  computed as\footnote{For simplicity, 
we use the abbreviation such as $\epsilon \bar{\gamma}^{I} u = 
\epsilon^{\dot{a}} \bar{\gamma}^{I}_{\dot{a} b} u^{b}$,~$\epsilon u = \epsilon^{\dot{a}} u_{\dot{a}}$,~
$u S = u^{a} S_{a}$,~$u \bar{\gamma}^{I} S = u^{\dot{a}} \bar{\gamma}^{I}_{\dot{a} b} S^{b}$, etc. 
and  omit the functional dependence on $w$ in the integrand.}
\begin{align}
&\left[ \epsilon Q ,\, V_{F}(u) \right] = \int[dw]~ e^{i k\cdot X}~ 
\biggl[ -i\, \Pi_{I} \left(\epsilon \bar{\gamma}^{I} u\right) 
- i\, \partial\left(\frac{1}{\sqrt{\Pi^{+}}}\right) \sqrt{\Pi^{+}}\, k_{I} 
\left(\epsilon \bar{\gamma}^{I} u\right) \notag\\
& -~ \frac{i}{\sqrt{2}}\frac{\Pi_{I} \Pi^{I}}{\Pi^{+}} \left(\epsilon u\right) 
- i \sqrt{2}\, \partial\left(\frac{1}{\sqrt{\Pi^{+}}}\right)\frac{k_{I} \Pi^{I}}{\sqrt{\Pi^{+}}} 
\left(\epsilon u\right) - i\, 2\sqrt{2}\, \partial^{2}\left(\frac{1}{\sqrt{\Pi^{+}}}\right) \frac{1}{\sqrt{\Pi^{+}}} 
\left(\epsilon u\right) \notag \\
& +~ \left(\frac{i}{12\sqrt{2}}\right) 
\frac{\Pi_{I} R^{I}}{\Pi^{+}} \left(\epsilon u\right) + \left(\frac{i}{12\sqrt{2}}\right) 
\frac{\Pi_{I} R_{J}}{\Pi^{+}} \left(\epsilon \bar{\gamma}^{I J} u\right)  \nn\\
& ~
+ \calA_1 +\calA_2 + \calA_3 + \calA_4 + \calA_5\biggr] \comma \label{OPEepsVf} 
\end{align}
where 
\begin{align}
\calA_1 &=i\, k_{I} \left(\epsilon \bar{\gamma}^{I} S\right) \left(u S \right)    \comma \label{epQVFone}\\ 
\calA_2 &= \frac{i}{\sqrt{2}} \frac{k_{I} \left(\epsilon \bar{\gamma}^{I} S\right) 
\Pi_{J} \left(u \bar{\gamma}^{J} S\right)}{\Pi^{+}}    \comma \label{epQVFtwo}\\
\calA_3 &= - \left(\frac{i \sqrt{2}}{12}\right)\frac{\Pi_{I}\left(u \bar{\gamma}^{J} S\right) k_{L} 
\left(\epsilon \bar{\gamma}^{I} \gamma^{J L} S\right)}{\Pi^{+}}  \comma  \label{epQVFthree}\\
\calA_4 &= \frac{i}{\sqrt{2}}\, \partial\left(\frac{\left(\epsilon \bar{\gamma}^{I} S\right)}{\sqrt{\Pi^{+}}}\right)
\frac{\left(u \bar{\gamma}^{I} S\right)}{\sqrt{\Pi^{+}}}\comma \label{epQVFfour}\\
\calA_5 &= 
 -~ \left(\frac{i}{12\sqrt{2}}\right)\frac{k_{I}\left(\epsilon \bar{\gamma}^{I} S\right) R_{J} 
\left(u \bar{\gamma}^{J} S\right)}{\Pi^{+}}   
  \period \label{epQVFfive} 
\end{align}
(We have singled out  the last 5 terms for later convenience.) 
In this computation, we have made appropriate use of the equation of motion for the fermion wave function (\ref{onshellu}). 

On the other hand, by inserting the form of the polarization vector $\tilde{\tilde{\zeta}}^{\mu}$ given in (\ref{epssusyzeta}), the boson  vertex operator 
$V_B(\tilde{\tilde{\zeta}})$ with the coefficients (\ref{CorrectAns})
becomes
\begin{align}
V_{B}(\tilde{\tilde{\zeta}}) &= \int[dw]~ e^{i k\cdot X}~ 
\left[ -i \sqrt{2} \left(\epsilon u\right) \left( \hat{\Pi}^{-} + \frac{1}{4} \frac{\Pi_{I} R^{I}}{\Pi^{+}}
- \frac{1}{96} \frac{R_{I} R^{I}}{\Pi^{+}} \right. \right. \notag\\
& \Biggl. \Biggl. +~ Y\, k^{-}\, k_{I}\Pi^{I} + Z\, \frac{\left(k_{I} \Pi^{I}\right)
\left(k_{J}\Pi^{J}\right)}{\Pi^{+}}\Biggr) +~ i \left(\epsilon \bar{\gamma}^{I} u\right) 
\left( \Pi_{I} - \frac{1}{4} R_{I} \right) \Biggr] \period \label{transformVb}
\end{align}  

We now wish to compare the results (\ref{OPEepsVf}) and (\ref{transformVb}). 
This turned out to require heavy use of  various $\ga$-matrix identities 
summarized in appendix C.1 to rewrite the terms 
 denoted by $\calA_1 \sim \calA_5$ in (\ref{OPEepsVf}).
 Below we sketch this procedure. 

First, focus on  the expression $\calA_1$, which is 
quadratic in the fermion operator $S$. By using the  Fierz identity (\ref{Fierz1}), it can be rewritten as
\begin{align}
i\, k_{I} \left(\epsilon \bar{\gamma}^{I} S\right) \left(u S\right) &= 
- \frac{i}{16}\left(S \gamma^{K L} S\right) k_{I} 
\left(\epsilon \bar{\gamma}^{I} \gamma^{K L} u\right) \notag \\
&= - \frac{i}{16}\left(S \gamma^{K L} S\right) k_{I} 
\left(\epsilon \left[ \bar{\gamma}^{I}, \gamma^{K L}\right] u\right) 
- \frac{i}{16}\left(S \gamma^{K L} S\right) k_{I} 
\left(\epsilon \bar{\gamma}^{K L} \bar{\gamma}^{I} u\right). 
\end{align}
Recalling the definition of the operator $R_{I}$ given in (\ref{defRI}), 
applying the Fierz identity (\ref{Fierz2}),  and making use of 
the on-shell condition (\ref{onshellu}) for the fermion, this expression 
can be simplified to  
\begin{align}
\calA_1 = 
\frac{i}{4} R_{I} \left(\epsilon \bar{\gamma}^{I} u\right) + \frac{i}{2 \sqrt{2}} 
k^{-} \left(\epsilon \bar{\gamma}^{I} S\right) \left(u \bar{\gamma}^{I} S\right) \period
\label{quadraticfermion1}
\end{align}

In a similar manner, the numerator of the expression $\calA_2$ can be transformed,  with the help of  (\ref{GammaId2}), (\ref{Fierz1}), (\ref{cubicgamma com}) and (\ref{onshellu}), into 
\begin{align}
k_{I} \left(\epsilon \bar{\gamma}^{I} S\right) \left(u \bar{\gamma}^{J} S\right) 
&= - \frac{1}{16} \left(S \gamma^{K L} S\right) k_{I} \left(\epsilon \bar{\gamma}^{I} 
\gamma^{K L} \gamma^{J} u\right) \notag\\ 
&= - \frac{1}{16} \left(S \gamma^{K L} S\right) k_{I} 
\left[ \epsilon \left( \left[\bar{\gamma}^{I},\, \gamma^{KL} \right] \gamma^{J} 
+ \bar{\gamma}^{KL} \left\{\bar{\gamma}^{I},\, \gamma^{J}\right\} - 
\bar{\gamma}^{KL} \bar{\gamma}^{J} \gamma^{I} \right) u \right] \notag\\ 
&= \frac{1}{4} R_{J} \left(\epsilon u\right) + \frac{1}{4} R_{I} 
\left(\epsilon \bar{\gamma}^{I J} u\right) + \frac{1}{2} k_{J} 
\left(\epsilon \bar{\gamma}^{I} S\right) \left(u \bar{\gamma}^{I} S\right) \period
\end{align}
Then $\calA_2$ becomes 
\begin{align}
\calA_2 &=
 \frac{i}{4 \sqrt{2}}\frac{\Pi^{J} R_{J}}{\Pi^{+}} \left(\epsilon u\right)
 + 
\frac{i}{4 \sqrt{2}}\frac{\Pi_{J} R_{I}}{\Pi^{+}} 
\left(\epsilon \bar{\gamma}^{I J} u\right) + \frac{i}{2 \sqrt{2}} 
\frac{k_{J} \Pi^{J}}{\Pi^{+}} \left(\epsilon \bar{\gamma}^{I} S\right) 
\left(u \bar{\gamma}^{I} S\right) . \label{quadraticfermion2}
\end{align}

Now we move on to the expression $\calA_3$. By using (\ref{Fierz1})
 it can be rewritten as 
\begin{align}
& k_{L} \left(u \bar{\gamma}^{J} S\right) \left(\epsilon \bar{\gamma}^{I} 
\gamma^{J L} S\right) = - \frac{1}{16} k_{L} \left(S \gamma^{M N} S\right) 
u \left(\bar{\gamma}^{J} \gamma^{M N} \gamma^{L J} 
\gamma^{I} \right) \epsilon \notag \\
& = - \frac{1}{16} k_{L} \left(S \gamma^{M N} S\right) 
u \left( \bar{\gamma}^{J} \left[\gamma^{M N},\, \gamma^{L J}\right] 
\gamma^{I} + \bar{\gamma}^{J} \gamma^{L J} \gamma^{M N} \gamma^{I} \right) \epsilon \comma 
\end{align}
where, actually, the last term vanishes due to the identity $\bar{\gamma}^{J} \gamma^{L J} 
= - 7\, \bar{\gamma}^{L}$ and (\ref{onshellu}).
Using the $\gamma$-commutator (\ref{quarticgamma com}), the first term becomes 
\begin{align}
&- \frac{1}{16} k_{L} \left(S \gamma^{M N} S\right) 
\left(u\, \bar{\gamma}^{J} \left[\gamma^{M N},\, \gamma^{L J}\right] \gamma^{I} \epsilon\right) 
\notag\\
&= -\frac{1}{4} R_{M} \left(u\, \bar{\gamma}^{J} \gamma^{M J} \gamma^{I} \epsilon\right)
- \frac{1}{4} k_{L} \left(S \gamma^{J M} S\right) 
\left(u\, \bar{\gamma}^{J} \gamma^{M L} \gamma^{I} \epsilon\right)
= R_{J} \left(u\,\bar{\gamma}^{J} \gamma^{I} \epsilon\right)
\period
\end{align}
For the last equality, the fermionic on-shell condition (\ref{onshellu}) was used. In this way, $\calA_3$ can be re-expressed as 
\begin{align}
\calA_3
&= - \left(\frac{i \sqrt{2}}{12}\right) \frac{\Pi_{I}}{\Pi^{+}} R_{J} 
\left(u\,\bar{\gamma}^{J} \gamma^{I} \epsilon\right) 
= \left(\frac{i \sqrt{2}}{12}\right)\frac{\Pi_{I} R^{I}}{\Pi^{+}} \left(\epsilon u\right)
+ \left(\frac{i \sqrt{2}}{12}\right)\frac{\Pi_{I} R_{J}}{\Pi^{+}} 
\left(\epsilon \bar{\gamma}^{I J} u\right)
\period  \label{quadraticfermion3}
\end{align}

Next consider the expression $\calA_4$. For this term, we perform an integration by parts appropriately and rewrite terms in a manner similar to the 
manipulation leading from (\ref{cubicS Id1}) to (\ref{cubicS Id2}). 
We then get
\begin{align}
\calA_4
= \frac{i}{\sqrt{2}}\,\frac{S \partial S}{\Pi^{+}} \left(\epsilon u\right)
- \frac{i}{2 \sqrt{2}} 
\frac{k^{-} \Pi^{+} + k_{J} \Pi^{J}}{\Pi^{+}}
\left(\epsilon \bar{\gamma}^{I} S\right) \left(u \bar{\gamma}^{I} S\right) 
\period 
\label{quadraticfermion4}
\end{align}

Finally consider $\calA_5$. The numerator of this expression  
can be rearranged by the Fierz identity (\ref{Fierz1}) and the (anti-)commutators of $\gamma$-matrices (\ref{defgamma8}) and (\ref{cubicgamma com})
in the following way:
\begin{align}
&k_{I}\left(\epsilon \bar{\gamma}^{I} S\right) R_{J} 
\left(u \bar{\gamma}^{J} S\right) = -\frac{1}{16} k_{I} R_{J} \left(S\gamma_{KL}S\right) 
\left( \epsilon \bar{\gamma}^{I} \gamma^{KL} \gamma^{J} u \right) \notag \\
&= -\frac{1}{16} k_{I} R_{J} \left(S\gamma_{KL}S\right) 
\epsilon \left( \left[\bar{\gamma}^{I},\, \gamma^{KL} \right] \gamma^{J} 
+ \bar{\gamma}^{KL} \left\{\bar{\gamma}^{I},\, \gamma^{J}\right\} - 
\bar{\gamma}^{KL} \bar{\gamma}^{J} \gamma^{I} \right) u = 
\frac{1}{4} R_{I} R^{I} \left( \epsilon u \right) \comma
\end{align}
Here we have used the identity $k_{I} R^{I} = 0$ and the on-shell condition (\ref{onshellu}) for the last equality. This leads to the simple result 
\begin{align}
\calA_5= -i \sqrt{2} 
\left( \epsilon u \right) \left(\frac{1}{96} \frac{R_{I} R^{I}}{\Pi^{+}}\right). \label{quarticfermion}
\end{align}

Substituting the reults   (\ref{quadraticfermion1}), 
(\ref{quadraticfermion2}), (\ref{quadraticfermion3}), (\ref{quadraticfermion4})  and 
(\ref{quarticfermion}) for $\calA_1 \sim \calA_5$ into the commutator  (\ref{OPEepsVf}) and using the on-shell condition (\ref{onshellu}) for the fermion,  
we obtain 
\begin{align}
&\left[ \epsilon Q ,\, V_{F}(u) \right] = \int[dw]~ e^{i k\cdot X}~ 
\left[ -i \left(\epsilon \bar{\gamma}^{I} u\right) \left(\Pi_{I} - \frac{1}{4} R_{I}\right)
+ i \sqrt{2} \left(\epsilon u\right) \left(\frac{1}{4} \frac{\Pi_{I} R^{I}}{\Pi^{+}}  
- \frac{1}{96} \frac{R_{I} R^{I}}{\Pi^{+}}\right) \right. \notag \\
& \left. -~ i \sqrt{2} \left(\epsilon u\right) \left\{\frac{1}{2}\left(\frac{\Pi_{I} \Pi^{I}}{\Pi^{+}} - 
\frac{S \partial S}{\Pi^{+}}\right) + \partial\left(\frac{1}{\sqrt{\Pi^{+}}}\right) 
\frac{k^{-} \Pi^{+} + k_{I} \Pi^{I}}{\sqrt{\Pi^{+}}} 
+ \frac{2}{\sqrt{\Pi^{+}}} \partial^{2}\left(\frac{1}{\sqrt{\Pi^{+}}}\right) 
\right\} \right] \period \label{OPEepsVf2} 
\end{align}
We see that the terms  in the first line constitute a part of  $-V_B(\tilde{\tilde{\zeta}})$, where $V_B(\tilde{\tilde{\zeta}})$ is  given in (\ref{transformVb}). 
This means that the relation (\ref{epscomVf}) holds  if the remaining 
 terms in the second line supply   the rest of $-V_B(\tilde{\tilde{\zeta}})$, 
plus an expression  of the form   $\acom{Q}{\Psi_F(\ep, u)}$ for some 
fermionic operator $\Psi_F(\ep, u)$.  In the next subsection we shall see  that 
 this is indeed the case for a particular choice of the coefficients $Y$ and $Z$. 
\subsection{BRST-exact terms in $\ep$-SUSY relations}
Inspection of some of the terms in the second line of (\ref{OPEepsVf2})
 suggests that the candidate  fermionic  operator $\Psi_F(\ep, u)$ would be 
\begin{align}
\Psi_{F}(\ep, u) = \int[dw]~ \left(- i \sqrt{2}\, \frac{b(w)}{\Pi^{+}(w)}\, e^{ik\cdot X(w)} \right) 
\left(\epsilon u\right) \comma 
\label{FermionOp1}  
\end{align}
where $b(w)$ is the reparametrization anti-ghost.  The anticommutator $\acom{Q}{\Psi_F(\ep, u)}$ can then be computed straightforwardly and gives 
\begin{align}
\left\{ Q,\, \Psi_{F}(\ep,u) \right\} &= \int[dw]~ e^{ik\cdot X} \left(\epsilon u\right) 
\left[ - \frac{i}{\sqrt{2}}\, \frac{\Pi_{I}\Pi^{I}(w)}{\Pi^{+}} +   
\frac{i}{\sqrt{2}}\, \frac{S \partial S}{\Pi^{+}} \notag \right.\\
& \left. -~ i \sqrt{2}\, \Pi^{-} + \frac{i}{\sqrt{2}}\, \frac{\left(\partial \Pi^{+}\right)^{2}}
{\left(\Pi^{+}\right)^{3}} - i \sqrt{2}\, \frac{\left(k^{-} \partial\Pi^{+} + 
k_{I} \partial\Pi^{I}\right)}{\Pi^{+}} \right]. \label{QexactVf} 
\end{align}
Note  that the first two terms are precisely those  that appear in (\ref{OPEepsVf2}),  while the third term can be recognized as a part of $V_B(\tilde{\tilde{\zeta}})$. 
This indicates that indeed the operator $\Psi_F(\ep, u)$ above should be relevant. 
(We should remark  that the terms which depend on the ghosts 
together form a total derivative inside the integral and drop out. )

To make the statement more precise,   let us  compute the difference of the left and the right 
 hand sides  of the equation (\ref{epscomVf}),  using the results (\ref{OPEepsVf2}) and (\ref{QexactVf}) above. After performing a partial  integration\footnote{For instance we use the formula  $\partial\left(e^{ik\cdot X}\right) = \left(k^{-} \Pi^{+} + k_{I} \Pi^{I}\right)\, e^{ik\cdot X}$.} we obtain 
\begin{align}
&\left[ \epsilon Q,\, V_{F}(u) \right] + V_{B}(\tilde{\tilde{\zeta}}) 
- \left\{ Q,\Psi_{F}(\ep, u) \right\} \notag \\
& = \left(-i \sqrt{2}\right) \int[dw]~ e^{ik\cdot X} \left(\epsilon u\right) 
\left[\partial\left(\frac{1}{\sqrt{\Pi^{+}}}\right)
\frac{\left(k^{-} \Pi^{+} + k_{I} \Pi^{I}\right)}{\sqrt{\Pi^{+}}} 
+ \frac{2}{\sqrt{\Pi^{+}}} \partial^{2}\left(\frac{1}{\sqrt{\Pi^{+}}}\right) \right. \notag\\
& \left.  +  \frac{3}{2}\frac{\left(\partial \Pi^{+}\right)^{2}}{\left(\Pi^{+}\right)^{3}} 
- \frac{1}{2} \frac{\partial^{2} \Pi^{+}}{\left(\Pi^{+}\right)^{2}} - 
\frac{\left(k^{-} \partial\Pi^{+} + k_{I} \partial\Pi^{I}\right)}{\Pi^{+}} + 
Y\, k^{-}\, k_{I}\Pi^{I} + Z\, \frac{\left(k_{I} \Pi^{I}\right) 
\left(k_{J}\Pi^{J}\right)}{\Pi^{+}}\right] \notag \\
& = \left(-i \sqrt{2}\right) \int[dw]~ e^{ik\cdot X} \left(\epsilon u\right) 
\left[ (Y + 1) k^{-}\, k_{I}\Pi^{I} + (Z + 1) \frac{\left(k_{I} \Pi^{I}\right) 
\left(k_{J}\Pi^{J}\right)}{\Pi^{+}}\right] . \label{remainingsVf}
\end{align}
This expression vanishes as desired  if and only if we make the choice 
\begin{align}
Y = - 1 \quad \textrm{and} \quad Z = - 1 \period 
\end{align}
This fixes the boson vertex operator completely with the result 
\begin{align}
V_{B}(\zeta) &= \int[dz]~ e^{i k\cdot X(z)}\, \Biggl[\zeta^{-}\, \Pi^{+}(z) + 
\zeta^{I}\,\left(\Pi_{I}(z) -\frac{1}{4} R_{I}(z)\right)\Bigr. \notag\\
& \left. +~\zeta^{+}\left(\hat{\Pi}^{-}(z) + \frac{1}{4} \frac{\Pi^{I}\,R_{I}(z)}{\Pi^{+}} 
- \frac{1}{96} \frac{R^{I}\,R_{I}(z)}{\Pi^{+}} - k^{-}\, k_{I}\Pi^{I}(z) - \frac{\left(k_{I} \Pi^{I}\right)
\left(k_{J}\Pi^{J}\right)(z)}{\Pi^{+}}\right) \right] \period \label{BVertex}
\end{align} 

Our last remaining task is to prove that the relation (\ref{epscomVb}) 
for the $\ep$-SUSY transformation holds for our vertices with some 
 choice of a fermionic  operator $\Psi_B(\ep, \zeta)$. 

After some non-trivial calculations similar to the above (some details are given 
 in  appendix C.3), we find that indeed the desired $\ep$-SUSY relation 
(\ref{epscomVb}) holds, with 
 the fermionic  operator $\Psi_B(\ep, \zeta)$ is given by 
\begin{align}
\Psi_{B}(\ep, \zeta) = - 2^{1/4}\, \zeta^{+}\, \int[dw]~ 
\left(\frac{k_{I} \left(\epsilon\, \bar{\gamma}^{I} S\right) b(w)}
{\left(\Pi^{+}\right)^{3/2}}\, e^{ik\cdot X(w)} \right). \label{BosonOp}
\end{align}

Note that the fermionic operators $\Psi_{B}(\ep, \zeta)$ and $\Psi_{F}(\ep, u)$ are
proportional to $\zeta^{+}$ and $\tilde{\tilde{\zeta}}^{+}$ respectively. 
This shows that these anomalous terms are indeed unphysical, since 
$\zeta^{+}$ and hence $\tilde{\tilde{\zeta}}^{+}$ vanish  in the physical 
light-cone gauge in  the super-Maxwell theory. 

As a further check of our vertex operators, we have explicitly computed 
the following double commutators, which show  that the SUSY algebras (\ref{Qab}) and (\ref{Qadotb}) 
are properly realized on our vertex operators:
\begin{align}
&\bigl[ \eta_{1}Q,\,\left[\eta_{2}Q,\, V_{B (F)}\right]\bigr] - 
\bigl[ \eta_{2}Q,\,\left[\eta_{1}Q,\, V_{B (F)}\right]\bigr] = 0 = - 2\sqrt{2}\, (\eta_{1}\eta_{2}) 
\left[p^{+},\, V_{B (F)}\right] , \\
&\bigl[ \eta Q,\,\left[\epsilon Q,\, V_{B (F)}\right]\bigr] - 
\bigl[ \epsilon Q,\,\left[\eta Q,\, V_{B (F)}\right]\bigr] = - 2\, (\eta \gamma^{I} \epsilon) 
\left[p^{I},\, V_{B (F)}\right] \period \label{etaepscom}
\end{align} 
For the verification of  (\ref{etaepscom}), the following identities have been 
used:
\begin{align}
\left[ \eta Q,\, \Psi_{B}(\ep, \zeta) \right] = \Psi_{F}(\ep, \tilde{u}) \comma \quad 
\left[ \eta Q,\, \Psi_{F}(\ep, u) \right] = 0 = \Psi_{B}(\ep, \tilde{\zeta}) .
\end{align} 
We have also checked  that the remaining double commutators are consistent with the closure relation (\ref{Qdotadotb}):
\begin{align}
&\bigl[ \epsilon_{1}Q,\,\left[\epsilon_{2}Q,\, V_{B (F)}\right]\bigr] - 
\bigl[ \epsilon_{2}Q,\,\left[\epsilon_{1}Q,\, V_{B (F)}\right]\bigr] \notag \\
& = 2\sqrt{2}\, (\epsilon_{1}\epsilon_{2}) \left[p^{-},\, V_{B (F)}\right] 
- 2\sqrt{2}\, (\epsilon_{1}\epsilon_{2}) \left\{Q,\, 
\left[\int[dw]\frac{b(w)}{\Pi^{+}},\, V_{B (F)}\right]\right\} \comma 
\end{align}
For this calculation, we have used $\left[Q,\, V_{B (F)}\right] = 0$ 
and 
the following identities:
\begin{align}
\Bigl(\left[\epsilon_{1}Q,\, \Psi_{B}(\epsilon_{2}, \zeta)\right] 
+ \Psi_{F}( \epsilon_{1},\delta_{\epsilon_{2}}u) \Bigr)
 - ( 1 \leftrightarrow 2 ) &= - 2\sqrt{2}\, (\epsilon_{1}\epsilon_{2}) 
\left[\int[dw]~\left(\frac{b(w)}{\Pi^{+}}\right),\, V_{B}(\zeta)\right] \comma \\
\left[\epsilon_{1}Q,\, \Psi_{F}(\epsilon_{2},u)\right] &= 
\Psi_{B}(\epsilon_{1}, \delta_{\epsilon_{2}}\zeta ) \period
\end{align} 
Note that these double commutator relations confirm once more that the choice of the negative signs on the right-hand side of the SUSY relations, 
(\ref{susyreleta}) and (\ref{susyreleps}),  is the correct one. 

In summary, we have constructed the BRST-invariant vertex operators, 
(\ref{BVertex}) for $V_{B}(\zeta)$ and (\ref{FVertex}) for $V_{F}(u)$, of the Green-Schwarz superstring at the lowest (massless) level, which 
describe  the photons and photinos in the ten-dimensional super-Maxwell theory. Also, we have shown that these vertex operators indeed form the representation of the quantum super-Poincar\`e algebra on the physical Hilbert space. 

Finally, let us make an important  remark on how one can use 
 these vertex operators 
 for the computation of the scattering amplitudes. Although the formulation 
in the SLC gauge has the advantage that the conformal symmetry is preserved  and 
the BRST invariance can be utilized,  some of the features remain to be  the same as in 
 the case of the LC gauge.  Notably,   the  vertex operators can be 
 constructed so far only for special  external momenta  satisfying $k^+=0$.  Also the inverse powers of $\del X^+$  appear in some parts of the vertex operators, which 
 are well-defined only when the zero mode part of $\del X^+$  has non-vanishing eigenvalue.  Therefore to compute the amplitudes, we should follow the 
scheme used in the LC gauge calculation\cite{Green:1981xx, Green:1987sp}. 
Namely,  the $N$-point amplitude $A_N$ should be computed,
schematically, 
as  $A_N \sim \langle k_1
| V_2(k_2) V_3(k_3) \cdots V_{N-1}(k_{N-1}) |  k_N \rangle$  for  
special momentum  cofiguration where $k_1^+ = k^+_N \ne 0$ for the initial and the final momenta and  $k_i^+=0$ for 
$i=2,3,\ldots, N-1$.  (Note that  $ k_i^I $ should  be allowed to be complex due to the on-shell condition.)  In such a configuration 
 $(\del X^+)^{-1}$'s in $V_i$  are well-defined and the amplitude can be computed properly.  Finally, for low values of $N$,  the amplitude so obtained for  special configuration can be  uniquely Lorentz-covariantized to yield the result valid for general configuration. 
It is an important future problem to improve on this situation by removing the 
 restriction $k^+=0$ for the vertex operators. 

\section{Similarity transformation  to the LC gauge and construction 
 of the DDF operators}
In this section, we will discuss the relation between  the formulation in the 
SLC-conformal  gauge we have  developed  to the conventional (\ie full)
LC  gauge formulation.  Explicitly, we will be able to find {\it the 
quantum similarity transformation which connects the operators in the two 
formulations.}  As an application,  the spectrum-generating DDF(Del Giudice-Di Vecchia-Fubini)  operators\cite{DelGiudice:1971fp} in the SLC gauge will be constructed 
using this method. 
\subsection{Similarity transformation to the LC gauge}
The most basic connection  between the two formulations is of course that 
the physical spectrum is the same. In other words, the cohomology of the 
BRST operator  $Q$ of the SLC gauge given in (\ref{Q}) coincides with  
the states  generated by the physical oscillators,  $\al^I_{-n}$ and $S^a_{-n}$, 
 of  the LC gauge.  One can prove this  by adapting the standard argument, 
 given  for example in the  Polchinski's book\cite{Polchinski:1998rq}
 for the bosonic  string, to the case of the superstring. 
 An  apparent complication, however,   is that our  Virasoro operator 
$T$  contains  a  complicated operator  proportional to $\del^2 \ln \Pi^+$, 
  which is not quadratic  in the fields as in the bosonic string case. 
 Essentially similar situation was already encountered  for the GS 
 formulation of the superstring  in the plane wave background in the SLC gauge
studied in Ref.~\citen{Kazama:2008as}. 
There it was shown that as long as the 
 non-linear term (such as $\del^2 \ln \Pi^+$)  contains at least one non-zero 
 mode of $\Pi^\pm$, the cohomology of $Q$ is independent of such a term 
and coincides  with the light-cone Hilbert space.  As this argument directly applies 
 to the present case, we will not repeat the proof. 
For more details see Ref.~\citen{Kazama:2008as}. 

The purpose of the present subsection is to demonstrate  that actually 
a much more concrete  relation between the SLC gauge and LC gauge formulations 
 can be established.  Namely, we will  be able to construct 
 an explicit  quantum similarity transformation which maps  the corresponding 
operators in the two gauges.  
A powerful technique for constructing such a 
 similarity transformation was developed in Ref.~\citen{Aisaka:2004ga} first  for the 
 bosonic string case and then extended for the superstring case to prove 
 the equivalence between a variant of the pure spinor  formalism 
and the LC gauge GS formalism.  This method can be applied to the 
 present case as well, although the construction will require  a new rather non-trivial 
twist  due to the presence of the term $\del^2 \ln \Pi^+$. 

\subsubsection{Description of the method}
The main idea of the method developed in Ref.~\citen{Aisaka:2004ga} is to 
systematically remove the set of unphysical modes 
$(b_n, c_n, \al^+_n, \al^-_n)_{n\ne 0}$  from the BRST operator $Q$  in the form of  a  similarity transformation.   This guarantees that the cohomology 
 is unchanged. As these modes are absent in the LC formalism, they  must form a ``quartet" with respect to an appropriate nilpotent operator, to be called $\delta$, and decouple from the cohomology of $Q$. 
Such a $\delta$ can easily be found within $Q$. 
When $p^+\ne 0$, the term $\sum_{n\ne 0} c_{-n} L_n$ in $Q$ contains the operator 
\begin{eqnarray}
\delta &\equiv & p^+ \sum_{n\ne 0} c_{-n} \al^-_n \comma 
\end{eqnarray}
and it satisfies the relations 
\begin{eqnarray}
\acom{\delta}{\delta} &=& 0 \comma \\
\com{\delta}{\al^+_{-n}} &=& p^+ n c_{-n} \comma \qquad 
 \acom{\delta}{c_{-n}} =0 \comma \\
\acom{\delta}{b_{-n}} &=&  p^+\al^-_{-n} \comma \qquad 
\com{\delta}{\al^-_{-n}} = 0 \period
\end{eqnarray}
This shows that indeed $(b_n, c_n, \al^+_n, \al^-_n)_{n\ne 0}$ form a quartet 
 (\ie two doublets) 
 with respect to $\delta$.  To remove these modes systematically, 
 it is convenient to assign  non-vanishing degrees
  to these unphysical modes  in such a way that (i) $\delta$ will carry 
 degree $-1$ and (ii) the remaining part of $Q$ will carry 
 non-negative degrees.  Separating   the zero-mode and the non-zero mode  parts (indicated by the overcheck)  of the  unphysical fields as
\begin{align}
\Pip(z) &= {\pplusup \over z} + \Pichk(z)  \comma \qquad \Pim(z) = {\pminusup \over z}
 + \Pichk^-(z) \comma  \\
c(z) &= c_0 z  + \cchk(z)  \comma \qquad b(z) ={ b_0 \over z^2} + \bchk(z) 
\comma 
\end{align}
it is achieved by the assignment\footnote{ 
This grading is essentially a refined version of the one used in 
Ref.~\citen{KatoOgawa}. }
\begin{align}
\deg (\Pichk^+) &= 2 \comma \qquad \deg (\Pichk^-) = -2 \\
\deg (\cchk) &= 1 \comma \qquad \deg (\bchk) = -1 \comma \qquad \deg ({\rm rest}) =0 \period
\end{align}

Unlike the case of the bosonic string, due to the presence of the term 
$\varpi\equiv \half \del^2 \ln \Pi^+$ in the Virasoro operator, our $Q$ contains 
 $\Pichk^+$ to aribitrary high powers and hence we will have to deal with infinitely 
 high degrees. Explicitly, we have 
\begin{align}
\varpi &= \half \del^2 \ln \left({\pplusup\over z} + \Pichk^+\right) 
= {1\over 2z^2} +\half \sum_{n \ge 1} {(-1)^{n-1} \over n} \del^2 \left( {z \Pichk^+ \over \pp}\right)^n \comma 
\end{align}
and it contributes to   $Q$  as
\begin{align}
\int[dz] c \varpi &= e_0 + \sum_{n=1}^\infty e_{2n+1} \comma 
\end{align}
where $e_m$, carrying degree $m$, is  given by 
\begin{align}
e_0 &= \int[dz] c_0 z \varpi  = \half c_0 \comma \\
e_{2n+1} &= 
  {(-1)^{n-1} \over 2n} 
\int [dz] \del^2 \cchk \left( {z\Pichk^+ \over \pp}\right)^n\comma   \qquad (n \ge 1) \period
\end{align}
For later purposes, let us express $e_{2n+1}$ in terms of the modes $\al^+_n$ of 
$\Pichk^+$. Expanding $\cchk(z)$ and $\Pichk^+(z)$ into modes and performing
 the integral we obtain 
\begin{align}
e_{2n+1} &\equiv  {(-1)^{n-1} \over 2n (\pp)^n} 
\sump m(m+1) c_{-m} [(\al^+)^n]_m \comma
\end{align}
where we introduced a convenient notation
\begin{align}
[(\al^+)^n]_m &\equiv \sum_{\sum_i^n k_i=m} \al^+_{k_1} \al^+_{k_2}
\cdots \al^+_{k_n}  \period \label{alpnm}
\end{align}

Let us now decompose the BRST operator $Q$ 
according to the degree.  Indicating the degree by the subscripts (except 
for $\delta$, which carries degree $-1$), the result is 
\begin{align}
Q &= \delta + Q_0 + d_1 + d_2 + d_3 + e_{\ge 3} \comma 
\end{align}
where 
\begin{align}
\delta &= \pplusup \int [dz] {1\over z} \cchk \Pichk^- =p^+ \sum_{n\ne 0} c_{-n} \al^-_n \comma \\
Q_0 &= c_0 \left( \half +\int [dz] z (  T\tozero  -\bchk \del \cchk ) \right)
\comma \label{Qzero}\\
d_1 &= \int [dz] ( \cchk T^{(0)} + \bchk\cchk\del \cchk )\comma \\
d_2 &= b_0 \int[dz] {1\over z^2} \cchk \del \cchk \comma\\
d_3 &= \pminusup \int[dz] {1\over z} \cchk \Pichk^+  \comma \\
e_{\ge 3} &\equiv  \sum_{n\ge 1} e_{2n+1} \period 
\end{align}
In the above, $T^{(0)}$ is the degree 0 part of $T$ and is given by 
\begin{align}
T^{(0)} &= z^{-2}\left( p^+p^- + \half p^I p^I \right) 
+ \Pichk^+\Pichk^- 
 + \half \Pichk^I \Pichk^I -\half S_a\del S_a \period
\end{align}
Since the physical fields are contained in $Q_0$, the first and the main task is 
 to remove the unphysical components with positive degrees, namely $d_m$ and $e_m$,  by a suitable similarity transformation. 
Once this is done, the rest of the unphysical  fields in $Q_0$ can be removed 
easily by another similarity transformation, to be discussed later. 

Therefore, we  seek the similarity transformation of the form 
\begin{align}
Q &= e^{-\mfR}(\delta + Q_0) e^\mfR = \delta + Q_0 + 
\com{\delta+Q_0}{\mfR} + \half \com{\com{\delta +Q_0}{\mfR}}{\mfR}+ \cdots 
  \period
\end{align}
Hereafter, we will write the graded commutator $[A,B\}$ simply as $AB$ or $(AB)$. 
In this notation the graded Jacobi identity  reads $A(BC) = (AB) C + (-1)^{AB} B(AC)$, where 
$(-1)^{AB}$ is $-1$ if $A$ and $B$ are both fermionic and $1$ otherwise. 
Then, decomposing the operator $\mfR$ according to the degree as
\begin{align}
\mfR=R_2+R_3 + \cdots \comma 
\end{align}
 the equation above becomes 
\begin{align}
Q &=  \delta + Q_0 + d_1 + d_2 + d_3 + e_{\ge 3} \nn\\
&= \delta + Q_0 + \delta R_2 + \delta R_3 + Q_0 R_2 + \delta R_4 + \half (\delta R_2)R_2 
+ \cdots \period \label{simeq}
\end{align}
From this we get an equation at each degree, like  $d_1=\delta R_2$, $d_2
 = \delta R_3 + Q_0 R_2$, etc., and the question is whether we can determine 
$R_n$'s  which consistently solve these infinite number of equations. 

There are two important ingredients for solving these equations. One is the 
 nilpotency of $Q$.  By decomposing  $Q^2=0$ in degrees, we obtain a relation 
 at every degree, starting from degree $-2$. Up to degree $7$, these equations read
\begin{align}
(E_{-2}) \quad & \delta \delta  =0\comma  \label{Emtwo}\\
(E_{-1}) \quad & \delta Q_0 =0 \comma \label{Emone} \\ 
(E_0) \quad & \half Q_0 Q_0 +  \delta d_1 =0\comma  \label{Ezero} \\
(E_1)\quad & Q_0d_1 +\delta d_2 =0\comma \label{Eone} \\
(E_2)\quad &  \delta  (d_3+e_3)   + Q_0 d_2  +\half d_1 d_1  =0 \comma \\
(E_3)\quad &  Q_0 (d_3 + e_3) + d_1 d_2 =0\comma \\
(E_4)\quad&  \delta e_5 + d_1( d_3 +   e_3) + \half d_2 d_2 =0\comma  \\
(E_5)\quad&  Q_0 e_5 + d_2 (d_3 +  e_3) =0 \comma\\
(E_6)\quad& \delta e_7 +  d_1 e_5 + \half (d_3+e_3) (d_3+e_3) =0\comma
\\
(E_7) \quad&  Q_0 e_7  + d_2 e_5  =0
\period
\end{align}
The other ingredient is the well-known fact about  the (co)homology of the nilpotent operator $\delta$. Consider the (homotopy) 
 operator $\Khat$ of degree 1 given by
\begin{eqnarray}
\Khat &\equiv & {1\over p^+}\sum_{n\ne 0}{1\over n}  \al^+_{-n}b_n
 \period
\end{eqnarray}
Further define
\begin{eqnarray}
\Nhat &\equiv & \delta \Khat
 = \sum_{n\ne 0} :( c_{-n} b_n +{1\over n} \al^+_{-n} \al^-_n ) : \period
\end{eqnarray}
This  is an extension of the ghost number operator and 
 assigns the ``$\Nhat$-number'' $(1,-1,1,-1)$ to the quartet 
$(c_n, b_n, \al^+_n, \al^-_n)$. Now let $\calO$ be a $\delta$-closed 
 operator carrying $\Nhat$-number $n$,
\ie $\delta\calO =0$ and  $\Nhat \calO =n \calO$. 
If $n \ne 0$, we can solve for $\calO$ as an $\delta$-exact form as
\begin{eqnarray}
\calO = {1\over n} \Nhat \calO = {1\over n} (\delta \Khat) \calO 
=\delta \left( {1\over n}  \Khat\calO\right) \period \label{deltahom}
\end{eqnarray}
This formula will be utilized repeatedly. 

To illustrate the method, let us briefly review the construction of $R_n$ 
 up to degree 3, which is 
 formally  the same as for the bosonic string described in Ref.~\citen{Aisaka:2004ga} \  
The first two nilpotency equations $(E_{-2})$ and $(E_{-1})$ simply say that 
$\delta $ is nilpotent and $\delta$ and $Q_0$ anticommute. The first non-trivial equation  is $(E_0)$. Since $Q_0$ itself is nilpotent by inspection, it says $\delta d_1=0$. Since the $\Nhat$-number of $d_1$ is 1, we can use the formula (\ref{deltahom}) to get $d_1 = \delta R_2$, where $R_2\equiv  \Khat d_1$. This solves the degree 1 part 
 of the basic equation (\ref{simeq}). Next, look at the relation $(E_1)$. Since 
$\delta d_2=0$ holds by inspection, $(E_1)$ actually splits into 
 two separate equations $\delta d_2=0$ and $Q_0 d_1=0$. Now since $\Nhat d_2 = 2d_2$, one can again use (\ref{deltahom}) to write $d_2 = \delta R_3$, 
 where $R_3 \equiv \half \Khat d_2$. For this to be consistent with 
 the degree 2 part of (\ref{simeq}), we must show  that $Q_0 R_2=0$. 
Using  the form of $R_2$ and the Jacobi identity, we can write this as 
$Q_0 R_2 = Q_0 (\Khat d_1) = (Q_0 \Khat) d_1   -\Khat(Q_0 d_1)$. While we already  know that $Q_0d_1=0$, it is easy to check explicitly that the anti-commutator  $Q_0 \Khat$ vanishes.  

This completes the construction of $R_2$ and $R_3$. As they play a crucial 
 role in the construction of the DDF operator in section 7.2, we shall give its explicit form:
\begin{align}
R &\equiv  R_2+R_3 = {1\over p^+} \sum_{k\ne 0} {1\over k}
\al^+_{-k} \Ltil^{tot}_k \period  \label{defR}
\end{align}
Here $\Ltil^{tot}_k$ is the degree 0 part of the Virasoro 
generator and consists of the bosonic, the fermionic and the ghost part given by%
\begin{align}
\Ltil_k^{tot} &= \Ltil_k^b+\Ltil^f_k  + \Ltil^g_k  \comma \\
\Ltil_k^b &= p^I \al^I_k + \half \sum_{n \ne 0} \al^\mu_{k-n} \al_{\mu, n}
 \comma \\
\Ltil^f_k&= \half \sum_{n \ne 0} n S^a_{k-n} S^a_n \comma 
\label{Ltilf}\\
 \Ltil^g_k &=\sum_{n \ne 0} n c_{-n} b_{k+n} \period
\end{align}
In the case of the bosonic string ({\it i.e.} with  $\Ltil^f_k$ absent )
it was shown in Ref.~\citen{Aisaka:2004ga} that $R$ is actually the exact exponent $\mfR$. 
In other words $R_n$ for $n \ge 4$ all vanish.  In the present case, however, 
they do not vanish due to the presence of the terms $e_{\ge 3}$ in $Q$ with 
 arbitrarily  high degrees.  We shall now describe the construction of $R_n$ for $n \ge 4$. 
\subsubsection{Construction of the similarity transformation at higher degrees}
In order to construct $R_{n\ge 4}$ by using the  method above, we must first 
examine  the structure of the nilpotency relations at higher degrees.  Although 
 the relations $(E_n)$ displayed previously up to  degree 7  look rather complicated, many of the terms actually vanish and in particular the relations above degree 4 
 turned out to be quite simple. 

Let us list the anticommutators that  vanish. 
First, the following ones among $ Q_0$ and $d_i$ 
can be shown to vanish, either directly or by the reasoning 
similar to those made in Ref.~\citen{Aisaka:2004ga}:
\begin{align}
Q_0 d_3 & = d_1 d_2 = d_1 d_3 =d_2 d_2 = d_2 d_3 =0 \period
\end{align}
Next,  by inspection we can immediately check the relations
\begin{align}
d_2 e_{2n+1} = d_3 e_{2n+1}  =e_{2n+1} e_{2m+1} =0 \period
\end{align}
Another useful relation is 
\begin{align}
Q_0 e_{2n+1} = 0 \period \label{Qezero}
\end{align}
A proof, which is slightly non-trivial, will be  provided  in appendix D.1. 

If we substitute these vanishing relations, we see that the nilpotency relations 
at higher degrees simplify drastically. $(E_3)$ becomes trivial and above $(E_4)$
it reduces simply to the following important relations
\begin{align}
(E_{2n+2})\quad d_1 e_{2n+1} &= -\delta e_{2n+3}\comma \qquad n \ge 1
\period \label{dedeltae}
\end{align}

We  now discuss  how one can determine $R_n$ for $n \ge 4$.  
Although we have explicitly constructed $R_n$ up to $R_{10}$, we will not
 show all the details of the computations  as they are rather involved and not so 
illuminating. What we will do is to describe in some detail  the construction of $R_4$
  and a number of important  general formulas, 
 which will also be needed  in the {\it all order analysis }
 to be presented in the next subsection. Then we will  quote the results  
up to $R_{10}$. 

To determine $R_4$, we must look at the degree 3 part of the basic equation 
 (\ref{simeq}). It reads
\begin{align}
d_3 + e_3 &= 
 \delta R_4+ Q_0 R_3 +
 \half (\delta R_2)R_2  \period \label{degthreeeq}
\end{align}
 First focus on the double commutator $\half (\delta R_2)R_2$ on the right hand side. 
By using the already established relations, we can rewrite it as 
\begin{align}
\half (\delta R_2)R_2 &= \half d_1 R_2 = \half d_1 (\Khat d_1) = \half (d_1 \Khat)d_1 -\half \Khat (d_1d_1) 
= -{1\over 4} \Khat (d_1d_1) \period 
\end{align}
Now applying  the relation $(E_2)$ to the expression ${1\over 4}(d_1d_1)$, 
 this  becomes 
\begin{align}
\half (\delta R_2)R_2 &= \half \Khat (\delta(d_3+e_3)) + \half \Khat (Q_0d_2) 
\period \label{deltaR2R2}
\end{align}
The second  term on the right hand side  equals   $-\half Q_0 (\Khat d_2) = - Q_0 R_3$ and    cancels $Q_0 R_3$  in (\ref{degthreeeq}) . 
On the other hand, the first term can be rewritten as 
\begin{align}
\half \Khat (\delta(d_3+e_3))  &= \half \Nhat (d_3+e_3)
- \half \delta (\Khat (d_3+e_3)) =  d_3+e_3 -\half \delta( \Khat e_3) \comma 
\end{align}
where we used $\Khat d_3=0$, which can be easily checked directly. 
Since the first two terms on the right hand side match the left hand side of (\ref{degthreeeq}), the degree 3 equation (\ref{degthreeeq}) reduces  to 
$\delta (R_4 -\half \Khat e_3) =0 $ and this gives $R_4 = \half \Khat e_3$. 

This type of expression turns out to be quite basic and it is  very useful to 
define the following quantity:
\begin{align}
r_{2n} &\equiv  {1\over n} \Khat e_{2n-1}  
= {(-1)^n \over 2n(n-1) (\pp)^{n}} [(\al^+)^{n}]_0 \comma \qquad n \ge 2
\period
\end{align}
In this notation our result for $R_4$ is  simply   $R_4=r_4$. 
The quantity  $r_{2n}$ has many nice  properties:\\ 
(i)\ Since $\Khat^2=0$, we have 
 $\Khat r_{2n} =0$.\\
(ii)\ Since  $Q_0 \Khat =0$ and $Q_0 e_{2n-1} =0$, we have  $Q_0 r_{2n} =0$. \\
(iii)\ By explicit computation, it is easy to prove 
$\delta r_{2n+2} = -d_1 r_{2n} $. \\
(iv)\ Applying $\Khat$ to the relation $(E_{2n})$ shown  in (\ref{dedeltae}), 
 {\it i.e.} $d_1 e_{2n-1} =  -\delta e_{2n+1}$,  and using the Jacobi identities, we easily obtain $
R_2 e_{2n-1} = (n+1) (\delta  r_{2n+1} - e_{2n+1}) $. 
Applying $\Khat$ again to this equation we establish an important equation
\begin{align}
r_{2n} R_2 &= (n+1) r_{2n+2} \period \label{rR}
\end{align}

In the case of the bosonic string, the reason why $\mfR=R_2+R_3$ 
was the exact answer was because various higher multiple commutators of 
$\delta$ and  $R_i$ ($i=2,3$)   vanished.  For the present case, it is no longer 
true in general and in particular the following relation plays an important role 
 in the higher degree analysis. It reads 
\begin{align}
{1\over n!} \delta \leftad{R_2}^n &= e_{2n-1} -\delta r_{2n} \comma 
\qquad (n \ge 2) \comma \label{multiRtwo}
\end{align}
where the multiple commutator is denoted by 
$ \delta \leftad{R_1}^n \equiv (((( ( \delta R_2) R_2 \cdots )R_2)$. 
The proof of this formula is given in appendix D.1. 

Although rather tedious, making use of these formulas, one can analyse the equation (\ref{simeq}) at each higher degree,  similarly to the case of degree 3,  and determine 
 the   operators $R_n$.  The results  up to degree 10  turned out to be 
\begin{align}
&R_4 = r_4 \comma \quad R_5=0 \comma \quad R_6 = \half r_6\comma \quad
R_7=0 \comma \nn\\
&   R_8 =0 \comma \quad R_9 =0 \comma \quad 
R_{10} = -{1\over 6} r_{10}  \period \label{Rten}
\end{align}
One immediately notes that the operators at odd degrees  vanish. Actually  we can prove that this is a general property.  In contrast,  it is apparently not possible to guess the  pattern for $R_{2n}$. Nevertheless, in what follows we will be able to construct 
 the exact similarity transformation to all degrees by a different approach and confirm that  it reproduces the result up to degree 10 given above. 
\subsubsection{Exact form of the similarity transformation}
We now describe a new idea which lets us construct the similarity transformation 
 exactly to all degrees.  
Consider an   ansatz of the form
\begin{align}
Q &= e^{-R_2-R_3} e^{-\Rtil} (\delta + Q_0) e^{\Rtil} e^{R_2+R_3}
\comma 
\end{align}
where $\Rtil$ is taken to be of the form 
\begin{align}
\Rtil &= \sum_{n \ge 2}  \xi_{2n} r_{2n}\comma 
\end{align}
with  $\xi_{2n}$  being appropriate coefficients  to be determined. 
Then, since $Q_0$ and $\delta r_{2n}$ commute with $r_{2n}$ , the similarity transformation by $\Rtil$ simply gives 
\begin{align}
e^{-\Rtil} (\delta + Q_0) e^{\Rtil} &= \delta + Q_0 + 
\sum_{n \ge 2} \xi_{2n} \delta r_{2n}  \period
\end{align}

For the $\delta +Q_0$ part, the subsequent  similarity transformation by 
$R_2+R_3$
 is almost the  same as for the  bosonic string case. The only difference is that, 
while $\delta \leftad{R_2}^n$ vanished for $n\ge 3$ for the bosonic string,   in the  present case  it  is non-vanishing  for all $n$ and is given by (\ref{multiRtwo}) 
 for $n \ge 2$. Therefore we have 
\begin{align}
&e^{-R_2-R_3} (\delta +Q_0) e^{R_2+R_3} \nn\\
&\quad = \delta + Q_0 + d_1 +d_2 +d_3 +
\sum_{n \ge 2} e_{2n-1}  -\sum_{n\ge 2} \delta r_{2n}  
= Q -\sum_{n\ge 2} \delta r_{2n} \period \label{simdelQz}
\end{align}
where $Q$ is the full BRST operator we want. 

The remaning contribution is 
$e^{-R_2-R_3} \sum_{n \ge 2} \xi_{2n} \delta r_{2n}  e^{R_2+R_3}
= \sum_{n \ge 2}e^{-R_2} \delta r_{2n} e^{R_2}$, where we used the fact that 
$R_3$ commutes with both $R_2$ and $\delta r_{2n}$.  If this contribution cancels 
the  term $-\sum_{n\ge 2} \delta r_{2n}$ in (\ref{simdelQz}) we get the desired result.  Now as we  have already encountered in (\ref{delrR}) in appendix D.1,  
there is a formula 
\begin{align}
(\delta r_{2n}) R_2 &= n \delta r_{2n+2} \period
\end{align}
Applying this  repeatedly $m$ times with the weight factor $1/m!$, we get 
 the multiple commutator needed in the computation of $e^{-R_2} \delta r_{2n} e^{R_2}$ as 
\begin{align}
{1\over m!}(\delta r_{2n}) \leftad{R_2}^m &= {1\over m!} 
 n (n+1) \cdots (n+m-1) \delta r_{2(n+m)}  \nn\\
&= {(n+m-1)! \over m! (n-1)!} \delta r_{2(n+m)} 
= \binomial{n+m-1}{m} \delta r_{2(n+m)}  \period 
\end{align}
Altogether the total similarity transformation gives 
\begin{align}
e^{-R} e^{-\Rtil} (\delta + Q_0) e^{\Rtil} e^R 
&= Q + \sum_{n \ge 2}  \xi_{2n} \sum_{m\ge 0} \binomial{n+m-1}{m} \delta r_{2(n+m)}  - \sum_{k \ge 2} \delta r_{2k}  \nn\\
&= Q + \sum_{k \ge 2} \left( \sum_{n=2}^k  \xi_{2n} \binomial{k-1}{k-n} -1\right) \delta r_{2k}  \period 
\end{align}
So we get the desired result, namely  $Q$,   if $\xi_{2n}$ can be  chosen to satisfy the relation 
\begin{align}
\sum_{n=2}^k  \xi_{2n} \binomial{k-1}{k-n} =1 \period \label{xieq}
\end{align}
Inspection of  this equation for low values of $k$ suggests that the general answer is 
$\xi_{2n} =(-1)^n$. In fact such a formula is easily proved by evaluating the 
 the integral $\int_0^1dx (1-x)^l = 1/(l+1)$ 
by expanding in powers of $x$. 

Thus, we have obtained a remarkably simple answer for the exact similarity transformation:
\begin{align}
Q &= e^{-R_2-R_3} e^{-\Rtil} (\delta + Q_0) e^{\Rtil} e^{R_2+R_3} \comma \\
\Rtil &= \sum_{n\ge 2} (-1)^n r_{2n}  \period
\end{align}
Actually, $\Rtil$ can be written  in a closed form. Going back to the definition of $r_{2n}$, it can be expressed in terms of a contour integral in the $z$-plane as 
\begin{align}
\Rtil &= \sum_{n \ge 2} (-1)^n r_{2n} = \int [dz] \sum_{n \ge 2} 
{1\over 2n (n-1)} \left( {z \Pichk^+ \over \pp}\right)^n \nn\\
&= \half \int [dz] \bigl[ f(z)+ (1-f(z))\ln (1-f(z)) \bigr] \comma 
\end{align}
where $f(z)$ is defined as  $f(z) \equiv  z\Pichk^+(z)/\pp $. 

As a check of the formula above, we wish to  compare it with the (partial)
answer obtained by the degree-wise analysis. To do this, we must 
rewrite the product $e^{\Rtil} e^{R_2+R_3}$ into the 
 form $e^\mfR$, using the Baker-Campbel-Hausdorff (BCH) formula. 
(Actually, as $R_3$ commutes with $R_2$ and $\Rtil$, we only 
need to deal with the product  $e^{\Rtil} e^{R_2}$.)  Since this computation 
 is rather non-trivial and technical, we shall only quote the result 
 here and relegate its derivation to appendix D.2.  
The exact result for $\mfR$ can be written rather compactly as 
\begin{align}
\mfR &= R_2 + R_3 + \Rhat \comma \\
\Rhat &= r_4 + \sum_{n\ge 1}(-1)^{n-1} (2n+1)B_n r_{2(2n+1)}
\comma \label{Rhat}
\end{align}
where  $B_n$ are the  Bernoulli numbers. Substituting $B_1=1/6, B_2=1/30, B_3=1/42$, the first few  terms of  $\Rhat$ read 
\begin{align}
\Rhat &= r_4 + \half r_6  -{1\over 6}r_{10} + {1\over 6} r_{14} + \cdots 
\period \label{expRhat}
\end{align}
Remarkably, this precisely  reproduces  the rather sporadic-looking result of the 
degree-wise analysis given in 
(\ref{Rten}).  Also, the formula (\ref{Rhat}) is consistent  with the general feature 
$R_{2n+1}=0$ deduced previously.

Our  final  task is to remove the unphysical oscillators still remaining in 
 $Q_0$,  given in (\ref{Qzero}),  by another similarity transformation and connect to the light-cone gauge formulation. Here we have to be a bit careful. 
First, in the ghost sector we have to change from the $sl(2)$-invariant 
 vacuum $\ket{0}_{inv}$, tacitly used 
 in the plane coordinate treatement adopted so far, to the so-called ``physical"   or ``down" vacuum 
given by $\ket{\downarrow}=c_1 \ket{0}_{inv}$. This adds the well-known
 intercept $-1$ to the zero mode of the Virasoro operator. After that 
$Q_0$ can be decomposed as
$Q_0 = Q_{lc} + c_0 \Ntil $, where\footnote{Note that the normal-ordering 
 for the ghost part in $\Ntil$ is now appropriate for the  down vacuum.}
\begin{align}
Q_{lc} &=c_0 \left( \half -1 +p^+p^-+ \int[dz] z T_{lc}(z) \right) 
   \comma \\
T_{lc}(z) &= -\half \del X^I \del X^I(z) -\half S^a \del S^a(z) \comma \\
\Ntil &= \sum_{n \ge 1} \left(\al^+_{-n} \al^-_n + \al^-_{-n} \al^+_n +
n c_{-n}b_n + nb_{-n} c_n \right) \period 
\end{align}
Here $T_{lc}(z)$ is the energy-momentum tensor for the physical matter fields in the plane coordinate, which carries the central charge $C_{lc}=12$. 
Now since the LC gauge formulation is constructed  in the cylinder coordinate 
$\zeta = \tau + i\sig$, we must make the familiar conformal transformation 
 $z \rightarrow \zeta  = \ln z$. Then we get  $T_{lc}(z) 
 = (1/z^2) (T_{lc}(\zeta) + (C_{cl}/24)) = (1/z^2) (T_{lc}(\zeta) + (1/2))$,
yielding an additional intercept of $1/2$.  Altogether the intercepts add up to 
 zero and we obtain the form of $Q_{lc}$ in the cylinder frame as 
\begin{align}
Q_{lc} &= c_0 \left( \half p^\mu p_\mu + \sum_{n \ge 1} \al^I_{-n} \al^I_n + \sum_{n\ge 1} n S^a_{-n} S^a_n \right) \period
\end{align}
We  now construct the similarity transformation which removes the $c_0\Ntil$ part consisting of  unphysical oscillators. To this end 
 introduce  the following  operator 
$\Ktil$ which is  similar to but different from $\Khat$:
\begin{align}
\Ktil &\equiv {1\over p^+} \sum_{n \ne 0} \al^+_{-n} b_n \period
\end{align}
Using this operator, it is easy to check  that $\Ntil$ can be expressed as the anticommutator
\begin{align}
\Ntil &= \acom{\Ktil}{\delta} \period
\end{align}
Then the following similarity transformation does the required job:
\begin{align}
e^{-c_0 \Ktil} (\delta + Q_0) e^{c_0 \Ktil} &= \delta + Q_0  -\com{c_0\Ktil}{\delta} = \delta + Q_0 -c_0 \Ntil = \delta + Q_{cl} \period
\end{align}
Finally, by a rather standard argument  one can show that the cohomology 
 of the operator $\delta + Q_{cl}$ indeed coincides with  the physical space 
of the light-cone  theory. We will not reproduce the argument here and refer the 
 reader to the description in Ref.~\citen{Aisaka:2004ga}. 
\subsection{Construction of the DDF operators}
Although there should be many applications of the similarity transformation 
that  connects the SLC and the LC gauges,  let us discuss below one direct application, 
namely  the construction  of the so-called DDF operators in the SLC gauge
  using this machinary. 
\subsubsection{Basic idea} 
In the case of the bosonic string, such a method was developed 
in Ref.~\citen{Aisaka:2004ga} and the bosonic DDF operator  $A^I_n$, 
 which commutes with the Virasoro generators  in the conformal gauge,  was generated  from the transverse physical oscillator $\al^I_n$ in the LC gauge
 by the similarity transformation. More precisely, it was demonstrated that the 
 following relation holds:
\begin{align}
A^I_n &=e^{inx^+/p^+}  \check{A}^I_n \comma \label{DDFA}\\
\check{A}^I_n&=  \int[d\tau] e^{in\tau} \del_\tau X^I(\tau) e^{i(n/p^+) \Xchk^+(\tau)} 
= e^{-R} \al^I_n e^{R} \period \label{Achk}
\end{align}
Here,  the operator $R$ is  the same as  $R_2+R_3$ given in 
(\ref{defR}) (without  fermions), $\tau$ is a variable in the interval 
 $[0,  2\pi]$, the integration measure is  $[d\tau] \equiv d\tau/2\pi$
 and $X^\mu(\tau)$ is given by 
\begin{align}
X^\mu(\tau) &= x^\mu + p^\mu \tau + \Xchk^\mu(\tau) \comma  \qquad 
\Xchk^\mu(\tau)=  i \sum_{n\ne 0} 
{1\over n} \al^\mu_n e^{-in\tau} \period
\end{align}

What we wish to do is to construct the fermionic DDF operator corresponding 
 to the LC oscillator $S^a_n$, to be 
 denoted by $\mbS^a_n$, by using our similarity transformation constructed 
 in the previous subsection.  Due to the 
presence of the $\del^2 \ln \Pi^+$ term in the Virasoro operator, the 
similarity transformation is more involved than in the bosonic string case. 
However, since the additional transformation, which consists solely of $\Pi^+$, does not affect $S^a$,  we should be able to  compute 
the main part\footnote{To get the genuine DDF operator, we need to 
multiply by  a zero-mode factor as in (\ref{DDFA}).} 
 of the DDF operator $\check{\mbS}^a_n$ by 
\begin{align}
\check{\mbS}^a_n &= e^{-R} S^a_n e^{R} \comma 
\end{align}
where $R$  is given by $R_2+R_3$ of (\ref{defR}). 
The  only difference from the bosonic string case is the presence of the 
fermionic part $\Ltil^f_k$ given in (\ref{Ltilf}). 
Therefore, the computation of the bosonic 
DDF operator via $e^{-R} \al_n^I e^R$ quoted in (\ref{Achk}) is exactly
 the same as in Ref.~\citen{Aisaka:2004ga}. 

In fact, it is useful to recall the salient feature of the calculation 
in Ref.~\citen{Aisaka:2004ga} 
 before we launch on the computation 
 of $ e^{-R} S^a_n e^{R}$.  The important point was that 
as one successively evaluates the terms in  the expansion 
$e^{-R} \al_n^I e^R= \al^I_n - \com{R}{\al^I_n} + \half 
\com{R}{\com{R}{\al^I_n}} + \cdots$, the oscillator $\al^I_n$ gets 
 multiplied  by powers of $\Xchk^+$ precisely in such a way to get dressed
 exponentially.  The simplicity of this process, in turn, was  due to 
 the fact that $\Xchk^+ $ and 
$\del_\tau \Xchk^I$ are primaries of dimension 0 and 1, respectively, 
 with respect to $\Ltil^{tot}_k$. 
 Consequently, the commutators 
of $R$ and these fields take the form 
\begin{align}
\com{R}{\Xchk^+(\tau)} &
= {1\over 2p^+} \del_\tau (\Xchk^+)^2 \comma \label{RXp}\\
\com{R}{\del_\tau \Xchk^I(\tau)} &= {1\over p^+} 
\del_\tau \left( \Xchk^+ \del_\tau\Xchk^I\right) \comma
\end{align}
which are both total derivatives. This allows one to perform integration by parts 
repeatedly  to reach the simple result (\ref{Achk}). 

The situation for the case of the  operator $S^a_n$ is qualitatively different. 
If we form  the field $S^a(\tau) = \sum_n S^a_n e^{-in\tau}$, it is also 
 a primary field with respect to $\Ltil^{tot}_k$ but is of dimension 
$1/2$, \ie 
\begin{align}
 \com{\Ltil^{tot}_k}{S^a(\tau)} &= e^{ik\tau} \left( {1\over i} \del_\tau + 
\half k \right) S^a(\tau) \period
\end{align}
Because of this the commutator with $R$ is given by 
\begin{align}
\com{R}{S^a(\tau)} &= {1\over p^+} 
\left( \Xchk^+ \del_\tau S^a + \half \del_\tau \Xchk^+ S^a 
\right) \comma \label{RS}
\end{align}
which is {\it not a total derivative}. This makes 
the computation of $e^{-R} S^a_ne^R$ much more involved. 

Nevertheless, by using the formula (\ref{RS}) as well as (\ref{RXp}) and 
integrating  by parts  appropriately, we can perform 
 the computation at low orders.  Up to the second order in $R$  the result is
\begin{align}
\check{\mbS}^a_n &= \int[d\tau] e^{in\tau} 
\left( S^a  -\com{R}{S^a} + \half \com{R}{\com{R}{S^a}} 
 + \cdots \right)  \nn\\
& =\int[d\tau] e^{in\tau}S^a(\tau)  \biggl(1+ {in \over p^+} \Xchk^+ + {1 \over 2p^+} \del_\tau \Xchk^+ 
 -\half \left({n\over p^+}\right)^2 ( \Xchk^+)^2  \nn\\
&-{1\over 8{p^+}^2} (\del_\tau \Xchk^+)^2  
+{in \over 4{p^+}^2 } \del_\tau (\Xchk^+)^2 + \cdots 
\biggr) \period \label{calSRR}
\end{align}
Although it looks a little  complicated at first sight, 
one can actually guess what this series should develop into. 
An important  property of the DDF operator is that it should commute with 
the Virasoro generators.  Thus, it should be given by an integral of a 
conformal primary of dimension 1. In the case of $\del_\tau{X}^I $, it is already 
 of dimension 1 and hence we only needed to dress it  by 
 dimension 0 operators, as in (\ref{Achk}).  In the present case,  as $S^a$ is 
 of dimension $1/2$, the dressing factor must carry dimension $1/2$, not zero. 
The natural guess is to first promote  $S^a$ into a dimension 1 primary 
by multiplying by    $\sqrt{\del_\tau X^+}$ and then 
 dress it further with the dimension zero operator. 
To define  $\sqrt{\del_\tau X^+} S^a$   properly, we must expand it around the 
 zero mode. Therefore more precise form of the guess would be 
\begin{align}
\check{\mbS}^a_n &=e^{-R} S^a_n e^R =
 \int[d\tau] e^{in\tau} e^{i(n/p^+)\Xchk^+}
\left( 1+ {\del_\tau \Xchk^+ \over p^+}\right)^{1/2} S^a(\tau) 
\period 
\label{guess}
\end{align}
Indeed it is easy to see that,  to the second order in $R$, 
  the expansion of  the square root
 in powers of $\del_\tau \Xchk^+$ exactly reproduces  
the expression (\ref{calSRR}). 
\subsubsection{A theorem on finite operatorial conformal transformation}
We now wish to prove the formula (\ref{guess}) to all orders in $R$. 
Clearly the brute force computation of $e^{-R} S^a_n e^R$ to higher orders 
 would not be useful for this purpose.  It turned out that understanding of the 
 meaning of our similarity transformation is crucial for  achieving  the goal. 
It leads to {\it a theorem on   finite  operatorial conformal transformation}. 

To state the theorem precisely, we first need to describe the  set-up. 
 Let the mode operators $\phi_n$ and $\chi_n$ enjoy  the following 
 commutation relations  with a set of operators $L_k$: 
\begin{align}
\com{L_k}{\phi_n} &= -(n+(1-h)k) \phi_{n+k}  \comma \label{Lkphi}\\
\com{L_k}{\chi_n} &= -(n+k)\chi_{n+k} \period \label{Lkchi}
\end{align}
These  relations are  of the same form as 
the commutation relations of the modes of Virasoro generators $L_k$ and 
 the modes of primary fields $\phi(\tau)$ and $\chi(\tau)$ 
 of  dimensions $h$ and $0$ respectively. 
 However, we will not require the algebra of $L_k$ themselves. 
For our present problem,  we need to consider the situation where 
the field $\chi(\tau)$ is formed without the zero mode. Namely we define 
\begin{align}
\phi(\tau)&\equiv \sum_n \phi_n e^{-in\tau} \comma \qquad 
\chi(\tau) \equiv \sum_n{}' \chi_n e^{-in\tau} \comma 
\end{align}
where the prime on the summation symbol signifies  that the zero mode is 
omitted.  Then, multiplying (\ref{Lkphi}) and (\ref{Lkchi})  by $e^{-in\tau}$ and summing  over the respective range as above, we get 
\begin{align}
\com{L_k}{\phi(\tau)} &= e^{ik\tau} \left( {1\over i} \del_\tau + hk \right) \phi(\tau) \comma \label{comLkphi}\\
\com{L_k}{\chi(\tau)} &= e^{ik\tau}  {1\over i} \del_\tau  \chi(\tau) 
\period
\label{comLkchi}
\end{align}

Now consider an infinitesimal   conformal transformation of $\phi(\tau)$ 
by $\tau' = \tau -\ep(\tau)$, or equivalently $\tau = \tau' + \ep(\tau')$, where 
 the parameter $\ep(\tau)$ has no zero mode part. 
Then, from the basic equation $\phi'(\tau') (d\tau')^h = \phi(\tau) (d\tau)^h $ 
one easily derives  the change of the functional form as 
\begin{align}
\delta \phi(\tau) &= \phi'(\tau) -\phi(\tau) =  \ep(\tau) \del_{\tau}\phi (\tau)
+ h \del_{\tau}\ep(\tau) \phi(\tau) \period
\end{align}
In terms of modes, this reads
\begin{align}
\delta \phi_n &= -i \sum_k{}' (n+(1-h)k) \ep_{-k} \phi_{n+k} \period 
\end{align}
Using (\ref{Lkphi}) this can be written as the commutator
\begin{align}
\delta \phi_n &= \com{i \sum_k{}' \ep_{-k} L_k}{\phi_n}  \period
\end{align}
Therefore, the operator which generates an infinitesimal conformal 
 transformation associated to $\tau\rightarrow \tau' = \tau-\ep(\tau)$ is given by 
\begin{align}
T_\ep  &= i \sum_k{}' \ep_{-k} L_k \period
\end{align}
In this formula, $\ep(\tau)$ is a function, not a field.  {\it In what follows, we  replace it 
 by a primary field $\chi(\tau)$ of dimension $0$} and consider the 
operator 
\begin{align}
T_\chi  &= i \sum_k{}' \chi_{-k} L_k \period \label{Tchi} 
\end{align}
We can now state the theorem.  \\
\underline{Theorem:}\quad Let $T_\chi$ be given by (\ref{Tchi}), where 
{\it  $\chi$ is finite, not  infinitesimal}. Then we can generate a finite 
 conformal transformation  associated with  $\tau' = \tau-\chi(\tau)$
 in the form of the similarity transformation 
\begin{align}
e^{T_\chi} \phi(\tau) e^{-T_\chi} &= \phi' (\tau)  \comma \label{thmconf}
\end{align}
where the functional form of $\phi'$ is given  by the formula for the 
{\it  finite} conformal transformation
\begin{align}
\phi'(\tau') (d\tau')^h &= \phi(\tau) (d\tau)^h \period \label{conftrphi}
\end{align}
It is extremely important that $\chi(\tau)$ is 
a primary field of dimension $0$, not just a parameter. The reason is that otherwise 
 $\com{T_\chi}{\phi}, \com{T_\chi}{\com{T_\chi}{\phi}}$, etc will not
 be proper conformal descendants and the theorem does not hold.  Despite the 
 simplicity of the form of  the theorem,  the proof is rather involved and we will 
 describe it in appendix D.3. 
\subsubsection{Exact form of the DDF operator}
Now we  apply this theorem with the identification 
$L_k= \Ltil^{tot}_k, \phi(\tau) =S^a(\tau), \chi(\tau) = -\Xchk^+(\tau)/p^+$
and $h=1/2$. 
Then,  $T_\chi $  is identified with   $-R$, where $R$ is given  in (\ref{defR}) and
 the associated  conformal transformation is given by 
$\tau' = \tau +\Xchk^+(\tau)/p^+$.  Then from the theorem we obtain
\begin{align}
e^{-R} S^a(\tau) e^R &= {S'}^a(\tau)  \period
\end{align}
What remains to be shown  is that ${S'}^a(\tau) $ is exactly the same function as 
$\check{\mbS}^a(\tau) \equiv \sum_n \check{S}^a_n e^{-in\tau}$, where 
$\check{\mbS}^a_n$ was given in (\ref{guess}).  In the following manipulation, we set $p^+=1$ for notational simplicity. 
 From (\ref{guess}) we can compute $\check{\mbS}^a(\rho)$ at an arbitrary argument $\rho$ as 
\begin{align}
\check{\mbS}^a(\rho) &= \int[d\tau] \sum_n e^{in (\tau + \Xchk^+(\tau) -\rho)}
\left( 1+ \del_\tau \Xchk^+\right)^{1/2} S^a(\tau) \nn\\
&= \int [d\tau] 2\pi \delta(\tau + \Xchk^+(\tau) -\rho) \left( 1+ \del_\tau \Xchk^+\right)^{1/2} S^a(\tau)  \period \label{Schk}
\end{align}
Let us set  $\rho=\taubar+\Xchk^+(\taubar)$, where $\taubar$ is an arbitrary new variable. Then, the $\delta$-function 
 can be rewritten as 
\begin{align}
\delta(\tau + \Xchk^+(\tau) -(\taubar+\Xchk^+(\taubar)))
= {\delta(\tau -\taubar) \over{ d(\tau + \Xchk^+(\tau)) \over d\tau}}
=  \delta(\tau -\taubar)\left( 1+\del_\tau \Xchk^+ \right)^{-1} 
\period 
\end{align}
Put this into (\ref{Schk}),  integrate it over $\tau$ and then rename $\taubar$ as $\tau$. This gives 
\begin{align}
\check{\mbS}^a(\tau+\Xchk^+(\tau)) &=\left( 1+\del_{\tau} \Xchk^+(\tau) \right)^{-1/2} S^a(\tau) \period 
\end{align}
Note that this  is nothing but the conformal transformation of $S^a(\tau)$ and 
 hence  the left-hand side equals ${S'}^a(\tau+\Xchk^+(\tau))$. In other words, 
 as functions, $\check{\mbS}^a(\tau) = {S'}^a(\tau)$.  Reinstating $p^+$, we have 
proved (\ref{guess}) as an identity.

Just as in the case of the bosonic string (\ref{DDFA}), the actual DDF operator $\mbS^a_n$ which commutes with the Virasoro generators is obtained by adjoining 
the zero-mode piece, namely, 
\begin{align}
\mbS^a_n &= e^{inx^+/p^+} \check{\mbS}^a_n  \period
\end{align}
Including this factor, the formula for the DDF operator can be written more 
 compactly as 
\begin{align}
e^{-R}  e^{inx^+/p^+} S^a_n e^R &= \mbS^a_n =  {1\over \sqrt{p^+}} \int[d\tau] e^{i(n/p^+)X^+}
\sqrt{\del_\tau X^+}\,  S^a(\tau) \comma 
\end{align}
using the full field $X^+(\tau)$ including the zero mode pieces. This is our final answer for the DDF operator. It is clear that $\mbS^a_n$'s  satisfy the same anti-commutation  relations as $S^a_n$'s, namely $\acom{\mbS^a_m}{\mbS^b_n} = \delta^{ab} \delta_{m+n, 0}$. Also, the operator $\mbS^a_n$ commutes with 
 the Virasoro algebra.
\section{Discussions}
In this article, starting from the basic action we have developed the 
operator formulation of the Green-Schwarz superstring in the SLC gauge 
 in a fairly systematic manner. We have clarified  the structure 
 of the quantum symmetry algebras  of the theory, constructed the 
vertex operators for the massless excitations and obtained the exact quantum
 map between the operators of the SLC gauge formulation and those 
 of the familiar LC gauge formulation. 

We now discuss some  issues to be investigated in  the future. 
One issue  is how to remove the restriction on  the choice of the momentum frame, namely the condition $k^+=0$,  for the vertex operators.  When we started our investigation, our  hope was that,  as we need not  identify  $X^+$ with the worldsheet time in the SLC gauge,  we woud be able to relax such a condition in contrast to the conventional LC gauge. In the LC gauge, as one imposes the  condition $X^+ = x^++p^+\tau$,  the remaining longitudinal field 
$X^-$ is expressed,  through Virasoro constraints, as bilinears  in the 
physical fields. This makes it extremely difficult to include  the 
term $ik^+X^-$ in the exponential $\exp (ik_\mu X^\mu)$ and hence 
one is forced to adopt the  $k^+=0$ frame.  This problem does not exist 
 in the  SLC gauge because  $X^-$ is  a genuine  independent field. 
Unfortunately, however, a similar problem arises from a different origin. 
Due to the existence of the quantum correction $\half \del^2 \ln \Pi^+$ in the  Virasoro generator,  the operator $\exp(i k_\mu X^\mu)$ is no longer
 a primary field if we include the $ik^+X^-$ term.  Therefore to avoid complication  we have decided to impose  $k^+=0$ condition in this work. In fact, 
 as indicated  in section  6,  at least for the consistency of the vertex operators 
 constructed in this work,  this condition appears to be necessary. 

There seem to be several possibilities to cure this problem. One is that, 
 as we have the  quantum Lorentz generator $\calM^{I-}$  at hand, it might be possible  to perform a finite Lorentz transformation $\exp(\xi \calM^{I-}) V \exp(-\xi \calM^{I-})$ on the vertex operator $V$ to  go to the frame where $k^+$ is non-vanishing. 

Another possibility is suggested by the recent work \cite{Baba:2009ns}, which 
introduced a conformal field theory composed of the longitudinal 
fields $X^\pm$,  for the purpose of defining the light-cone gauge string field theory in non-critical dimensions in the path  integral formalism. This system 
is essentially the same as the one that occured in our work and the Virasoro generator contains  terms made up of $\Pi^+$'s  with two derivatives. Consequently, the conformal property of $X^-$ becomes rather complicated as in our case. Nevertheless the authors of Ref.~\citen{Baba:2009ns} formally 
succeeded in computing the amplitudes  containing the vertex operators of the type $e^{ik^+\hat{X}^-}$ with non-vanishing $k^+$, where $\hat{X}^-$ 
 is the modified field mentioned in setion 6.1. As discussed there, $\hat{X}^-$, while having a good  conformal property, has a severe singularity with itself. 
Although it is not straightforward  to relate the path integral formulation they adopted and our operator formalism, this indicates that  there may be a way to construct the vertex operator for non-zero $k^+$  in the operator language as well.   This is left as a  problem for the future. 

Another problem  worth studying in the future 
 is the construction of the $Dp$-brane boundary state in the SLC gauge. 
Within the GS formalism, the construction of such a boundary state
 was performed  in Ref.~\citen{Green:1994iw, Green:1996um} in the LC gauge. 
In that gauge, however, since the ``time" coordinate $X^+$ must 
satisfy the Dirichlet boundary condition, what one obtains  
 is the ``$(p+1)$-instanton" rather than the  $Dp$-brane. One must perform 
 the double Wick rotation to get the genuine  $Dp$-brane. 
In constrast, in the SLC-conformal gauge, it should be possible to construct the 
genuine $Dp$-brane with the Neumann boundary condition for $X^0$ much more naturally (at least for $p \ge 1$).  For this study, the structure of the super-Poincar\'e and the BRST symmetry clarified in the present work will be indispensable. 
\section*{Acknowledgements}
The research of  Y.K. is supported in part by the
 Grant-in-Aid for Scientific Research (B)
No.~12440060  from the Japan
 Ministry of Education, Culture, Sports,  Science and Technology, while
that of
 N.Y. is supported in part by the JSPS Research Fellowships for Young
Scientists.
\appendix
\section{Conventions} 
%
\subsection{Metrics and light-cone coordinates}
The metric of  the target 10 dimensional Minkowski space is taken to be 
$\eta_{\mu\nu} = (-1, +1,  \ldots , +1)$, where $\mu,\nu=0 \sim 9$. 
The light-cone components of  a vector $A^\mu$ are defined by
$A^\pm \equiv \ovsqtwo (A^9 \pm A^0)$. 
Accordingly  the contraction of two vectors is expressed as 
$A^\mu B_\mu = A^+B^-+A^-B^+ + A^I B^I $, 
where the  subscript $I$ for the transverse components runs from 1 to 8. 

The worldsheet coodinates are denoted by $\xi^i = (\xi^0, \xi^1) = (t,\sig)$. 
The metric $\eta_{ij}$, the antisymmetric  tensor $\ep_{ij}$,   the light-cone coordinates $\sig_{\pm}$ and the derivatives $\del_\pm$ are taken as 
$ \eta_{ij} = (-1,+1)$, $\ep_{01} = -\ep^{01} \equiv  1$,  
$\sig_{\pm}= \xi^0\pm \xi^1$ and  $\del_\pm = \half (\del_0 \pm \del_1)$. 
Therefore $\del_i A \del^i B = -2(\del_+A \del_-B +\del_-A  \del_+B)$ and 
 $\del_i \del^i = -4\del_+\del_-$. 
\subsection{Gamma matrices and spinors}
$ 32 \times 32$ $SO(9,1)$ Gamma matrices are denoted by 
$\Ga^\mu, (\mu =0 \sim 9)$ and they obey the  Clifford algebra
$ \acom{\Ga^\mu}{\Ga^\nu} = 2\eta^{\mu\nu} $. 
The 10 dimensional chirality operator is taken to be 
$\chiten \equiv \Ga^0 \Ga^1 \cdots \Ga^9$ and it satisfies 
$(\chiten)^2 =1$. We use the Majorana basis, where $\Ga^\mu$ are all real 
and unitary. Within the Majorana basis, we define the Weyl basis to be the one 
 in which $\chiten = {\rm diag}\,  (1_{16}, -1_{16})$, where $1_{16}$ denotes 
the $16\times 16$ unit matrix. In this basis, a general $32$-component spinor $\Lam$ is written as $\vecii{\lam^\al}{\lam_\al} $, where $\lam^\al$ and $\lam_\al$ are chiral and anti-chiral respectively, with $\al=1 \sim 16$. Correspondingly, 
 $\Ga^\mu$, which flips chirality, can be expressed in terms of  $16\times 16$ 
 matrices $\ga^\mu$ and $\gabar^\mu$ as 
\begin{align}
\Ga^\mu &= \matrixii{0}{(\ga^\mu)^{\al\be}}{(\gabar^\mu)_{\al\be}}{0} 
\period
\end{align}
From the symmetry property of $\Ga^\mu$, we have $\ga^m = \gabar^m\,  (m=1 \sim 9)$   and $\ga^0 = -\gabar^0$,  as matrices. 
However, we shall distinguish $\ga^\mu$ and $\gabar^\mu$ in order to keep track of the 10D chirality structure. 
From the Clifford algebra satisfied by $\Ga^\mu$, we obtain the algebra 
 for $\ga^\mu$ and $\gabar^\mu$ as
\begin{align}
\ga^\mu \gabar^\nu + \ga^\nu \gabar^\mu &= 2\eta^{\mu\nu} \comma 
\qquad 
\gabar^\mu \ga^\nu + \gabar^\nu \ga^\mu = 2\eta^{\mu\nu}  \period
\end{align}
Antisymmetrized products of two $\ga^\mu$'s are defined with appropriate 
 index positions as 
\begin{align}
\left(\ga^{\mu\nu}\right)^\al{}_\be &= \half \left( \ga^\mu \gabar^\nu 
 -\ga^\nu \gabar^\mu \right)^\al{}_\be 
\comma \qquad \left(\gabar^{\mu\nu}\right)_\al{}^\be 
= \half \left( \gabar^\mu \ga^\nu 
 -\gabar^\nu \ga^\mu \right)_\al{}^\be  \period
\end{align}

$SO(8)$ decomposition of the $\ga$-matrices and 
spinors is taken as follows. The  $SO(8)$ chirality operator $\Gabar_8$, 
which anticommutes with $\Ga^I$ and satisfies $\Gabar_8^2=1$, 
 is defined  by 
\begin{align}
\Gabar_8 &\equiv  \Ga^1\Ga^2 \cdots \Ga^8 = \matrixii{\chi_8}{0}{0}{\chibar_8} 
\comma 
\end{align}
where $\chi_8$ and $\chibar_8$,  which act  repectively  on the chiral and the anti-chiral 
 sectors,  are given 
\begin{align}
(\chi_8)^\al{}_\be  &\equiv  (\ga^1 \gabar^2 \cdots \ga^7 
\gabar^8)^\al{}_\be\comma \qquad \chi_8^2 =1 \comma   \\
(\chibar_8)_\al{}^\be  &\equiv  (\gabar^1 \ga^2 \cdots \gabar^7 
\ga^8)_\al{}^\be\comma \qquad \chibar_8^2 =1 \period
\end{align}
Since $\chi_8^2=\chibar_8^2=1$, we can choose the basis where they are 
diagonal, namely 
\begin{align}
(\chi_8)^\al{}_\be &= \matrixii{1_8}{0}{0}{-1_8} 
= \matrixii{\delta^a{}_b}{0}{0}{-\delta^\adot{}_\bdot} \comma \\
(\chitil_8)_\al{}^\be  &= \matrixii{1_8}{0}{0}{-1_8} 
= \matrixii{\delta_a{}^b}{0}{0}{-\delta_\adot{}^\bdot} \comma 
\end{align}
where  $SO(8)$ chiral and anti-chiral sectors are indexed  by the letters $a,b, \ldots$ and 
 $\adot, \bdot, \ldots$ respectively. These indices run from 1 to 8. 
In this basis each of the  chiral and anti-chiral spinors 
$\lam^\al$ and $\lam_\al$ is decomposed into $SO(8)$-chiral and anti-chiral 
 components as
\begin{align}
\lam^\al &= \vecii{\lam^a}{\lam^\adot} \comma \qquad 
\lam_\al = \vecii{\lam_a}{\lam_\adot} \period
\end{align}
In $16$ component notation, these decompositions are effected by the 
following two types of  $SO(8)$ chiral projection operators:
\begin{align}
(P^\pm_8)^\al{}_\be &= \half (1\pm \chi_8)^\al{}_\be \comma \qquad 
(\Pbar^\pm_8)_\al{}^\be = \half (1\pm \chibar_8)_\al{}^\be  \period
\end{align}
Correspondingly, we can make the $SO(8)$ decomposition of 
$\ga^\mu$ matrices in the following way. First since $\ga^I$ and $\gabar^I$ ``anti-commute" with 
the $SO(8)$ chirality operators  in the manner 
$\ga^I \chibar_8 = -\chi_8 \ga^I$ and $\gabar^I \chi_8 = -\chibar_8 \gabar^I$, 
they must be  block off-diagonal as 
\begin{align}
\ga^I &= \matrixii{0}{(\ga^I)^{a\bdot}}{(\ga^I)^{\adot b}}{0}  \comma \qquad 
\gabar^I =  \matrixii{0}{(\gabar^I)_{a\bdot}}{(\gabar^I)_{\adot b}}{0}  \period
\end{align}
Next we examine  the $SO(8)$ content of the light-cone components 
$\ga^\pm$ and $\gabar^\pm$, which are  defined by 
\begin{align}
\Ga^\pm &= \ovsqtwo (\Ga^9\pm \Ga^0) 
= \matrixii{0}{\ga^\pm}{\gabar^\pm}{0} \comma \\
\ga^\pm &= \ovsqtwo (\ga^9 \pm \ga^0) \comma \qquad 
\gabar^\pm = \ovsqtwo(\gabar^9 \pm \gabar^0) \period
\label{defgapm}
\end{align}
The relations $\ga^+\gabar^- + \ga^-\gabar^+=\gabar^+\ga^- + \gabar^-\ga^+ = 2 $, which follow from the Clifford algebra, are often useful. 
We now fix the explicit  form of $\ga^\pm$ and $\gabar^\pm$ by 
adopting a suitable convention. 
From the relation $\Gabar_{1,9} = \Ga^0 \Gabar_8 \Ga^9$, we 
 get $\ga^9 =\ga^0 \gabar^1 \ga^2 \cdots \gabar^8$ and 
$\gabar^9 = -\gabar^0 \ga^1 \gabar^2 \cdots \ga^8$. They are  actually 
  equal since, as was already mentioned,  $\ga^m = \gabar^m$ for $m=1\sim 9$ and $\ga^0 =-\gabar^0$. One consistent convention we adopt  is 
$\ga^0=1= -\gabar^0$. Then  from the definition (\ref{defgapm}) 
 the non-vanishing $SO(8)$ components of 
$\ga^\pm$ and $\gabar^\pm$ take the form
\begin{align}
(\ga^+)^{ab} &= \sqtwo \delta^{ab} \comma
 \qquad (\ga^-)^{\adot\bdot} = -\sqtwo\delta^{\adot\bdot}\comma  \\
(\gabar^+)_{\adot\bdot} &= -\sqtwo\delta_{\adot\bdot} \comma 
\qquad (\gabar^-)_{ab} = \sqtwo\delta_{ab} \period
\end{align}
\section{Some details of the Lorentz algebra  }
\subsection{Illustration of OPE method for Lorentz algebra}
Let us illustrate our OPE method for the calculation of the Lorentz 
 algebra  by giving an  example  from the computation of 
$\com{L^{\mu\nu}}{L^{\rho\sig}}$ using the form of $L^{\mu\nu}$ given 
 in (\ref{Lmunu}).  It suffices to discuss the basic commutator 
\begin{align}
\com{\int[dz] \Xcirc^\mu \Pi^\nu(z)}{\int[dw] \Xcirc^\rho\Pi^\sig(w)} 
\end{align}
and  we shall  focus on the contribution coming from the contraction of $\Xcirc^\mu(z)$ and $\Xcirc^\rho(w)$. As this produces a logarithm that  violates  the  condition (i) described in the main text,  we go back to the definition (\ref{comAB}) and denote the integrals 
 for the two regions as $I_\zgw$ and $I_\wgz$. 
Then, $I_\zgw$ is given by 
\begin{align}
I_\zgw &=  \int_\zgw [dz][dw]  \left[ \left(-\eta^{\mu \rho} \ln \left( 1-{w \over z}\right) \right) :\Pi^\nu(z) \Pi^\sig(w):
 \right] \period
\end{align}
Although a logarithm is present, the cut in the $z$-plane is from $0$ to $w$ and 
 there is no singularity along the $z$-contour. Therefore we can substitute $\Pi^\nu(z) = i\del \Xcirc^\nu(z) + (p^\nu/z)$ and perform integration by parts to rewrite it as 
\begin{align}
I_\zgw &= \eta^{\mu\rho}  \int_\zgw [dz][dw]  \del_z  \ln \left( 1-{w \over z}\right)i :\Xcirc^\nu(z) \Pi^\sig(w): \nn\\
&- \eta^{\mu\rho}  \int_\zgw [dz][dw]   \ln \left( 1-{w \over z}\right)
{p^\nu \over z}  \Pi^\sig(w) \period
\end{align}
The second term actually vanishes since the $z$-contour can be moved to 
 infinity and then we have  $\ln 1=0$. Therefore
\begin{align}
I_\zgw &= i \eta^{\mu\rho}  \int_\zgw [dz][dw] \left( {1\over z-w} -{1\over z}
\right):\Xcirc^\nu(z) \Pi^\sig(w):  \nn\\
&=  i \eta^{\mu\rho}  \int_\zgw [dz][dw] {1\over z-w} 
:\Xcirc^\nu(z) \Pi^\sig(w):  -2i\eta^{\mu\rho}x^\nu \pi^\sig \period 
\label{XXzw}
\end{align}
Note that we  pick up an additional zero mode contribution. 
As for the $I_\wgz$ integral, by similar reasoning we can rewrite it as 
\begin{align}
I_\wgz 
&= - \int_{\wgz} [dz][dw] \eta^{\mu\rho}  \ln \left( 1-{z \over w}\right) 
\Pi^\sig(w)\left( i \del \Xcirc^\nu(z) + {p^\nu \over z} \right) \nn\\
&= i  \eta^{\mu\rho}  \int_{\wgz} [dz][dw] \del_z  \ln \left( 1-{z \over w}\right) \Pi^\sig(w)\Xcirc^\nu(z)  \nn\\
&\qquad  - \int_{\wgz} [dz][dw] \eta^{\mu\rho}  \ln \left( 1-{z \over w}\right) 
\Pi^\sig(w) {p^\nu \over z} \period
\end{align}
Again the $z$ integral in  the second term vanishes since $z$ contour is around $0$ and we have the factor $\ln 1=0$.  Therefore we obtain
\begin{align}
I_\wgz&=  i  \eta^{\mu\rho}  \int_{\wgz} [dz][dw]  {1\over z-w} \Pi^\sig(w)\Xcirc^\nu(z)  \period \label{XXwz}
\end{align}
We can now put  the two contributions (\ref{XXzw}) and (\ref{XXwz}) together. 
The parts containing the simple pole can be combined in the usual way and we easily obtain 
\begin{align}
 I_\zgw-I_\wgz &= i \eta^{\mu\rho} \int[dw] :\Xcirc^\nu \Pi^\sig(w):  -2i\eta^{\mu\rho}x^\nu \pi^\sig \period 
\end{align}
Contributions from all the other contractions can be computed in a similar manner and we find\footnote{Double contraction contributions are easily seen to vanish.}
\begin{align}
&\com{\int[dz] \Xcirc^\mu \Pi^\nu(z)}{\int[dw] \Xcirc^\rho \Pi^\sig(w)}\nn\\&\quad =   i \int[dw] \left[ \eta^{\mu \sig} \Xcirc^\rho \Pi^\nu -\eta^{\nu \rho}
\Xcirc^\mu \Pi^\sig + i\eta^{\mu\rho}\Xcirc^\nu \Pi^\sig(w)
-\eta^{\nu\sig} \Xcirc^\rho \Pi^\mu  \right]  \nn\\
& \quad + 2i ( -\eta^{\mu \rho} x^\nu p^\sig+\eta^{\nu\sig} x^\rho p^\mu
+\eta^{\mu\sig} x^\rho p^\nu-\eta^{\nu\rho} x^\mu p^\sig) \period
\end{align}
Note that the extra zero mode part in the second line is symmetric under the interchange $\mu \leftrightarrow \nu$. Thus it vanishes upon antisymmetrization and we 
recover the usual closure for the algebra $\com{L^{\mu\nu}}{L^{\rho\sig}}$.

Having explained the basic method of computation, we now comment on 
 the calculation of the commutator between a term involving $\Xcirc^\mu$ and a term free of $\Xcirc^\mu$. As an illustration, let us consider the following 
 commutator which occurs in  $\com{\calM^{I-}}{\calM^{J-}}$: 
\begin{align}
&\left[\int[dz]~\Xcirc^{I}\Pi^{-}(z),\, \int[dw]~\frac{\left(\bar{\gamma}^{J} S\right)_{\dot{a}}
\left(\bar{\gamma}^{K} S\right)_{\dot{a}} \Pi^{K}(w)}{\Pi^{+}} \right] \notag \\
&= \int_{\zgw}\left(\Xcirc^{I}\Pi^{-}(z)~ 
\frac{\left(\bar{\gamma}^{J} S\right)_{\dot{a}}
\left(\bar{\gamma}^{K} S\right)_{\dot{a}} \Pi^{K}(w)}{\Pi^{+}}\right) - 
\int_{\wgz}\left(\frac{\left(\bar{\gamma}^{J} S\right)_{\dot{a}}
\left(\bar{\gamma}^{K} S\right)_{\dot{a}} \Pi^{K}(w)}{\Pi^{+}}~\Xcirc^{I}\Pi^{-}(z)\right) \period
\end{align}
In this case, since the relevant basic OPE's have  no logarithmic singularities, we only need to be careful about the zero mode parts coming from (\ref{defPichk}) and (\ref{prodPiXcirc}).  We then get
\begin{align}
&\left[\int[dz]~\Xcirc^{I}\Pi^{-}(z),\,\int[dw]~\frac{\left(\bar{\gamma}^{J} S\right)_{\dot{a}}
\left(\bar{\gamma}^{K} S\right)_{\dot{a}} \Pi^{K}(w)}{\Pi^{+}} \right] \notag \\
&= \int[dw] \left(i\,\frac{\left(\bar{\gamma}^{J} S\right)_{\dot{a}}
\left(\bar{\gamma}^{I} S\right)_{\dot{a}}\Pi^{-}(w)}{\Pi^{+}} -  
\frac{\left(\bar{\gamma}^{J} S\right)_{\dot{a}}
\left(\bar{\gamma}^{K} S\right)_{\dot{a}} \partial \Xcirc^{I} \Pi^{K}(w)}
{\left(\Pi^{+}\right)^{2}} \right) \notag \\ 
& \quad + i \int_{\wgz}[dz][dw] \left(\frac{1}{w}\right) 
\left(\frac{\left(\bar{\gamma}^{J} S\right)_{\dot{a}}
\left(\bar{\gamma}^{I} S\right)_{\dot{a}}(w)}{\Pi^{+}}\right) \Pi^{-}(z) \notag \\
&= - i \int[dw]~\left(\frac{\left(\bar{\gamma}^{I} S\right)_{\dot{a}}
\left(\bar{\gamma}^{J} S\right)_{\dot{a}}\Pi^{-}(w)}{\Pi^{+}} -  
\frac{\left(\bar{\gamma}^{J} S\right)_{\dot{a}}
\left(\bar{\gamma}^{K} S\right)_{\dot{a}} \Pi^{I} \Pi^{K}(w)}
{\left(\Pi^{+}\right)^{2}} \right) \notag \\
& \quad - i p^{-} \frac{\left(\bar{\gamma}^{I} S\right)_{\dot{a}}
\left(\bar{\gamma}^{J} S\right)_{\dot{a}}(0)}{\Pi^{+}}
- i p^{I} \frac{\left(\bar{\gamma}^{J} S\right)_{\dot{a}}
\left(\bar{\gamma}^{K} S\right)_{\dot{a}} \Pi^{K}(0)}
{\left(\Pi^{+}\right)^{2}}\comma \label{MIMJone}
\end{align}
where we have used the relation (\ref{defPichk}) for the last equality.
Another expression which pairs with the previous one  in $\com{\calM^{I-}}{\calM^{J-}}$ is 
\begin{align}
&\left[\int[dz]~\Xcirc^{-}\Pi^{I}(z),\,\int[dw]~\frac{\left(\bar{\gamma}^{J} S\right)_{\dot{a}}
\left(\bar{\gamma}^{K} S\right)_{\dot{a}} \Pi^{K}(w)}{\Pi^{+}} \right] \notag\\
&= i \int[dw]~\left(\frac{\left(\bar{\gamma}^{I} S\right)_{\dot{a}}
\left(\bar{\gamma}^{J} S\right)_{\dot{a}}\Pi^{-}(w)}{\Pi^{+}} -  
\frac{\left(\bar{\gamma}^{J} S\right)_{\dot{a}}
\left(\bar{\gamma}^{K} S\right)_{\dot{a}} \Pi^{I} \Pi^{K}(w)}
{\left(\Pi^{+}\right)^{2}} \right) \notag \\
& \quad - i p^{-} \frac{\left(\bar{\gamma}^{I} S\right)_{\dot{a}}
\left(\bar{\gamma}^{J} S\right)_{\dot{a}}(0)}{\Pi^{+}}
- i p^{I} \frac{\left(\bar{\gamma}^{J} S\right)_{\dot{a}}
\left(\bar{\gamma}^{K} S\right)_{\dot{a}} \Pi^{K}(0)}
{\left(\Pi^{+}\right)^{2}} \period \label{MIMJtwo}
\end{align}
We see that the 
zero mode parts exactly cancel in the difference of (\ref{MIMJone}) and (\ref{MIMJtwo}) 
and the result is the same as what one obtains by applying the standard OPE formula (\ref{XPiOPE}). 
This phenomenon  occurs also  in other commutators of similar types.  
\subsection{Calculation of $\com{{\cal M}^{I -}}{{\cal M}^{J -}}$ }
Here we display some details of the non-trivial computation  leading 
from (\ref{boost com1}) to (\ref{boost com2}) in the calculation 
 of the commutator $\com{{\cal M}^{I -}}{{\cal M}^{J -}}$. 

First, using the formula  (\ref{quarticgamma com}) for the commutator 
 of $\ga^{IJ}$'s, we can rewrite the 7th term of (\ref{boost com1}) as 
\begin{align}
& - \frac{1}{4}\left(\frac{\Pi^{K} \Pi^{L}}{\left(\Pi^{+}\right)^2}\right)
\left\{S_{a}\left(\gamma^{IK}\,\gamma^{JL}\right)^{a}{}_{b}S^{b}\right\} = 
- \frac{1}{8}\left(\frac{\Pi^{K} \Pi^{L}}{\left(\Pi^{+}\right)^2}\right)
\left\{S_{a}\left[\gamma^{IK},\,\gamma^{JL}\right]^{a}{}_{b}S^{b}\right\} \notag \\
& = \frac{1}{4}\frac{\Pi^{I} \Pi^{K}(\bar{\gamma}^{J} S)_{\dot{a}}
(\bar{\gamma}^{K} S)_{\dot{a}}}{\left(\Pi^{+}\right)^{2}} 
-~\frac{1}{4}\frac{\Pi^{J} \Pi^{L}(\bar{\gamma}^{I} S)_{\dot{a}}
(\bar{\gamma}^{L} S)_{\dot{a}}}{\left(\Pi^{+}\right)^{2}} \notag \\
& \quad +~\frac{1}{4}\frac{\Pi^{K} \Pi^{K}(\bar{\gamma}^{I} S)_{\dot{a}}
(\bar{\gamma}^{J} S)_{\dot{a}}}{\left(\Pi^{+}\right)^{2}}. 
\end{align}   
Next, the 8th term of (\ref{boost com1}) can be rewritten, up to total derivative with respect to $w$, 
as\,\footnote{The symbol ``$\simeq$'' represents the equality up to total derivatives.}, 
\begin{align}
& \frac{1}{16} \left\{\partial \left(\frac{S_{a}}{\Pi^{+}}\right)\right\}
\left[\gamma^{IK},\,\gamma^{JK}\right]^{a}{}_{b}\left\{\partial \left(\frac{S^{b}}
{\Pi^{+}}\right)\right\} \notag \\
& = -\frac{3}{4}\left\{\partial \left(\frac{
\left(\bar{\gamma}^{I}S\right)_{\dot{a}}}{\Pi^{+}}\right)\right\} \left\{\partial \left(\frac{
\left(\bar{\gamma}^{J}S\right)_{\dot{a}}}{\Pi^{+}}\right)\right\} \notag \\
& \simeq ~\frac{3}{4}\left\{\partial^{2} \left(\frac{\left(\bar{\gamma}^{I}S\right)_{\dot{a}}}
{\Pi^{+}}\right)\right\} 
\left(\frac{\left(\bar{\gamma}^{J}S\right)_{\dot{a}}}{\Pi^{+}}\right) \notag \\
& \simeq ~\frac{3}{4}\left\{\partial^{2}\left(\frac{1}{\Pi^{+}}\right)\right\} 
\left(\frac{\left(\bar{\gamma}^{I}S\right)_{\dot{a}}\left(\bar{\gamma}^{J}S\right)_{\dot{a}}}
{\Pi^{+}}\right) \notag \\
& \quad +~\frac{3}{4}\frac{\left(\bar{\gamma}^{I} \partial^{2}S\right)_{\dot{a}}
\left(\bar{\gamma}^{J}S\right)_{\dot{a}}}{\left(\Pi^{+}\right)^{2}} 
-~\frac{3}{2} \frac{\partial \Pi^{+}
\left(\bar{\gamma}^{I}\partial S\right)_{\dot{a}}\left(\bar{\gamma}^{J}S\right)_{\dot{a}}}
{\left(\Pi^{+}\right)^{3}}. 
\end{align} 
Thirdly, by using the formula ${\rm Tr}\left(\{\gamma^{IK},\,\gamma^{JL}\}\right) = 
16 ~\left(\delta^{IL}\delta^{JK}-\delta^{IJ}\delta^{KL}\right)$, which follows  from the  identity
(\ref{quarticgamma acom}), the last term of (\ref{boost com1})
can be simplified, again up to total derivatives,  to 
\begin{align}
&- \frac{1}{8}\left\{\partial \left(\frac{\Pi^{K}}{\Pi^{+}}\right)\right\}
\left(\frac{\Pi^{L}}{\Pi^{+}}\right) {\rm Tr}\left(\gamma^{IK}\,\gamma^{JL}\right) \notag \\
& = - \frac{1}{16}\left\{\partial \left(\frac{\Pi^{K}}{\Pi^{+}}\right)\right\}
\left(\frac{\Pi^{L}}{\Pi^{+}}\right) 
{\rm Tr}\left(\left\{\gamma^{IK},\,\gamma^{JL}\right\}\right) \notag \\
& \simeq - \left\{\partial \left(\frac{\Pi^{J}}{\Pi^{+}}\right)\right\} 
\left(\frac{\Pi^{I}}{\Pi^{+}}\right) 
\simeq - \frac{1}{2} \frac{ \Pi^{I} \partial \Pi^{J} - \Pi^{J} \partial \Pi^{I}}
{\left(\Pi^{+}\right)^{2}}.  
\end{align}
Finally, we can rewrite the 6th term of (\ref{boost com1}), which is quartic in 
 fermions,  as (up to a total derivative) 
\begin{align}
& -\frac{1}{16} \left\{\partial \left(\frac{(\bar{\gamma}^I S)_{\dot{a}}
(\bar{\gamma}^{K} S)_{\dot{a}}}{\Pi^{+}}\right)\right\}
\left(\frac{(\bar{\gamma}^J S)_{\dot{b}}(\bar{\gamma}^{K} S)_{\dot{b}}}
{\Pi^{+}}\right) \notag \\
& \simeq -\frac{1}{32}\left(\frac{1}{\left(\Pi^{+}\right)^{2}}\right) 
\Bigl[\partial_{w}\left\{(\bar{\gamma}^{I} S)_{\dot{a}}(\bar{\gamma}^{K} S)_{\dot{a}}\right\}\cdot
(\bar{\gamma}^{J} S)_{\dot{b}}(\bar{\gamma}^{K} S)_{\dot{b}} - (\bar{\gamma}^I S)_{\dot{a}}
(\bar{\gamma}^{K} S)_{\dot{a}}\cdot\partial_{w}\left\{(\bar{\gamma}^{J} S)_{\dot{b}}
(\bar{\gamma}^{K} S)_{\dot{b}}\right\}\Bigr] \notag \\
&= -\frac{1}{4}\frac{S_{a}\partial S_{a}\, (\bar{\gamma}^{I}S)_{\dot{a}}(\bar{\gamma}^{J}S)_{\dot{a}}}
{\left(\Pi^{+}\right)^{2}}.
\end{align}
In this process we have used the basic $\ga$-matrix identity (\ref{GammaId1}) and the following formula ( which itself can be verified  with  repeated use of  (\ref{GammaId1}) and (\ref{GammaId2}). ):
\begin{align}
(\bar{\gamma}^{I}\partial S)_{\dot{a}}(\bar{\gamma}^{K} S)_{\dot{a}}
(\bar{\gamma}^{J}S)_{\dot{b}}(\bar{\gamma}^{K}S)_{\dot{b}} - 
(\bar{\gamma}^{I}S)_{\dot{a}}(\bar{\gamma}^{K}S)_{\dot{a}}
(\bar{\gamma}^{J}\partial S)_{\dot{b}}(\bar{\gamma}^{K} S)_{\dot{b}} = 
6\, S_{a}\partial S_{a}\,(\bar{\gamma}^{I}S)_{\dot{a}}(\bar{\gamma}^{J}S)_{\dot{a}} \period
\end{align}

Using these rearrangements, (\ref{boost com1}) can be  brought  to 
 the simpler form (\ref{boost com2}).

\section{Construction of vertex operators }
In the construction of the vertex operators, one needs various identities involving 
$SO(8)$ $\ga$-matrices and spinors, some of which are rather non-trivial. In this 
 appendix, we  record and discuss such identities. 
\subsection{Useful $SO(8)$ $\ga$-matrix identities  }
In this appendix, we collect some useful identities involving the $SO(8)$ $\ga$-matrices, which are frequently used  in the calculations described in the main text and in other parts of the appendices. 

First we give the definitions of the anti-symmetrized products of the $SO(8)$ $\ga$-matrices:
\begin{align}
 & \gamma^{IJ} \equiv \frac{1}{2}\left(\gamma^{I}\bar{\gamma}^{J} - 
\gamma^{J}\bar{\gamma}^{I}\right)\comma  \qquad
\bar{\gamma}^{IJ} \equiv \frac{1}{2}\left(\bar{\gamma}^{I}\gamma^{J} 
- \bar{\gamma}^{J}\gamma^{I}\right), \notag \\
&\gamma^{I J K} \equiv  \frac{1}{3\,!} \left(\gamma^{I} \bar{\gamma}^{J} \gamma^{K} 
\pm (\textrm{cycl.})\right) \comma \qquad 
\bar{\gamma}^{I J K} \equiv  \frac{1}{3\,!} \left(\bar{\gamma}^{I} \gamma^{J} \bar{\gamma}^{K} 
\pm (\textrm{cycl.})\right) \comma 
\nn\\
& \gamma^{I J K L} \equiv  \frac{1}{4\,!} 
\left(\gamma^{I} \bar{\gamma}^{J} \gamma^{K} \bar{\gamma}^{L} 
\pm (\textrm{cycl.})\right) \comma \qquad \bar{\gamma}^{I J K L} \equiv  \frac{1}{4\,!} 
\left(\bar{\gamma}^{I} \gamma^{J} \bar{\gamma}^{K} \gamma^{L} 
\pm (\textrm{cycl.})\right) \period
\end{align}
(As  remarked earlier, once we decompose  the spinors and the $\ga$-matrices into $SO(8)$ components, we can raise and lower the indices by $\delta^{ab}$, $\delta_{ab}$ and so on.  Below we use this freedom to lower all the indices. )
The $SO(8)$ Clifford relations read
\begin{align}
\left\{\gamma^{I},\,\bar{\gamma}^{J}\right\}_{ab} &\equiv 
(\gamma^{I} \bar{\gamma}^{J} + \gamma^{J} \bar{\gamma}^{I} )_{ab}= 2\,\delta_{ab} \delta^{I J}, \quad 
\left\{\bar{\gamma}^{I},\,\gamma^{J}\right\}_{\dot{a} \dot{b}}
 \equiv 
(\bar{\gamma}^{I} \gamma^{J} + \bar{\gamma}^{J} \gamma^{I})_{\adot\bdot} = 2\,\delta_{\dot{a}\dot{b}} \delta^{I J} 
\period \label{defgamma8}
\end{align}
Often used fundamental identities are 
\begin{align}
2 \delta_{a b} \delta_{\dot{a} \dot{b}} &= 
\gamma^{I}_{a \dot{a}} \gamma^{I}_{b \dot{b}} + 
\gamma^{I}_{a \dot{b}} \gamma^{I}_{b \dot{a}} \comma \label{GammaId1} \\
\left(\gamma^{I} \bar{\gamma}^{J}\right)_{a b} &= 
\delta^{I J}\delta_{ab} + \left(\gamma^{I J}\right)_{ab} \comma 
\quad \left(\bar{\gamma}^{I} \gamma^{J}\right)_{\dot{a} \dot{b}} 
= \delta^{I J} \delta_{\dot{a} \dot{b}} 
+ \left(\bar{\gamma}^{I J}\right)_{\dot{a} \dot{b}} \period \label{GammaId2}
\end{align}
Fundamental Fierz identities read
\begin{align}
S_{a} S_{b} &= \frac{1}{16} \left(S\gamma_{KL}S\right) \left(\gamma^{KL}\right)_{a b}  \comma \label{Fierz1}\\
\gamma^{I}_{a \dot{a}} \gamma^{I}_{b \dot{b}} &= \delta_{a b} \delta_{\dot{a} \dot{b}} 
+ \frac{1}{4} \gamma^{K L}_{a b} \bar{\gamma}^{K L}_{\dot{a} \dot{b}} . \label{Fierz2}
\end{align}
Useful formulas for various commutators and anti-commutators are given by 
\begin{align}
\left[\bar{\gamma}^{I},\, \gamma^{J K} \right] &\equiv 
\bar{\gamma}^{I} \gamma^{J K} - \bar{\gamma}^{J K} \bar{\gamma}^{I} = 
2 \left(\delta^{I J} \bar{\gamma}^{K} - \delta^{I K} \bar{\gamma}^{J}\right) , \label{cubicgamma com}\\ 
\left\{ \bar{\gamma}^{I},\, \gamma^{J K} \right\} &\equiv \bar{\gamma}^{I} \gamma^{J K} + 
\bar{\gamma}^{J K} \bar{\gamma}^{I} = 2 \bar{\gamma}^{I J K} , \label{cubicgamma acom}\\
\left[\gamma^{I J},\, \gamma^{K L}\right] &= 2 \left( \delta^{I L}\, \gamma^{J K} 
+ \delta^{J K}\, \gamma^{I L} - \delta^{I K}\, \gamma^{J L} - \delta^{J L}\, \gamma^{I K} \right) , 
\label{quarticgamma com}\\
\left\{\gamma^{I J},\, \gamma^{K L}\right\} &= 
2 \left( \delta^{I L} \delta^{J K} - \delta^{I K} \delta^{J L}\right) + 2 \gamma^{I J K L} \period
\label{quarticgamma acom}
\end{align}

\subsection{Identities involving $SO(8)$ spinors  }
\def\dotc{{\dot{c}}}
\def\fdot{{\dot{f}}}
First we discuss the identity involving a product of five $S_a$'s of the  form
\begin{align}
T &\equiv k_A k_B k_C (S\ga^{JA}S) (S\ga^{JB} S) (S\ga^C \ep) =0 \comma \label{5S Id}
\end{align}
which holds  for $ k_Ak_A=0$. All the capital Roman indices  refer to the
 transverse components, running from 1 to 8.  We will give a sketch of the proof, 
which requires some amount  of computation. 

In terms of components, we have $T=k_A k_B k_C S_a (S_b S_c) (S_d S_e) \ga^{JA}_{ab} \ga^{JB}_{cd} \ga^C_{e \fdot} \ep_\fdot$. The basic idea is to expand
the antisymmetric products  $S_b S_c$ and $S_d S_e$ in terms of 
 $\ga^{AB}$, which  span the space of $8\times 8$ antisymmetric quantities. 
Explicitly, $S_b S_c = (1/16) (\ga^{PQ})_{bc} (S\ga_{PQ} S)$ and similarly for $S_dS_e$. Substituting them into $T$, we get
\begin{align}
T &= \left({1\over 16}\right)^2 k_A k_B k_C (\ga^{JA}\ga^{PQ} \ga^{JB} \ga^{MN} \ga^C )_{a \fdot}
(S\ga_{PQ} S)  (S\ga_{MN} S) S_a \ep_\fdot \period
\end{align}
The product of five $\ga$ matrices appearing in  this expression can be reduced  by 
repeated use of the familiar 
identity $\ga^{AB}\ga^{CD} = \ga^{ABCD} + \delta^{BC} \ga^{AD}
+ \delta^{AD} \ga^{BC} -\delta^{AC} \ga^{BD} - \delta^{BD}\ga^{AC}$. 
In the course of this reduction, because of the contraction with $k_Ak_Bk_C$, some of the terms drop out due to 
the condition $k_A k_A=0$.  Simplification also occurs when a symmetric quantity 
 is contracted with an  antisymmetric quantity such as $S\ga_{PQ}S$. After some straightforward but tedious computation one proves $T=0$. 

Another non-trivial identity one needs involves products of three $S_a$'s. It  reads
\begin{align}
  k^B R^A(\ep \gabar^I \ga^{AB} S) +2k^I R^C (\ep \gabar^C S)
-6k^C R^I  (\ep \gabar^C S)  =0 \comma  \label{idtwo}
\end{align}
where $R$ is an expression quadratic in $S$ given by 
\begin{align}
R^A &= k^D S \ga^{AD} S  \period 
\end{align}
Just as before, this identity holds only when   $k_A k_A=0$. To prove it, we first make it  into a slightly better form by applying  the simple formula
$\gabar^I \ga^{AB}= \gabar^A \ga^{BI} + 2\delta^{IA} \gabar^B -\delta^{IB} \gabar^A -\delta^{AB} \gabar^I$ to the first term of  (\ref{idtwo}). 
Then, what we need to prove becomes $A_1=4A_3-A_2$, where 
\begin{align}
A_1 &= k^B R^A(\ep \gabar^A \ga^{BI} S) 
\comma \qquad A_2 =  k^I R^C (\ep \gabar^C S) \comma \qquad 
A_3=k^C R^I  (\ep \gabar^C S) \period
\end{align}
We now make use of  the basic $SO(8)$ $\ga$-matrix identity (\ref{Fierz2}). 
Contract this with \break $k^B k^C S_a S_c \gabar^C_{\adot c} \ga^{BI}_{bd} \ep_\bdot S_d$. The left hand side precisely yields  $A_1$. The right hand side gives 
$A_3+  B$, where $B=(1/4) k^B k^C (S \ga^{PQ} \ga^{BI} S) (\ep \gabar^{PQ}\gabar^C S)$. We can  reduce the product of $\ga$-matrices in this expression 
by using the familiar formulas. Again in the course of this reduction, some of the terms vanish due to 
$k_Ak_A=0$. After some compuation, we obtain $B=(1/2)(A_1+2A_3-A_2)$. 
Together we find $A_1=A_3+(1/2)(A_1+2A_3-A_2)$, which gives the desired 
relation $A_1=4A_3-A_2$. 
\subsection{Some details of  the $\epsilon$--SUSY transformation of $V_{B}(\zeta)$}
In this appendix we provide  some details of the $\ep$-SUSY 
 transformation property of the bosonic vertex operator $V_B(\zeta)$ given 
in  (\ref{BVertex}).  
\subsubsection{Calculation of the commutator $
\left[ \epsilon^{\dot{a}} Q_{\dot{a}},\, V_{B}(\zeta) \right]$}
Using the OPE technique, it is tedious but straightforward 
 to compute  the commutator $
\left[ \epsilon^{\dot{a}} Q_{\dot{a}},\, V_{B}(\zeta) \right]$, with the result
\begin{align}
&\left[ \epsilon Q,\, V_{B}(\zeta) \right] =  
\int[dw]~ e^{i k\cdot X} \left\{\left(- 2^{1/4} \zeta^{-}\right) \sqrt{\Pi^{+}}\, 
k_{I} \left(\epsilon \bar{\gamma}^{I} S\right) + \left(2^{1/4} \zeta_{I}\right) k^{-} \sqrt{\Pi^{+}} 
\left(\epsilon \bar{\gamma}^{I} S\right) \right. \notag \\
& \quad + \left(2^{1/4} \zeta_{I}\right) 
\frac{k_{J} \Pi^{J} \left(\epsilon \bar{\gamma}^{I} S\right)}{\sqrt{\Pi^{+}}} 
+ \left(- 2^{1/4} \zeta^{J}\right) \frac{\Pi_{J} k_{I} \left(\epsilon \bar{\gamma}^{I} S\right)}{\sqrt{\Pi^{+}}}
+ \left(2^{- 3/4} \zeta_{J}\right)\frac{\Pi_{I} k_{L} 
\left(\epsilon \bar{\gamma}^{I} \gamma^{J L} S\right)}{\sqrt{\Pi^{+}}} \notag \\
& \quad + \left(2^{- 3/4} \zeta^{+}\right) \partial \left(\frac{\Pi_{I} 
\left(\epsilon \bar{\gamma}^{I} S\right)}{\left(\Pi^{+}\right)^{3/2}}\right) 
+ \left(2^{- 3/4} \zeta^{+}\right) \frac{k_{I} \Pi^{I} \Pi_{J} 
\left(\epsilon \bar{\gamma}^{J} S\right)}{\left(\Pi^{+}\right)^{3/2}}\notag \\ 
& \quad + \left(- 2^{1/4} \zeta^{+}\right) \frac{\hat{\Pi}^{-} k_{I} 
\left(\epsilon \bar{\gamma}^{I} S\right)}{\sqrt{\Pi^{+}}}
+ \left(2^{- 3/4} \zeta_{J}\right) \partial \left(\frac{1}{\sqrt{\Pi^{+}}}\right) 
k^{J} k_{L} \left(\epsilon \bar{\gamma}^{L} S\right) \notag \\
& \quad + \left(2^{- 7/4} \zeta^{+}\right) \partial^{2} \left(\frac{k_{I} 
\left(\epsilon \bar{\gamma}^{I} S\right)}{\left(\Pi^{+}\right)^{3/2}}\right) 
+ \left(- 2^{- 3/4} \zeta^{+}\right) \partial \left(\frac{1}{\sqrt{\Pi^{+}}}\right) 
\frac{k_{J} \Pi^{J} k_{L} \left(\epsilon \bar{\gamma}^{L} S\right)}{\Pi^{+}} \notag \\ 
& \quad + \left(- 2^{- 3/4} \zeta^{+}\right) \frac{\Pi_{J} \Pi^{J} k_{L} 
\left(\epsilon \bar{\gamma}^{L} S\right)}{\left(\Pi^{+}\right)^{3/2}} 
+ \left(- 7\cdot 2^{- 7/4} \zeta^{+}\right) \partial^{2} \left(\frac{1}{\sqrt{\Pi^{+}}}\right) 
\frac{k_{L} \left(\epsilon \bar{\gamma}^{L} S\right)}{\Pi^{+}} \notag \\
& \quad + \left(2^{1/4} \zeta^{+}\right) \frac{k^{-} k_{J} \Pi^{J} 
k_{I}\left(\epsilon \bar{\gamma}^{I} S\right)}{\sqrt{\Pi^{+}}} + 
\left(2^{1/4} \zeta^{+}\right) k^{-}\, \partial \left(\frac{k_{I}\left(\epsilon 
\bar{\gamma}^{I} S\right)}{\sqrt{\Pi^{+}}}\right) \notag \\
& \quad + \left(2^{1/4} \zeta^{+}\right) \frac{\left(k_{J}\Pi^{J}\right)
\left(k_{L}\Pi^{L}\right)k_{I}\left(\epsilon \bar{\gamma}^{I} S\right)}{\left(\Pi^{+}\right)^{3/2}} 
+ \left(2^{5/4} \zeta^{+}\right) \partial \left(\frac{k_{I}
\left(\epsilon \bar{\gamma}^{I} S\right)}{\sqrt{\Pi^{+}}}\right) 
\frac{k_{J}\Pi^{J}}{\Pi^{+}} \notag \\ 
& \quad + \left(- 2^{- 7/4} \zeta^{+}\right) \partial \left(\frac{\left(\epsilon \bar{\gamma}^{I} 
S\right)}{\sqrt{\Pi^{+}}}\right) \frac{R_{I}}{\Pi^{+}} 
+ \left(\frac{2^{1/4}}{24} \zeta^{+}\right) \frac{\Pi_{I} R_{J} k_{L} 
\left(\epsilon \bar{\gamma}^{I} \gamma^{J L} S\right)}{\left(\Pi^{+}\right)^{3/2}} \notag \\
& \quad + \left(2^{- 7/4} \zeta^{J}\right) 
\frac{R_{J} k_{I} \left(\epsilon \bar{\gamma}^{I} S\right)}{\sqrt{\Pi^{+}}}
+ \left(- 2^{- 7/4} \zeta^{+}\right) \frac{\Pi_{J} R^{J} k_{I} 
\left(\epsilon \bar{\gamma}^{I} S\right)}{\left(\Pi^{+}\right)^{3/2}} \notag \\
& \Biggl. \quad + \left(\frac{2^{1/4}}{96} \zeta^{+}\right) \frac{R_{J} R^{J} k_{I} 
\left(\epsilon \bar{\gamma}^{I} S\right)}{\left(\Pi^{+}\right)^{3/2}} \Biggr\} . \label{OPEepsVb} 
\end{align} 
In this calculation, we have performed an integration by parts for the first line 
and made use of the on-shell condition $k_{I} k^{I} = 0$ 
and the $\gamma$-matrix identities (\ref{cubicgamma com}), (\ref{cubicgamma acom}) appropriately.  Another identity we used is 
\begin{align}
\epsilon \left(\bar{\gamma}^{I} \gamma^{J L}\right) S = \frac{1}{2}~ \epsilon \left(
\left\{ \bar{\gamma}^{I},\, \gamma^{J L} \right\} + \left[ \bar{\gamma}^{I},\, \gamma^{J L} \right]\right) S  \period
\end{align}
For example a term proportional to $k_{I} R_{J} k_{L} \left(\epsilon \bar{\gamma}^{I} \gamma^{J L} S\right)$ was shown to vanish 
 due to the above identity  as well as the relation  $k_{I} R^{I} = 0$.
\subsubsection{Fermion vertex $V_{F}(\tilde{\tilde{u}})$} 
We now want to compare the above commutator with $V_{F}(\tilde{\tilde{u}})$. Using the form 
of $\tilde{\tilde{u}}^{\alpha}$, 
given in (\ref{epssusyu}) (with $k^+$ set to zero), we obtain 
\begin{align}
V_{F}(\tilde{\tilde{u}}) =& \int[dw]~ e^{i k\cdot X} 
\Biggl\{\left(2^{1/4} \zeta_{I}\right) \sqrt{\Pi^{+}}\, k^{-} \left(\epsilon \bar{\gamma}^{I} S\right)
+ \left(- 2^{1/4} \zeta^{-}\right) \sqrt{\Pi^{+}}\, k_{I} \left(\epsilon \bar{\gamma}^{I} S\right) \Biggr. \notag \\
& + \left(- 2^{- 3/4} \zeta_{L}\right) \frac{\Pi_{I}}{\sqrt{\Pi^{+}}} k_{J} 
\left(\epsilon \bar{\gamma}^{JL} \bar{\gamma}^{I} S\right)
+ \left(- 2^{- 3/4} \zeta^{+}\right) \frac{\Pi_{I}}{\sqrt{\Pi^{+}}} k^{-} 
\left(\epsilon \bar{\gamma}^{I} S\right) \notag \\
&\left. + \left(\frac{2^{- 3/4}}{12} \zeta_{L}\right) \frac{R_{I}}{\sqrt{\Pi^{+}}} k_{J} 
\left(\epsilon \bar{\gamma}^{JL} \bar{\gamma}^{I} S\right)
+ \left(\frac{2^{- 3/4}}{12} \zeta^{+}\right) \frac{R_{I}}{\sqrt{\Pi^{+}}} k^{-} 
\left(\epsilon \bar{\gamma}^{I} S\right) \right\} . \label{transform Vf}
\end{align}
First, one easily sees  the terms in the first line are consistent with the corresponding terms in the commutator 
(\ref{OPEepsVb}). Next, the third term of $V_{F}(\tilde{\tilde{u}})$ 
can be rewritten, using the $\gamma$-identity (\ref{cubicgamma com}), as
\begin{align}
\Pi_{I} k_{J} \zeta_{L} \left(\epsilon \bar{\gamma}^{J L} \bar{\gamma}^{I} S\right) = 
\Pi_{I} k_{J} \zeta_{L} \left(\epsilon \bar{\gamma}^{I} \gamma^{J L} S\right) 
- 2\, k_{J} \Pi^{J} \zeta_{L} \left(\epsilon \bar{\gamma}^{L} S\right) 
+ 2\, \zeta^{I} \Pi_{I} k_{J} \left(\epsilon \bar{\gamma}^{J} S\right)
\period 
\end{align} 
This is consistent with the terms in the second line of the commutator (\ref{OPEepsVb}). Now for the fifth term of $V_{F}(\tilde{\tilde{u}})$, we can make 
use of the same $\gamma$-identity (\ref{cubicgamma com}) and 
the property $k_{I} R^{I} = 0$ to bring it to the form 
\begin{align}
R_{I} k_{J} \zeta_{L} \left(\epsilon \bar{\gamma}^{J L} \bar{\gamma}^{I} S\right) = 
R_{I} k_{J} \zeta_{L} \left(\epsilon \bar{\gamma}^{I} \gamma^{J L} S\right) 
+ 2\, \zeta^{I} R_{I} k_{J} \left(\epsilon \bar{\gamma}^{J} S\right)
\period
 \label{triple fermion1}
\end{align}
This will be useful below. 
\subsubsection{Difference  as a BRST term}
We now wish to show that the difference $\left[ \epsilon^{\dot{a}} Q_{\dot{a}},\, V_{B}(\zeta) \right] 
- V_{F}(\tilde{\tilde{u}})$ 
is equal to the BRST-exact expression $\left\{ Q,\, \Psi_{B}(\ep, \zeta) \right\}$. 
Using the form of $\Psi_{B}(\ep, \zeta)$ given in (\ref{BosonOp}) we find 
\begin{align}
\left\{ Q,\, \Psi_{B}(\zeta) \right\} 
&=\left(- 2^{1/4} \zeta^{+}\right)\int[dw]~ e^{i k\cdot X} 
\left[ \frac{\Pi^{-} k_{I} \left(\epsilon \bar{\gamma}^{I} S\right)}{\sqrt{\Pi^{+}}} 
+ \frac{1}{2} \frac{\Pi_{J}\Pi^{J} k_{I} \left(\epsilon \bar{\gamma}^{I} S\right)}
{\left(\Pi^{+}\right)^{3/2}} \right. \notag\\
& -~ \frac{1}{2} \frac{\left(S \partial S\right) k_{I} \left(\epsilon \bar{\gamma}^{I} S\right)}
{\left(\Pi^{+}\right)^{3/2}} + \frac{3}{4} \frac{k_{I} \left(\epsilon \bar{\gamma}^{I} \partial^{2} S\right)}
{\left(\Pi^{+}\right)^{3/2}} + \frac{k^{-} \partial \Pi^{+} k_{I} \left(\epsilon \bar{\gamma}^{I} S\right)}
{\left(\Pi^{+}\right)^{3/2}} \notag\\
& \left. +~ \frac{k_{J} \partial \Pi^{J} k_{I} \left(\epsilon \bar{\gamma}^{I} S\right)}
{\left(\Pi^{+}\right)^{3/2}} - \frac{1}{4} \frac{\left(\partial^{2} \Pi^{+}\right) k_{I} 
\left(\epsilon \bar{\gamma}^{I} S\right)}{\left(\Pi^{+}\right)^{5/2}} 
-\, \frac{1}{2} \frac{\left(\partial \Pi^{+}\right)^{2} 
k_{I} \left(\epsilon \bar{\gamma}^{I} S\right)}{\left(\Pi^{+}\right)^{7/2}} \right] \period \label{exact term Vb} 
\end{align}   

We will organize the proof of the relation $\left[ \epsilon^{\dot{a}} Q_{\dot{a}},\, V_{B}(\zeta) \right] 
- V_{F}(\tilde{\tilde{u}}) = \left\{ Q,\, \Psi_{B}(\ep, \zeta) \right\}$
according to the number of fermions $S_a$.

First consider the term in the commutator consisting 
 of five fermions (the last term of (\ref{OPEepsVb}) ). Since no 
such term exists in $V_{F}(\tilde{\tilde{u}})$ nor in 
 the BRST-exact expression  (\ref{exact term Vb}), it must vanish 
 by itself. This is indeed the case due to the identity  (\ref{5S Id}) proved  in  appendix C.2. 

Next, consider the terms cubic in $S_a$. 
For the first such term  in (\ref{OPEepsVb}), 
we  perform the following partial integration, 
\begin{align}
\int[dw]~ \partial \left(\frac{\left(\epsilon \bar{\gamma}^{I} 
S\right)}{\sqrt{\Pi^{+}}}\right) \frac{R_{I}}{\Pi^{+}}\, e^{ik\cdot X} 
=& \int[dw]~ \left[\frac{1}{3}\frac{\left\{2 \left(\epsilon \bar{\gamma}^{I} 
\partial S\right) R_{I} - \left(\epsilon \bar{\gamma}^{I} S\right) \partial R_{I}\right\}}
{\left(\Pi^{+}\right)^{3/2}}\, e^{ik\cdot X} \right. \notag \\
&\left. -~ \frac{1}{3} \frac{\left( k^{-} \Pi^{+} + k_{J} \Pi^{J} \right)
\left(\epsilon \bar{\gamma}^{I} S\right) R_{I}}
{\left(\Pi^{+}\right)^{3/2}}\, e^{ik\cdot X} \right] \comma
\label{cubicterm}
\end{align}     
and rearrange the numerator of the first term on the RHS\footnote{Note 
that $R^{I} = k_{L} (\bar{\gamma}^I S)_{\dot{a}}(\bar{\gamma}^{L} S)_{\dot{a}}$.} 
by using the identity (\ref{cubicS Id2}). 
Then the RHS of (\ref{cubicterm}) becomes 
\begin{align}
- \int[dw] \left[\frac{2^{-7/4}}{3}\frac{\left\{2 \left(\epsilon \bar{\gamma}^{I} 
\partial S\right) R_{I} - \left(\epsilon \bar{\gamma}^{I} S\right) \partial R_{I}\right\}}
{\left(\Pi^{+}\right)^{3/2}}\right] = \int[dw] \left[2^{-3/4} \frac{S_{b} \partial S_{b}\, k_{L} 
\left(\epsilon \bar{\gamma}^{L} S\right)}{\left(\Pi^{+}\right)^{3/2}}\right]  \period 
\end{align}
This precisely matches  the three-fermion term in the BRST-exact expression (\ref{exact term Vb}). 

As there are no more three-fermion terms in (\ref{exact term Vb}), 
the rest of the three-fermion terms must match between those in 
the commutator  (\ref{OPEepsVb}) and the ones in $V_{F}(\tilde{\tilde{u}})$. This means that the following relation must hold:
\begin{align}
& \left(\frac{2^{1/4}}{24} \zeta^{+}\right) \frac{\Pi_{I} R_{J} k_{L} 
\left(\epsilon \bar{\gamma}^{I} \gamma^{J L} S\right)}
{\left(\Pi^{+}\right)^{3/2}}
+ \left(\frac{2^{- 7/4}}{3} \zeta^{+}\right) 
\frac{k^{-} \left(\epsilon \bar{\gamma}^{I} S\right) R_{I}}{\sqrt{\Pi^{+}}} \\ 
& + \left(\frac{2^{- 7/4}}{3} \zeta^{+}\right) \frac{k_{J} \Pi^{J} 
\left(\epsilon \bar{\gamma}^{I} S\right) R_{I}}{\left(\Pi^{+}\right)^{3/2}} 
+ \left(2^{- 7/4} \zeta^{J}\right) \frac{R_{J} k_{I} 
\left(\epsilon \bar{\gamma}^{I} S\right)}{\sqrt{\Pi^{+}}}\\ 
& + \left(- 2^{- 7/4} \zeta^{+}\right) \frac{\Pi_{J} R^{J} k_{I} 
\left(\epsilon \bar{\gamma}^{I} S\right)}{\left(\Pi^{+}\right)^{3/2}}\\
& = \left(\frac{2^{- 3/4}}{12} \zeta_{L}\right) \frac{R_{I}}{\sqrt{\Pi^{+}}} k_{J} 
\left(\epsilon \bar{\gamma}^{JL} \bar{\gamma}^{I} S\right)
+ \left(\frac{2^{- 3/4}}{12} \zeta^{+}\right) \frac{R_{I}}{\sqrt{\Pi^{+}}} k^{-} 
\left(\epsilon \bar{\gamma}^{I} S\right). 
\end{align}
Using (\ref{triple fermion1}) and the on-shell condition $\zeta^{+} k^{-} = - \zeta^{J} k_{J}$, 
we can reorganize  this  equation in powers of $\Pi^{+}$ in the 
following way:
\begin{align}
& \frac{\zeta^{J}}{(\Pi^{+})^{1/2}}\left\{ 4\, R_{J} k_{I} 
\left(\epsilon \bar{\gamma}^{I} S\right) - k_{J} R_{I} \left(\epsilon \bar{\gamma}^{I} S\right) 
+ R_{I} k_{L} \left(\epsilon \bar{\gamma}^{I} \gamma^{J L} S\right) \right\} \notag \\
& + \frac{\zeta^{+}}{\left(\Pi^{+}\right)^{3/2}}\left\{ 
\Pi_{I} R_{J} k_{L} \left(\epsilon \bar{\gamma}^{I} \gamma^{J L} S\right) 
+ 2\, k_{J} \Pi^{J} R_{I} \left(\epsilon \bar{\gamma}^{I} S\right) 
- 6\, \Pi_{J} R^{J} k_{I} \left(\epsilon \bar{\gamma}^{I} S\right) \right\} = 0 . 
\end{align} 
Indeed, we can show that the coefficient of  $1/\left(\Pi^{+}\right)^{1/2}$
and $1/\left(\Pi^{+}\right)^{3/2}$ vanishes separately, 
thanks to the identities (\ref{idtwo}) and the relation  $A_{1} - 4 A_{3} + A_{2} = 0$, where $A_1 \sim A_3$ are defined in appendix C.2. 
Therefore, all the three-fermion terms are consistent with the $\epsilon$-SUSY
 relation. 

Finally,  consider the single-fermion terms. 
First, we focus on the  two terms which constitute  the third line 
of the commutator (\ref{OPEepsVb}). By using an integration by parts, 
 they can be combined as
\begin{align}
& \int[dw]~ e^{i k\cdot X} \left\{ \left(2^{- 3/4} \zeta^{+}\right) \partial \left(\frac{\Pi_{I} 
\left(\epsilon \bar{\gamma}^{I} S\right)}{\left(\Pi^{+}\right)^{3/2}}\right)
+ \left(2^{- 3/4} \zeta^{+}\right) \frac{k_{I} \Pi^{I} \Pi_{J} 
\left(\epsilon \bar{\gamma}^{J} S\right)}{\left(\Pi^{+}\right)^{3/2}} \right\} \notag\\
& = \int[dw]~ e^{i k\cdot X} \left\{- \left(2^{- 3/4} \zeta^{+}\right) \frac{k^{-} \Pi_{I} 
\left(\epsilon \bar{\gamma}^{I} S\right)}{\sqrt{\Pi^{+}}} \right\} .
\end{align} 
This is recognized as the single-fermion term in $V_{F}\left(\tilde{\tilde{u}}\right)$. 
The remaining single-fermion terms in the commutator (\ref{OPEepsVb}) 
can be written in the form 
\begin{align}
& \zeta^{+}\, \int[dw]~ k_{I} \left(\epsilon \bar{\gamma}^{I} S\right)\, e^{i k\cdot X} 
\left\{ \left(- 2^{1/4}\right) \frac{\Pi^{-}}{\sqrt{\Pi^{+}}} 
+ \left(- 2^{- 3/4}\right) \frac{\Pi_{J} \Pi^{J}}{\left(\Pi^{+}\right)^{3/2}} 
+ \left(- 2^{- 3/4}\right) \frac{1}{\sqrt{\Pi^{+}}}\, \partial^{2} \left(\frac{1}{\Pi^{+}}\right) 
\right. \notag \\ 
& + \left(2^{- 7/4}\right) \frac{1}{\left(\Pi^{+}\right)^{3/2}}\, 
\partial^{2} \left( e^{i k\cdot X} \right) e^{- i k\cdot X} 
+ \left(- 2^{- 3/4}\right) k^{-}\, \partial \left(\frac{1}{\sqrt{\Pi^{+}}}\right) 
+ \left(- 2^{- 3/4}\right) \frac{k_{J}\Pi^{J}}{\Pi^{+}}\, 
\partial \left(\frac{1}{\sqrt{\Pi^{+}}}\right) \notag \\ 
& + \left(- 7\cdot 2^{- 7/4}\right) \frac{1}{\Pi^{+}}\, \partial^{2} 
\left(\frac{1}{\sqrt{\Pi^{+}}}\right) + \left(2^{1/4}\right) \frac{k^{-} k_{J} \Pi^{J}}{\sqrt{\Pi^{+}}} 
+ \left(- 2^{1/4}\right) \frac{k^{-}}{\sqrt{\Pi^{+}}}\, \partial 
\left( e^{i k\cdot X} \right) e^{- i k\cdot X} \notag \\ 
& \left. + \left(2^{1/4}\right) \frac{\left(k_{J}\Pi^{J}\right)
\left(k_{L}\Pi^{L}\right)}{\left(\Pi^{+}\right)^{3/2}} + \left(- 2^{5/4}\right) 
\frac{1}{\sqrt{\Pi^{+}}}\, \partial \left(\frac{k_{J}\Pi^{J}\, e^{i k\cdot X}}{\Pi^{+}}\right) 
e^{-i k\cdot X} \right\},
\end{align}
where  the on-shell condition $\zeta^{J} k_{J} = - \zeta^{+} k^{-}$ has been  used for the fifth term. 
After some further manipulations, this can be brought to the following form:
\begin{align}
& \zeta^{+}\, \int[dw]~ k_{I} \left(\epsilon \bar{\gamma}^{I} S\right)\, e^{i k\cdot X} 
\left\{ \left(- 2^{1/4}\right) \frac{\Pi^{-}}{\sqrt{\Pi^{+}}} 
+ \left(- 2^{- 3/4}\right) \frac{\Pi_{J} \Pi^{J}}{\left(\Pi^{+}\right)^{3/2}} \right. \notag \\
& + \left(- 37\cdot 2^{-15/4}\right)\frac{\left(\partial 
\Pi^{+}\right)^{2}}{\left(\Pi^{+}\right)^{7/2}} 
+ \left(11\cdot 2^{-11/4}\right)\frac{\partial^{2} \Pi^{+}}{\left(\Pi^{+}\right)^{5/2}} + 
\left(2^{-3/4}\right)\frac{k^{-}\, \partial \Pi^{+}}{\left(\Pi^{+}\right)^{3/2}} \notag \\
& \left. + \left(- 7\cdot 2^{-7/4}\right) \frac{k_{J}\, 
\partial \Pi^{J}}{\left(\Pi^{+}\right)^{3/2}} 
+ \left(9\cdot 2^{-7/4}\right) \frac{k_{J} \Pi^{J}\,\partial \Pi^{+}}{\left(\Pi^{+}\right)^{5/2}}
+ \left(- 3\cdot 2^{-7/4}\right) 
\frac{\left(k^{-} \Pi^{+} + k_{J} \Pi^{J}\right)^{2}}{\left(\Pi^{+}\right)^{3/2}} \right\} .
\label{remainings OPEVb}
\end{align}  
Now, again after some partial integrations, we find that this expression
 matches precisely with the remaining  single-fermion terms in the BRST-exact 
contribution (\ref{exact term Vb}), including all the non-trivial coefficients. 

This completes the demonstration that our vertex operators 
 satisfy the desired $\epsilon$-SUSY relation
\begin{align}
\left[ \epsilon^{\dot{a}}Q_{\dot{a}},\,V_{B}(\zeta) \right] = V_{F}(\tilde{\tilde{u}})~ 
+~ \left\{ Q,\, \Psi_B(\ep, \zeta) \right\}.
\end{align}

\section{Similarity transformation to the light-cone gauge and construction of DDF operators}
\subsection{Some details of  the  degree-wise analysis  }
First we give a proof of the relation (\ref{Qezero}), namely $Q_0e_{2n+1} =0$. 
For this purpose,  let us  list some useful formulas involving 
 the quantity $[(\al^+)^n]_m$  introduced in (\ref{alpnm}), 
which can be proved by appropriate symmetrization procedure:
\begin{align} 
&(A_1)\quad \al^-_l [(\al^+)^n]_m = nl [(\al^+)^{n-1}]_{l+m}\comma  \\
&(A_2)\quad \sump \al^+_{-m} [(\al^+)^n]_m  = [(\al^+)^{n+1}]_0 
\comma \\
&(A_3)  \quad \sum_{m\ne 0}  m \al^+_{-m} [(\al^+)^n]_m =0  \comma \\
&(A_4)\quad \sum_{m\ne 0}  m \al^+_{-m} [(\al^+)^n]_{m+k} =-{k\over n+1} [(\al^+)^{n+1}]_k \period
\end{align}
Now the part of $Q_0$ which acts on $e_{2n+1}$ 
 is $c_0\sump_k  (\al^+_{-k} \al^-_k + k c_{-k} b_k )$ and the main 
 part of $e_{2n+1}$ is $\sump m(m+1) c_{-m} [(\al^+)]_m$. 
As for the action of $\sump_k \al^+_{-k} \al^-_k$,  using 
 $(A_1)$ we obtain 
\begin{align}
&(\sump_k \al^+_{-k} \al^-_k) \sump m(m+1) c_{-m} [(\al^+)]_m  \nn\\
&\qquad = \sump m(m+1) c_{-m} \sump_k n k \al^+_{-k} [(\al^+)^{n-1}]_{k+m} 
\period
\end{align}
By applying  $(A_4)$ (with $m$ and $k$ interchanged)
this can be further simplified to 
\begin{align}
 -\sump m(m+1) c_{-m} m [(\al^+)]_m  \period \label{contal}
\end{align}
On the other hand, the action of $ k c_{-k} b_k $ part gives 
\begin{align}
&(\sump_k k c_{-k} b_k ) \sump m(m+1) c_{-m} [(\al^+)]_m \nn\\
&\qquad =\sump m(m+1) c_{-m} m [(\al^+)]_m  \period 
\end{align}
As this cancels (\ref{contal}) we get  $Q_0 e_{2n+1} =0$. 

Next, we give a proof of the formula (\ref{multiRtwo}) by  mathematical 
 induction.  Suppose the formula holds for $n$ commutators and apply $R_2$ from 
 right on both sides to form $n+1$ commutators. This gives 
\begin{align}
{1\over n!} \delta \leftad{R_2}^{n+1} &= e_{2n-1} R_2-(\delta r_{2n}) R_2 \period
\end{align}
Then by using various relations already established and the Jacobi identities, the term $e_{2n-1} R_2$ on the right hand side can be transformed  successively as 
\begin{align}
e_{2n-1} R_2 &= -R_2 e_{2n-1} = -(\Khat d_1) e_{2n-1} 
= -\Khat (d_1 e_{2n-1}) + (\Khat e_{2n-1}) d_1 \nn\\
&= \Khat (\delta e_{2n+1}) + n r_{2n} d_1 
= \Nhat e_{2n+1} -\delta (\Khat e_{2n+1}) + n r_{2n} \delta_{2n+2} \nn\\
&= (n+1) e_{2n+1} -\delta r_{2n+2} \period
\end{align}
As for the second term on the right hand side, we can rewrite it as 
\begin{align}
(\delta r_{2n})R_2 &= \delta (r_{2n} R_2) -d_1 r_{2n} = (n+1) \delta r_{2n+2}
 +\delta r_{2n+2} = n \delta r_{2n+2} \period \label{delrR}
\end{align}
Adding up  we get 
\begin{align}
(e_{2n-1} -\delta r_{2n}) R_2 &= (n+1) (e_{2n+1} -\delta r_{2n+2}) \period
\end{align}
Dividing by the factor $n+1$, it becomes the right hand side of  the formula (\ref{multiRtwo}) for $n+1$  case and hence the mathematical induction is completed. 
\subsection{Derivation of the formula (\ref{Rhat})
for $\Rhat$ }
Here we describe some details of the computation of $e^{\Rtil}e^{R_2}$, 
 which leads to  the formula (\ref{Rhat}) for $\Rhat$. 

We need to  make use of  the general form of  the BCH formula, which reads
\begin{align}
e^\lam e^\mu &= e^E  \comma \\
E &=  \lam + \int_0^1 dt 
\psi\left(e^{\lam} e^{t \mu}\right) \mu  \comma \label{eqE} \\
\psi(z) &= {z \ln z \over z-1} \comma 
\end{align}
where $\lam$ and $\mu$ are arbitrary operators. 
Here and hereafter, products of 
 $\lam$'s  and $\mu$'s  are to be  understood as  successive adjoint actions. 
For example, $\lam \mu$ means $\ad_\lam \mu = [\lam, \mu]$, 
and $e^\lam \mu$ means $e^{\ad_\lam} \mu = \mu 
 + \com{\lam}{\mu} + \half \com{\lam}{\com{\lam}{\mu}} + \cdots$, 
and so on. 
We shall  apply this BCH formula with $\lam = \Rtil, \mu = R_2$. 
One  simplifying feature for  the present case is that   $\lam$ commutes 
 with the commutator $\lam \mu$, and therefore we have  $\lam^n \mu =0$
 for $n\ge 2$. 

Let us introduce  the quantity $x\equiv e^{\lam}e^{t\mu} -1$. Then, 
 the $\psi$ function in (\ref{eqE}) becomes  $\psi(1+x)$, which 
 can  be readily expanded in powers of $x$.  It reads 
\begin{align}
\psi (1+x) &= {(1+x) \ln (1+x) \over x}  = 1+\sum_{n=1}^\infty 
{(-1)^{n+1} \over n(n+1)} x^n \period 
\end{align}
We need to  act this expression on $\mu$. 
Because $e^{t\mu}\mu =\mu$ and $\lam \lam \mu=0$,   we get
$x \mu = e^\lam \mu -\mu =(1 + \lam)\mu -\mu = \lam\mu$, etc. 
Then, it is easy to see that $x^n \mu$ can be written as 
\begin{align}
x^n \mu &= y^{n-1}  (\lam \mu) \comma 
\end{align}
where $y \equiv e^{t\mu}-1$. Combining, the exponent of the BCH formula 
is simplified  as 
\begin{align}
E&= \lam + \mu + \int_0^1 dt  \sum_{n=1}^\infty 
{(-1)^{n+1} y^{n-1} \over n(n+1)}
(\lam\mu)  \nn\\
&=  \lam + \mu + \int_0^1 dt  {1\over y} (\psi(1+y) -1) (\lam \mu) 
\period 
\end{align}
Furthermore, the integral over $t$ can be done explicitly. 
By  changing the  variable from $t$ to $t\mu$ and  further to $y$ itself, 
we can compute  it as 
\begin{align}
I &\equiv \int_0^1 dt  {1\over y} (\psi(1+y) -1)\nn\\
&= {1\over \mu} \int_0^{e^\mu -1} dy \left( {\ln (1+y) \over y^2} 
 -{1 \over y (1+y)} \right) 
= {1\over \mu} {e^\mu -1 -\mu \over e^\mu -1} \period 
\end{align}
This can be further rewritten in terms of $\coth (\mu / 2)$ 
and is expanded in powers of $\mu$ in the following way:
\begin{align}
I &= \half + {1\over \mu} -\half \coth {\mu \over 2} 
=  \half + \sum_{m=1}^\infty (-1)^m {B_m \mu^{2m-1} \over (2m)!} 
\period
\end{align}
Here $B_n$ are the Bernoulli numbers and $\mu$ really 
  means $\ad_\mu$. 
Now we apply this to $(\lam \mu)$. Being careful about the sign, we have
$\mu^{2m-1} (\lam\mu) = -\lam \leftad{\mu}^{2m}$. 
Substituting $\lam = (-1)^n r_{2n}, \mu= R_2$ and using the basic 
 formula $r_{2n} R_2 = (n+1) r_{2n+2} $ repeatedly, we easily get
\begin{align}
-{1\over (2m)!} \lam \leftad{\mu}^{2m}
& = -(-1)^n\binomial{n+2m}{n} r_{2(n+2m)} \period 
\end{align}
In this way, we finally obtain the full exponent $\mfR(=E)$ as 
\begin{align}
\mfR  &= R_2 + R_3 + \Rhat \comma \\
\Rhat &= \sum_{n \ge 2} (-1)^n r_{2n} 
+ \half \sum_{n\ge 2}(n+1) r_{2(n+1)} \nn\\
& + \sum_{n\ge 2} \sum_{m \ge 1} (-1)^{m-1} B_m 
(-1)^n\binomial{n+2m}{n} r_{2(n+2m)}  \period 
\end{align}

We can  rewrite $\Rhat$ into a more convenient form by grouping terms into 
$r_{2\cdot 2n}$ type and $r_{2\cdot (2n+1)}$ type.  We will call them ``even"
 and ``odd"  types. This separation gives 
\begin{align}
\Rhat &= \Rhat_{even} + \Rhat_{odd} \comma \\
\Rhat_{even} &= r_4 + \sum_{n\ge 2} \left( -(n-1) + \sum_{m=1}^{n-1} 
 (-1)^{m-1} B_m \binomial{2n}{2m}  \right) r_{2\cdot 2n} \comma \\
\Rhat_{odd} &= \half r_6 + \sum_{n\ge 2} \left( n-\half 
-\sum_{m=1}^{n-1} (-1)^{m-1} B_m \binomial{2n+1}{2m} 
\right) r_{2\cdot (2n+1)}  \period
\end{align}
By using appropriate  identities for the Bernoulli numbers  $B_m$, we may bring $\Rhat_{even}$ and 
$\Rhat_{odd}$  into much simpler forms. The relevant identities are 
\begin{align}
&(i)\qquad \sum_{m=1}^{n-1} 
 (-1)^{m-1} B_m \binomial{2n}{2m}   = n-1 \comma \\
&(ii)\qquad \sum_{m=1}^n (-1)^{m-1} B_m \binomial{2n+1}{2m}  = n-\half  
\period 
\end{align}
The first identity can be applied directly to $\Rhat_{even}$ and yields  a remarkable 
result $\Rhat_{even} = r_4$. As for $\Rhat_{odd}$,  the use of the second identity  reveals that all except $m=n$ term cancel in the sum. It turns out that the first term 
 $r_6/2$ can be identified with $n=1$ term of the remaning expression and hence 
 we can write the result succinctly as 
\begin{align}
\Rhat_{odd} &= \sum_{n\ge 1}(-1)^{n-1} (2n+1)B_n r_{2\cdot(2n+1)}
\period 
\end{align}
Summarizing,    $\Rhat$ is obtained as
\begin{align}
\Rhat &= r_4 + \sum_{n\ge 1}(-1)^{n-1} (2n+1)B_n r_{2(2n+1)}
\comma 
\end{align}
which was quoted in (\ref{Rhat}). 
\subsection{Proof of the theorem on finite conformal transformation }
In this appendix, we give a proof of the theorem on  fininte conformal transformation 
 stated in (\ref{thmconf}). 

To prove the theorem, it is convenient to rescale the  field $\chi$ in $T_\chi$ by  a parameter  $\lam$ and introduce  the following quantities:
\begin{align}
f(\lam,\tau) &\equiv e^{\lam T_\chi} \phi(\tau) e^{-\lam T_\chi}\comma  
\qquad 
g(\lam,\tau) \equiv e^{\lam T_\chi} \chi(\tau) e^{-\lam T_\chi} \period
\end{align}
Then, the theorem we want to show  is expressed as   $f(1,\tau) = \phi'(\tau)$.  Our strategy is to 
 make use of  a set of equations for $f(\lam,\tau)$ and $g(\lam,\tau)$,  obtained by  differentiating them with respect to $\lam$. Using the 
 commutation relations 
\begin{align}
\com{T_\chi}{\phi} &= \chi \del_\tau \phi + h \del_\tau \chi \phi \comma \qquad 
\com{T_\chi}{\chi} = \chi \del_\tau  \chi \comma 
\end{align}
which follow easily from (\ref{comLkphi}) and (\ref{comLkchi}),  
we readily obtain 
\begin{align}
&(i)\quad \del_\lam g = g\del_\tau g \comma \qquad g(0, \tau) = \chi(\tau)\comma  \\
&(ii) \quad \del_\lam f = g\del_\tau f + h  f \del_\tau g \comma
\qquad f(0, \tau) = \phi(\tau) \period
\end{align}

As the first step, we will  prove  the relation $g(1, \tau) = \chi'(\tau)$, 
namely the special case of the theorem for the field $\chi$ itself. 
Since the parameter field $\chi(\tau)$ of the conformal transformation is rescaled by $\lam$, 
the corresponding transformation of the  coordinate is given by 
$\tau' = \tau -\lam \chi(\tau)$. But since $\chi(\tau)$,  being  a primary of dimension $0$,  satisfies the property $\chi'(\tau') = \chi(\tau)$, we can rewrite this relation and  express $\tau$ in terms
 of $\tau'$ as $\tau = \tau' + \lam \chi'(\tau')$. To emphasize that  the form of the function $\chi'(\tau)$ actually depends on $\lam$, we will write this as
\begin{align}
\tau &=\tau' + \lam \chi'(\lam, \tau') \period \label{eqchi}
\end{align}
Let us now  substitute this expression of $\tau$   into the right-hand-side of the equation $ \chi'(\tau')= \chi(\tau)$. Then we get an equation where  all the  arguments are $\tau'$.  To make it look simpler, we rename $\tau'$ as $\tau$. Then, it reads 
\begin{align}
\chi'(\lam,\tau)&= 
\chi(\tau + \lam \chi'(\lam,\tau) )  \period  \label{chipeq}
\end{align}
Although it  looks like  a complicated nested functional equation characterizing $\chi'(\lam,\tau)$, we can convert it into a simpler differential equation. By 
differentiating with respect to $\tau$ and $\lam$, 
 we obtain 
\begin{align}
\del_\tau  \chi' &= \del_\tau  \chi (\tau+\lam \chi'(\lam,\tau)) (1+\lam \del_\tau  \chi')\comma  \\
\del_\lam \chi' &= \del_\tau \chi (\tau+\lam \chi'(\lam,\tau)) (\chi' 
+ \lam \del_\lam \chi') \period 
\end{align}
Taking their ratio, we get
\begin{align}
 (\del_\lam \chi') (1+\lam \del_\tau \chi') &=
(\del_\tau \chi') (\chi' + \lam \del_\lam \chi')
 \period
\end{align}
Finally, expanding this relation we easily obtain the equation 
\begin{align}
\del_\lam \chi' &= \chi'  \del_\tau  \chi'  \period
\end{align}
But this is nothing but  the differential equation $(i)$ for $g(\lam, \tau)$. Furthermore,  at $\lam=0$, we obviously have $\chi'(\tau) = \chi(\tau)$, 
which is the same initial condition satisfied by $g(\lam, \tau)$. 
Hence, we can identify  $g(\lam,\tau) =\chi'(\lam, \tau)$ and in particular $g(1, \tau) = \chi'(1, \tau) =\chi'(\tau)$. 

Next we consider the equation $(ii)$. The counterpart of the equation (\ref{chipeq}) 
for $\phi$ is 
\begin{align}
\phi'(\lam, \tau) &= \phi(\tau+\lam \chi'(\lam, \tau)) (1+\lam \del_\tau \chi')^h \period \label{eqphi}
\end{align}
Just as before, we compute $\del_\lam \phi'$ and $\del_\tau \phi'$, making use 
of the fact that $\chi' = g$. We then get
\begin{align}
\del_\lam \phi' &= (\del_\tau \phi) (g + \lam \del_\lam g) (1+\lam \del_\tau g)^h  
 + h \phi (1+\lam \del_\tau g)^{h-1} \del_\tau (g + \lam \del_\lam g)\comma 
 \\
\del_\tau \phi' &= (\del_\tau \phi) (1+\lam \del_\tau g)^{h+1}  
 + h\phi (1+\lam \del_\tau g)^{h-1}  \lam \del_\tau ^2g \period
\end{align}
Now we use the property of $g$ to rewrite $g+\lam \del_\lam g = g+\lam g \del_\tau
 g 
 = g(1+\lam \del_\tau  g)$. Also, we use (\ref{eqphi}) to identify 
$\phi (1+\lam \del_\tau  g)^{h-1}   = \phi' /(1+\lam \del_\tau  g)$.
 Then, the above equations simplify to
\begin{align}
\del_\lam \phi' &= (\del_\tau  \phi) (1+\lam \del_\tau  g)^{h+1}   g 
 +  {h\phi' \over 1+\lam \del_\tau  g}  \del_\tau (g (1+\lam \del_\tau  g)) \comma \\
\del_\tau  \phi' &= (\del_\tau  \phi) (1+\lam \del_\tau  g)^{h+1} 
 + {h\phi' \over 1+\lam \del_\tau  g}  \lam \del_\tau ^2 g \period
\end{align}
From these equations, it is not difficult to verify that the following equations hold
\begin{align}
\del_\lam \phi' &= h \del_\tau  g \phi' + g \del_\tau  \phi' \period 
\end{align}
But this is identical to the equation $(ii)$ satisfied by $f$, with the proper initial condition.  Thus we can identify $f(1, \tau) = \phi'(\tau)$, which  completes the proof of the theorem.

\end{document}